\documentclass[aphy,a4paper]{apjrnl}
\usepackage{amsmath}
\usepackage{amssymb}
\usepackage{epsfig}
\usepackage{rotating}
\equationsection
\begin{document}
%\bibliographystyle{unsrt}
%% Author, fill in article title here
\title{Semiclassical theory of quasiparticles in the superconducting state}
%% Fill in author list here
\author{K.P.Duncan and B.L.Gy\"orffy
\\H. H. Wills Physics Laboratory, 
\\University of Bristol,
\\Tyndall Avenue, 
\\Bristol BS8 1TL, UK   %As many lines as needed
\\E-mail: kevin.duncan@bris.ac.uk
}
\date{December 10, 2001} %please do not use \today, use actual date of version
%\date{\today} %please do not use \today, use actual date of version

\maketitle

\begin{abstract}  
We have developed a semiclassical approach to solving the Bogoliubov -
de Gennes equations for superconductors. It is based on the study of 
classical orbits governed by an effective Hamiltonian corresponding
to the quasiparticles in the superconducting state and includes an account
of the Bohr-Sommerfeld quantisation rule, the Maslov index, torus
quantisation, topological phases arising from lines of phase singularities 
(vortices), and semiclassical wave functions for 
multi-dimensional systems. The method is illustrated by studying the
problem of an SNS junction and a single vortex.
\end{abstract}

%%%% Authors begin text of article here %%%

%\paragraph{Paragraph Title.} This is a named \cmd{paragraph}.

%For references to the equations, use the command \cmd{eqnref} in order to obtain the correct format for the journal style, as in Eq. \eqnref{1.1} and \eqnref{eq.a1}.
%\tableofcontents
\section{Introduction}
A most convenient microscopic theory of superconductivity is provided by 
the Bogoliubov-de Gennes (BdG) equations ~\cite{degennes:89:0}. 
On the one hand it encapsulates all the
basic ideas of Bardeen, Cooper and Schrieffer and on the other hand it is
the general form of the Kohn-Sham Euler-Lagarange equations of the density
functional theory for superconductors.
Ever since its formulation semiclassical methods have been used, very 
successfully, for investigating its solutions under various 
circumstances ~\cite{degennes:89:0,andreev:64:0,andreev:66:0}. In this 
paper we wish to contribute to making this method even more effective by 
developing further the particular approach, based upon effective
classical orbits, pioneered by
Azbel' ~\cite{azbel:71:0}.

As in the case of the standard Schr\"odinger equation
there are two semiclassical approaches to the problem of solving the BdG
equations. 
The first, initiated by Andreev ~\cite{andreev:64:0,andreev:66:0}, 
de Gennes ~\cite{degennes:89:0}, and Bardeen and co workers ~\cite{bardeen:69:0},
utilises the WKB form
for the wave function and takes advantage of the slowly varying amplitudes
to convert the second order equation into one that is first order in the
derivatives and hence more readily soluble.
However the usual machinations of
Sch\"odinger quantum mechanics e.g. of matching logarithmic derivatives etc.
are still employed. 

The second approach, initiated by Azbel' ~\cite{azbel:71:0}, 
uses effective classical Hamiltonians and orbits together with 
Bohr quantisation
conditions to study the quasiparticle energy spectra and the 
corresponding wave functions.  We extend Azbel's approach by including an 
account of complex order parameters, the 
origin of the Maslov indices, the construction of a 3-dimensional wave 
function, and using torus quantisation, that is to say the full machinery of 
modern semiclassics ~\cite{gutzwiller:90:0,ozorio:88:0,brack:97:0,maslov:81:0}.

The principle motivation behind such systematic development of semiclassical
methods for superconductors is the need to adopt this powerful technique to 
the treatment of Type II superconductors and those with exotic, $p$- and 
$d$-wave, pairing. This need arises from the difficulty of solving numerically
differential equations, such as the BdG equations, which feature many 
wildly different length scales, such as lattice parameters, 
$a$, coherence length, $\xi$,
penetration depth, radii of Landau orbits, and flux lattice unit
cell sizes. We hope that the replacement of the numerical problems of
integrating differential equations by solving Hamiltons equations for 
classical orbits can alleviate some of these difficulties.

The layout of this paper is then as follows. In section 
\ref{sec:SemiclassicalTheoryForSuperconductors} we present a multicomponent
semiclassical theory for the BdG equations. This starts with a discussion of 
the general
form of the semiclassical ansatz before showing how the BdG equations can be
reduced,
at zeroth order, to the solution of a pair of Hamiltonian systems describing
the dynamics of quasiparticles along orbits comprised of both particle-like
and hole-like segments. We also
include a discussion of topological phases arising from singularities 
in the phase of pairing potential. We then solve the first order equations 
and construct the general multicomponent wave function.  We briefly discuss
torus quantisation and demonstrate that EBK quantisation conditions (a
generalisation of Bohr-Sommerfeld quantisation rules) can 
only be constructed in the absence of the above mentioned singularities. 
In section \ref{sec:EffectiveSemiclassicalTheoryForSuperconductors}
we construct an effective semiclassical theory whose Hamiltonian systems are
$\hbar$-dependent. By extending the semiclassical theory in this way we
remove the obsticles to constructing the EBK rules and consequently we 
present a generalised EBK rule which includes the contribution from both the
Maslov index and topological phases. We conclude the section with a discussion
of the interpretation of a theory based upon $\hbar$-dependent Hamiltonian
systems. The remaining parts of the paper are devoted to two applications. Thus
in section \ref{sec:SNS_junction} we apply the theory to a 
superconductor-normal metal-superconductor junction and include a discussion
of Andreev retro-reflection and the semiclassical spectrum. In an accompanying
appendix we calculate the Maslov index by analytically continuing the 
semiclassical wave functions taking into account of Stokes phenomenon. Then in
section \ref{sec:theSingleVortex} we investigate a single vortex which is
an ideal example for demonstrating the r\^ole of the $\hbar$-dependence
in the Hamiltonians and the consequences of the phase singularities in the 
pairing potential. The paper concludes with a discussion and summary.
%------------------------------------------------------------------------------
\section{Semiclassical theory for superconductors}
\label{sec:SemiclassicalTheoryForSuperconductors}
\subsection{General form for the multicomponent WKB ansatz}
To begin, let us recall the conventional
semiclassical ~\cite{andreev:64:0,andreev:66:0,degennes:89:0,bardeen:69:0} 
BCS theory of the superconducting state.
The context for the most general version of this is provided by the 
Bogoliubov-de Gennes equations for the two component wave function
for a quasiparticle:
\begin{equation}
	\left(
		\begin{array}{cc}
			H(
				\hat{\bf p},
				{\bf r}
			) &
			|\Delta({\bf r})| e^{i\phi({\bf r})} \\
			|\Delta({\bf r})| e^{-i\phi({\bf r})} &
			-H^*(
				\hat{\bf p},
				{\bf r}
			)
		\end{array}
	\right)
	\left(
		\begin{array}{c}
			u_\lambda ({\bf r}) \\
			v_\lambda ({\bf r}) 
		\end{array}
	\right)
	=
	E_\lambda
	\left(
		\begin{array}{c}
			u_\lambda ({\bf r}) \\
			v_\lambda ({\bf r}) 
		\end{array}
	\right),
\label{eqn:BdGequations}	
\end{equation}
where
\begin{equation}
	H(\hat{\bf p},{\bf r})
	=
	\frac{1}{2m}
	\left(
		\hat{\bf p} +e{\bf A}({\bf r})
	\right)^2
	+ V({\bf r}) -\epsilon_F ,
\label{eqn:HamiltonianOperator}
\end{equation}
$\epsilon_F$ is the Fermi energy, $V({\bf r})$ is an
external potential and
$\Delta({\bf r})=|\Delta({\bf r})| e^{i\phi({\bf r})}$ is the complex 
pairing potential satisfying the self-consistency condition
\begin{equation}
	\Delta({\bf r})
	=
	g\sum_\lambda u_\lambda ({\bf r})v^*_\lambda ({\bf r})
	\left( 1-2 f(E_\lambda) \right).
\label{eqn:DeltaSelfConsistency}
\end{equation}
(Here $g$ is the BCS coupling constant and $f(E_\lambda)$ is the Fermi
function.) As usual, the interpretation of the components 
$u_\lambda ({\bf r})$ and $v_\lambda ({\bf r})$ is that they are the 
amplitudes for the quasiparticle being a particle and a hole respectively.
Moreover, a state is superconducting if the solution of equations
(\ref{eqn:BdGequations}), (\ref{eqn:HamiltonianOperator}),
and (\ref{eqn:DeltaSelfConsistency}) is such that $\Delta({\bf r})\ne 0$.

In what follows, for a prescribed pairing potential $\Delta({\bf r})$, we
develop a multicomponent WKB approximation for solving equation
(\ref{eqn:BdGequations}). Asymptotically, as $\hbar \rightarrow 0$,
the order of the multicomponent differential equation changes abruptly, just
as for a single component equation, and therefore, we expect the solutions to
exhibit the familiar essential singularity in the form of the phase 
$e^{iS({\bf r}) / \hbar}$.
Thus a reasonable guess at the multicomponent generalisation of the WKB wave 
function would be
\begin{eqnarray}
	\left(
		\begin{array}{c}
			u_\lambda ({\bf r}) \\
			v_\lambda ({\bf r}) 
		\end{array}
	\right)
	=
	\left(
		\begin{array}{c}
			\tilde{u}_\lambda ({\bf r}) \\
			\tilde{v}_\lambda ({\bf r}) 
		\end{array}
	\right)
	e^{
	   \frac{i}{\hbar}
	   S_0({\bf r})
	}
	\left(
		1+{\cal O}(\hbar)
	\right)
	\quad
	(\mathrm{guess})
\label{eqn:guess}
\end{eqnarray}
where $\tilde{u}_\lambda ({\bf r})$ and $\tilde{v}_\lambda ({\bf r})$
are slowly varying amplitudes.
However, to be
systematic, it is more convenient to start with the general complex spinor:
\begin{eqnarray}
	\left(
		\begin{array}{c}
			u_\lambda ({\bf r}) \\
			v_\lambda ({\bf r}) 
		\end{array}
	\right)
	=
	\left(
		\begin{array}{c}
			e^{i
				\left(
					S({\bf r})+\Sigma({\bf r})
				\right)
			} \\
			e^{i
				\left(
					S({\bf r})-\Sigma({\bf r})
				\right)
			}
		\end{array}
	\right),
\label{eqn:complexspinor}
\end{eqnarray}
where both $S({\bf r})$ and $\Sigma({\bf r})$ are complex
quantities to allow for the amplitude and phase variation of each of
the components.
To proceed we expand them in powers of $\hbar$ by writing $S$ and $\Sigma$ as
\begin{align}
	S({\bf r})
	& =
	\frac{ S_0({\bf r}) }{\hbar}
	+S_1({\bf r}) + \hbar S_2({\bf r}) + \dotsb,
\label{eqn:expandedS} \\
\intertext{and}
	\Sigma({\bf r})
	& =\frac{ \Sigma_0({\bf r}) }{\hbar}
	+\Sigma_1({\bf r}) + \hbar \Sigma_2({\bf r}) + \dotsb.
\label{eqn:expandedSigma}
\end{align}
In keeping with the original definition of $S$ and $\Sigma$ as complex
quantities, at every order the real ($r$)
and imaginary ($i$) parts of $S_j$ and
$\Sigma_j$ must be determined. Up to and including the first order
in $\hbar$ the spinor in equation (\ref{eqn:complexspinor}) then contains the 
following quantities
\begin{eqnarray}
	\left(
		\begin{array}{c}
			u_\lambda ({\bf r}) \\
			v_\lambda ({\bf r}) 
		\end{array}
	\right)
	=
	\left(
		\begin{array}{c}
			e^{i
				\left(
					\Sigma_0^r({\bf r})/\hbar
					+\Sigma_1^r({\bf r})
				\right)
			}
			e^{-\Sigma_1^i({\bf r})} \\
			e^{-i
				\left(
					\Sigma_0^r({\bf r})/\hbar
					+\Sigma_1^r({\bf r})
				\right)
			}
			e^{+\Sigma_1^i({\bf r})}
		\end{array}
	\right) 
	e^{i
		S_0^r({\bf r})/\hbar
		+i S_1^r({\bf r})
	}
	e^{-S_1^i({\bf r})}.
\label{eqn:ComplexSpinorToOrderh}
\end{eqnarray}
The question of whether to include $\Sigma_0^r$
(i.e. a fast degree of freedom for the spinor components) is a subtle
one. If $\phi({\bf r})$ is not expanded
in $\hbar$ (and it is usual in semiclassical theory to regard potentials
such as $V({\bf r})$ and $\Delta({\bf r})$ as externally imposed potentials -
hence not to expand them) then it is found that the BdG equations have no
expansion in $\hbar$ if $\Sigma_0^r({\bf r})\neq 0$. We will return to this
point again once we have developed our formalism.

Setting $\Sigma_0^r({\bf r})=0$ in (\ref{eqn:ComplexSpinorToOrderh})
we see the wave function does indeed take the form (\ref{eqn:guess})
where however  
\begin{math}
	\left( \begin{smallmatrix}
			\tilde{u}_\lambda ({\bf r}) \\
			\tilde{v}_\lambda ({\bf r})
		\end{smallmatrix}
	\right)
\end{math}
is understood to be the {\it complex} slowly varying spinor:
\begin{eqnarray}
	\left(
		\begin{array}{c}
			\tilde{u}_\lambda ({\bf r}) \\
			\tilde{v}_\lambda ({\bf r}) 
		\end{array}
	\right)
	=
	\left(
		\begin{array}{c}
			u_{0,\lambda} ({\bf r})
			e^{i\Sigma^r_1({\bf r})}\\
			v_{0,\lambda} ({\bf r})
			e^{-i\Sigma^r_1({\bf r})}
		\end{array}
	\right)
	e^{iS_1^r({\bf r})}
	e^{-S_1^i({\bf r})},
\label{eqn:correctedguess}
\end{eqnarray}
where $u_{0,\lambda} ({\bf r})$, $v_{0,\lambda} ({\bf r})$, 
and $e^{-S_1^i({\bf r})}$ are amplitudes, 
and $S_0^r({\bf r})$, $S_1^r({\bf r})$ and 
$\Sigma^r_1({\bf r})$
are phases, to be determined by solving the appropriate zeroth order and 
first order equations obtained by substituting (\ref{eqn:correctedguess})
into the BdG equations and expanding the result in powers of $\hbar$.
%---------------------------------------------------------------------------%
\subsection{Semiclassical solution of the BdG equations to zeroth order}
Noting that $e^{iS_0({\bf r})/\hbar}$ may be regarded
%--------------beginfootnote----------------------------------------------
as a unitary operator\footnote{Since
we will not need the $r$ suffix on $S_0^r({\bf r})$ for what follows we
drop it from here onwards.}
%-------------endfootnote-------------------------------------------------
we may write, in the position representation,
\begin{equation}
	e^{-iS_0({\bf r})/\hbar}
	\hat{\bf p}
	e^{iS_0({\bf r})/\hbar}
	=
	\hat{\bf p} +
	\frac{\partial S_0}{\partial {\bf r}}.
\end{equation}
Applying this procedure to both sides of equation 
(\ref{eqn:BdGequations}) one readily finds
\begin{eqnarray}
	\left(
		\begin{array}{cc}
			H(
				\hat{\bf p} +
				\frac{\partial S_0}{\partial {\bf r}},
				{\bf r}
			) &
			|\Delta({\bf r})| e^{i\phi({\bf r})} \\
			|\Delta({\bf r})| e^{-i\phi({\bf r})} &
			-H^*(
				\hat{\bf p} +
				\frac{\partial S_0}{\partial {\bf r}},
				{\bf r}
			)
		\end{array}
	\right)
	\left(
		\begin{array}{c}
			\tilde{u}_\lambda ({\bf r}) \\
			\tilde{v}_\lambda ({\bf r}) 
		\end{array}
	\right)
	=
	E_\lambda
	\left(
		\begin{array}{c}
			\tilde{u}_\lambda ({\bf r}) \\
			\tilde{v}_\lambda ({\bf r}) 
		\end{array}
	\right).
\label{eqn:UnitaryTransformedBdGequations}	
\end{eqnarray}
Then, the zeroth order approximation in $\hbar$ is obtained by neglecting
terms containing $\hbar$ (i.e., $\hat{\bf p}=-i\hbar{\pmb \nabla}$, whose
action upon $\tilde{u}_\lambda ({\bf r})$, $\tilde{v}_\lambda ({\bf r})$ is 
small). Then we observe that $\phi({\bf r})$ can be removed from the
above equation at each
${\bf r}$-point by a unitary transformation
\begin{eqnarray}
	U_\phi({\bf r}) =
	\left(
		\begin{array}{cc}
			e^{-i\phi({\bf r})/2} & 0           \\
			0 	     & e^{i\phi({\bf r})/2}
		\end{array}
	\right).
\label{eqn:SingularUnitaryMatrix}
\end{eqnarray}
To zeroth order in $\hbar$ such transformations of equation 
(\ref{eqn:UnitaryTransformedBdGequations}) yield
{\setlength{\multlinegap}{0pt}
\begin{multline}
	\left(
		\begin{array}{cc}
			E_\lambda -H^e_0({\bf p}_0,{\bf r}) &
			-|\Delta ({\bf r})| \\
			-|\Delta ({\bf r})|  &
			E_\lambda+H^h_0({\bf p}_0,{\bf r})
		\end{array}
	\right)
	\left(
		\begin{array}{c}
			\tilde{u}_\lambda ({\bf r}) e^{-i\phi({\bf r})/2} \\
			\tilde{v}_\lambda ({\bf r}) e^{+i\phi({\bf r})/2}
		\end{array}
	\right)
	=0.
\label{eqn:RealZerothOrderBdGequations}	
\end{multline}}
Here $H^e_0({\bf p}_0,{\bf r})$ is an electron hamiltonian of the form
(\ref{eqn:HamiltonianOperator}) with the momentum operator $\hat{\bf p}$
replaced by ${\bf p}_0=\frac{\partial S_0}{\partial {\bf r} }$, and 
the same is true for the hole Hamiltonian,
$H^h_0({\bf p}_0,{\bf r})$, except $e\rightarrow -e$.
Since the transformed Hamiltonian matrix 
in equation (\ref{eqn:RealZerothOrderBdGequations}) is real, 
the spinor wave function is also constrained. Most generally, via equation
(\ref{eqn:correctedguess}), this implies that
\begin{equation}
	\Sigma_1^r ({\bf r})
	-
	\frac{\phi({\bf r})}{2}
	=
	n\frac{\pi}{2},
\label{eqn:SigmaInTermsOfPhiNonSelfConsistent}
\end{equation}
for $n$ an integer. However self-consistency, 
equation (\ref{eqn:DeltaSelfConsistency}), requires that the 
$\Delta({\bf r})$ with which we started our calculation, i.e.
$\Delta({\bf r})=\left| \Delta({\bf r}) \right| e^{i\phi({\bf r})}$,
and the $\Delta({\bf r})$ constructed from our solutions $u_\lambda({\bf r})$
and $v_\lambda({\bf r})$ be one and the same. Since 
$u_\lambda({\bf r}) v^*_\lambda({\bf r}) \propto e^{i\phi({\bf r})+in\pi}$,
we must have $n=2m$, $m$ an integer, for this to be true. Thus we find
\begin{equation}
	\Sigma_1^r ({\bf r})
	-
	\frac{\phi({\bf r})}{2}
	=
	m\pi.
\label{eqn:SigmaInTermsOfPhi}
\end{equation}
But now, since $m$ introduces the 
same sign change for both upper and lower components of the spinor it can be
factored out. Choosing the sign of the wave function appropriately $m$ is
eliminated and we have determined $\Sigma_1^r({\bf r})$.

In the context of a general solution of the BdG equations it has been observed
that if the order parameter, $\Delta({\bf r})$, contains vortices, 
${\pmb \nabla}\phi({\bf r})$ acquires
{\it global} curvature ~\cite{vafek:00:0} i.e
\begin{equation}
	\oint_c 
	{\pmb \nabla}\phi({\bf r})\cdot
	d{\bf r}
	\neq
	0,
\label{eqn:globalcurvature}
\end{equation}
for paths, $c$, containing vortices. To clarify the nature of 
the singularity associated with a vortex in three dimensions
observe that it is one where $|\Delta({\bf r})|=0$ along lines. 
$\phi({\bf r})$ is then 
indeterminate i.e. $|\Delta({\bf r})|=0$ defines 
a line of phase singularities ~\cite{pismen:99:0}.
If we continue $\phi({\bf r})$ along any path enclosing such a singularity
we obtain (\ref{eqn:globalcurvature}) with $2\pi$ on the right hand side.
A unitary transformation of the form $U_\phi$ is then
{\it multivalued}, as stressed by Anderson ~\cite{anderson:99:0}.
For instance $U_\phi$ has two branches and moves from one to the other
when continued around a vortex. The same is true for our spinor: if upon the
first branch it takes the values
\begin{equation}
	\left(
		\begin{array}{c}
			\tilde{u}_\lambda ({\bf r}) \\
			\tilde{v}_\lambda ({\bf r}) 
		\end{array}
	\right),
\nonumber 
\end{equation}
then, after a circuit around a vortex, it moves onto the second branch,
taking the values
\begin{equation}
	-\left(
		\begin{array}{c}
			\tilde{u}_\lambda ({\bf r}) \\
			\tilde{v}_\lambda ({\bf r}) 
		\end{array}
	\right).
\nonumber
\end{equation}
(A second trip around the vortex is required to return the spinor to 
the original branch.)
Thus, through (\ref{eqn:SigmaInTermsOfPhi}), we have discovered that 
$\Sigma_1^r({\bf r})$ may contain a contribution topological in 
%-------------footnote-----------------------------------------------------
origin\footnote{It is the singular part of $\phi({\bf r})$
which \protect{has been called  ~\cite{vafek:00:0} 
a ``Berry'' phase although}
it is in fact simpler depending only upon the topology not the geometry of
the path along which it is continued.}.
%----------endfootnote-----------------------------------------------------
This is, however, not a problem for our semiclassical theory.
With $\phi({\bf r})$ removed we must diagonalise the resulting matrix
subject to the physical condition that the original wave 
function in (\ref{eqn:guess}) is single-valued.
As we shall see presently, this constraint will modify the 
generalised Bohr-Sommerfeld quantisation 
rule which we will derive.

A non-trival solution to (\ref{eqn:RealZerothOrderBdGequations}) requires
the vanishing of the determinant. This yields two equations:
\begin{equation}
	E^\alpha_\lambda
	=
	E_0^{\alpha}({\bf p}_0,{\bf r}),
\label{eqn:HamiltonJacobiSemiclassical}
\end{equation}
where $\alpha=\pm$, and
\begin{equation}
	E_0^{\alpha}({\bf p}_0,{\bf r})
	=
	{\bf p}_0\cdot{\bf v}_0({\bf r})
	+\alpha
	\sqrt{
		\left(
			\frac{p^2_0}{2m}+\frac{1}{2}mv_0^2({\bf r}) 
				+ V({\bf r}) -\epsilon_F
		\right)^2
		+ |\Delta({\bf r})|^2
	},
\label{eqn:ClassicalHamiltonian}
\end{equation}
where ${\bf v}_0=e{\bf A}({\bf r})/m$. The constancy of 
$E_0^{\alpha}({\bf p}_0,{\bf r})$ defines, implicitly, the functions
${\bf p}_0^{\alpha}({\bf r})=\frac{\partial S_0^{\alpha}}{\partial {\bf r} }$.

We see immediately
that each of the equations (\ref{eqn:HamiltonJacobiSemiclassical})
is a (stationary)
{\it Hamilton-Jacobi equation} for  a (fictitous) classical mechanics. Its
solution (when it exists)
is a classical action function $S_0^\alpha({\bf r},{\bf I})$, where
$I_1,\dots,I_n$ is a set of action integrals and appears in place of
$\lambda$ which labelled the eigenvalues of the BdG equation.

\begin{figure}[htbp]
	\begin{center}
		\epsfig{figure=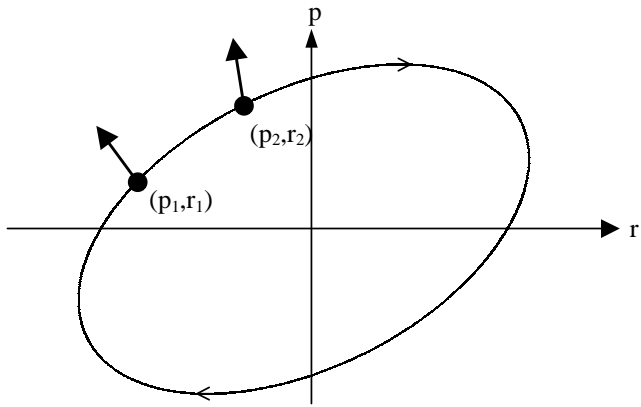,clip=,width=86mm}
		\caption{Phase space orbits 
			$({\bf p}_0({\bf r},{\bf I}),{\bf r})$
			defined implicitly by 
			$E_{\bf I}^\alpha=E_0^\alpha({\bf p}_0,{\bf r})$. 
			The vector attached to the orbit changes orientation
			as the spinor is carried along the trajectory.
			Notice that
			${\bf p}_0^\alpha({\bf r},{\bf I})$ 
			is a many-valued funciton of ${\bf r}$.}
		\label{fig:multivaluedP}
	\end{center}
\end{figure}

Thus viewing each $E_0^{\alpha}({\bf p}_0,{\bf r})$ as a classical Hamiltonian
governing the propagation of a quasiparticle excitation the restriction to a
constant energy shell in phase space 
($E_{\bf I}^\alpha=E_0^{\alpha}({\bf p}_0,{\bf r})$)
define the phase space orbits 
which we shall label $({\bf p}_0({\bf r},{\bf I}),{\bf r})$.
In general the functions ${\bf p}_0^{\alpha}({\bf r},{\bf I})$ 
are many-valued functions
of ${\bf r}$ as illustrated in FIG.~\ref{fig:multivaluedP}, whose branches
shall be indexed by $j$ when necessary.
%--------------Figure to be inserted here-------------------------------
%---------------endfigure-------------------------------------------------
Mathematically ~\cite{arnold:89:0} finding a complete integral to the 
Hamilton-Jacobi equation is equivalent to solving Hamilton's 
equations of motion. So we can write down Hamilton's equations for the 
quasiparticle excitation:
\begin{equation}
	\dot{\bf r}^\alpha
	=
	\left(
		\frac{\partial E_0^\alpha ({\bf p},{\bf r}) }
			{ \partial {\bf p}}
	\right),
	\quad
	\dot{\bf p}^\alpha
	=
	-\left(
		\frac{\partial E_0^\alpha ({\bf p},{\bf r}) }
			{ \partial {\bf r}}
	\right),
\label{eqn:HamiltonsEquations}
\end{equation}
where after differentiation we must substitute ${\bf r}={\bf r}^\alpha(t)$
and ${\bf p}={\bf p}_0^\alpha ({\bf r}^\alpha(t))$. Since our Hamiltonian is
conservative, time here is simply 
a parameter governing the evolution of a quasiparticle along the orbit.
Thus the $2\times 2$ differential matrix BdG equation (\ref{eqn:BdGequations})
has been reduced, at zeroth order, to
a pair of Hamiltonian systems for  a (fictitous) classical mechanics specified
by the appropriate Hamiltonian in (\ref{eqn:ClassicalHamiltonian}). 

%For each initial condition $S_0^{\alpha,j}({\bf r}_0,{\bf I})$ 
A solution 
to (\ref{eqn:HamiltonJacobiSemiclassical}) is obtained as
\begin{equation}
	S_0^{\alpha,j}({\bf r},{\bf I})
	=
	\int^{{\bf r}(t)}_{{\bf r}_0(t_0)}
	{\bf p}_0^{\alpha,j}({\bf r},{\bf I})\cdot d{\bf r},
\label{eqn:ActionFunctionSemiclassical}
\end{equation}
where integration is carried out along a ray of one of the Hamiltonian systems
(\ref{eqn:HamiltonsEquations}). (For convenience we have chosen the initial
condition $S_0^{\alpha,j}({\bf r}_0,{\bf I})=0$.)
This does not complete the zeroth order theory 
however, since our `particle' has an internal
structure which is represented by the complex spinor defined at every point
along the trajectory.

So let us now consider
the spinor solution of the BdG equation corresponding 
to one of these `classical' orbits. It is useful to 
view the slowly changing quasiparticle spinor 
of (\ref{eqn:RealZerothOrderBdGequations}) 
as an axis of pseudo spin quantisation transported along the 
trajectory defined by ${\bf p}_0^{\alpha}({\bf r},{\bf I})$.
Clearly, this axis can be represented by a vector attached
to the orbit (FIG.~\ref{fig:multivaluedP}) whose orientation changes as it
is carried along the path in phase space. This vector represents the 
internal particle-hole degree of freedom of the excitation.
To find how the eigenvectors of (\ref{eqn:RealZerothOrderBdGequations})
change as they are carried along the trajectory 
%defined by ${\bf p}_0^{\alpha}({\bf r},{\bf I})$ 
we diagonalise the Hamiltonian matrix, $H({\bf r})$, at every ${\bf r}$-point
\begin{eqnarray}
	U({\bf r})H({\bf r})U^\dagger({\bf r})
	=
	\left(
		\begin{array}{cc}
			E_0^+({\bf p}_0,{\bf r}) & 0 \\
			0 & E_0^-({\bf p}_0,{\bf r})		
		\end{array}
	\right),
\end{eqnarray}
so that in the locally diagonal frame (\ref{eqn:RealZerothOrderBdGequations})
becomes
\begin{eqnarray*}
	\left(
		\begin{array}{cc}
			E_0^+({\bf p}_0,{\bf r}) & 0 \\
			0 & E_0^-({\bf p}_0,{\bf r})		
		\end{array}
	\right)
	\left(
		\begin{array}{c}
			u^{\mathrm{diag}} \\
			v^{\mathrm{diag}}
		\end{array}
	\right)
	= E_{{\bf I}}
	\left(
		\begin{array}{c}
			u^{\mathrm{diag}} \\
			v^{\mathrm{diag}}
		\end{array}
	\right),
\end{eqnarray*}
giving
\begin{align}
	E^+_{{\bf I}}=E_0^+({\bf p}_0,{\bf r}),\qquad &
	\left(
		\begin{array}{c}
			u^{\mathrm{diag},+} \\
			v^{\mathrm{diag},+} 
		\end{array}
	\right)
	=
	\left(
		\begin{array}{c}
			1 \\
			0
		\end{array}
	\right), 
\label{eqn:localDiagonalResults} \\
	E^-_{{\bf I}}=E_0^-({\bf p}_0,{\bf r}),\qquad &
	\left(
		\begin{array}{c}
			u^{\mathrm{diag},-} \\
			v^{\mathrm{diag},-} 
		\end{array}
	\right)
	=
	\left(
		\begin{array}{c}
			0 \\
			1
		\end{array}
	\right).
\label{eqn:localDiagonalResults2}
\end{align}
Rotating back to the common laboratory frame we have two solutions
\begin{align}
	E^+_{{\bf I}}=E_0^+({\bf p}_0,{\bf r}),\qquad &
	\left(
		\begin{array}{c}
			u_{0,{\bf I}}^+ ({\bf r}) \\
			v_{0,{\bf I}}^+ ({\bf r}) 
		\end{array}
	\right)
	=
	\ U^\dagger({\bf r})
	\left(
		\begin{array}{c}
			1 \\
			0
		\end{array}
	\right), 
\nonumber \\
	E^-_{{\bf I}}=E_0^-({\bf p}_0,{\bf r}),\qquad &
	\left(
		\begin{array}{c}
			u_{0,{\bf I}}^- ({\bf r}) \\
			u_{0,{\bf I}}^- ({\bf r})
		\end{array}
	\right)
	=
	\ U^\dagger({\bf r})
	\left(
		\begin{array}{c}
			0 \\
			1	
		\end{array}
	\right).
\nonumber
\end{align}
Note that while
\begin{math}
	\left( \begin{smallmatrix}
			u_{0,{\bf I}}^+ ({\bf r}) \\
			v_{0,{\bf I}}^+ ({\bf r})
		\end{smallmatrix}
	\right)
\end{math}
is a particle in its local frame, (\ref{eqn:localDiagonalResults}), it is
both particle and hole, with amplitudes $u_{0,{\bf I}}^+ ({\bf r})$
and $v_{0,{\bf I}}^+ ({\bf r})$, in the laboratory frame.

If we now substitute $E^\alpha_0$ into
(\ref{eqn:RealZerothOrderBdGequations})
we have the ratio
\begin{equation}
	\frac{v_{0,{\bf I}}^\alpha({\bf r})}{u_{0,{\bf I}}^\alpha({\bf r})}
	=
	\frac{E_0^\alpha -H_0^e({\bf p}_0,{\bf r})}
		{-|\Delta({\bf r})|}.
\label{eqn:RatioOfAmplitudes}
\end{equation}
The electron Hamiltonian, $H_0^e({\bf p},{\bf r})$ is
\begin{align}
	H_0^e({\bf p}_0,{\bf r})
	&= 
	{\bf p}_0\! \cdot \! {\bf v}_0 +
	\frac{{\bf p}_0^2}{2m} + \frac{1}{2}m{\bf v}_0^2 
	+V({\bf r}) - \epsilon_F,
\nonumber \\
	&=
	{\bf p}_0\! \cdot \!  {\bf v}_0 +
	\beta
	\sqrt{
		\left(
			E_0^\alpha({\bf p}_0,{\bf r}) 
			- 
			{\bf p}_0\! \cdot \!{\bf v}_0
		\right)^2 
		-|\Delta({\bf r})|^2
	     },
\label{eqn:ElectronHamiltonianSNS}
\end{align}
by (\ref{eqn:ClassicalHamiltonian}), where $\beta=\pm$, and 
the Hamilton-Jacobi equation 
$E^\alpha_{\bf I}=E^\alpha_0({\bf p}_0,{\bf r})$ should be used.
Using this to eliminate $H_0^e({\bf p},{\bf r})$ from 
(\ref{eqn:RatioOfAmplitudes}) the amplitudes are expressed in terms 
of $E_0^\alpha$ and the momentum branches, ${\bf p}_0^j({\bf r})$.
It is clear from equation (\ref{eqn:ElectronHamiltonianSNS}) 
that each ${\bf p}_0^j$ determines a unique choice of $\beta$ (see 
also Appendix \ref{app:BetajDependence}) so that there is one pair of
amplitudes for each branch of the momentum. Then the 
normalised amplitudes are given by
\begin{align}
	u_{0,{\bf I}}^{\alpha,j} ({\bf r})
	&=
	\sqrt{
		\frac{1}{2}
		\left(\textstyle
			1+\beta
			  \frac{	
				\sqrt{ 
					\left( 
						E_{\bf I}^\alpha
						-
						{\bf p}_0^j\cdot{\bf v}_0
					\right)^2
					-
					\left| \Delta({\bf r}) \right|^2
				     }
			      }
			      {
				E_{\bf I}^\alpha-{\bf p}_0^j\cdot{\bf v}_0
			      }
		\right)
	     },
\nonumber \\
	v_{0,{\bf I}}^{\alpha,j} ({\bf r})
	&=
	\sqrt{
		\frac{1}{2}
		\left(\textstyle
			1-\beta
			  \frac{
				\sqrt{ 
					\left( 
						E_{\bf I}^\alpha
						-
						{\bf p}_0^j\cdot{\bf v}_0
					\right)^2
					-
					\left| \Delta({\bf r}) \right|^2
				     }
			      }
			      {
				E_{\bf I}^\alpha-{\bf p}_0^j\cdot{\bf v}_0
			      }
		\right)
	     }.
\label{eqn:NormalisedAmplitudes}
\end{align}
(By normalising we have of course 
redefined $e^{-S_1^i({\bf r},{\bf I})}$ but since we are yet to determine this
amplitude we shall not introduce new notation for it.) 
This completes the zeroth order theory.

To summarise our results so far we have shown 
that a general semiclassical solution 
for a multicomponent system can be written as a spinor multiplying the
familiar phase, $e^{iS_0/\hbar}$ (\ref{eqn:guess}), where however the 
spinor is in general a complex quantity (\ref{eqn:correctedguess}).
Using this solution we derived the zeroth order matrix equation 
(\ref{eqn:RealZerothOrderBdGequations}) and determined one of the spinor 
phases, $\Sigma^r_1$, which may have a topological contribution. 
We then diagonalised the matrix equations reducing the solution of the 
BdG equations at zeroth order to the problem of solving a pair
of Hamiltonian systems for our excitation specified by the 
classical Hamiltonians $E_0^\alpha({\bf p}_0,{\bf r})$. Our excitation
has an internal structure which, at zeroth order, is represented by the
amplitudes (\ref{eqn:NormalisedAmplitudes}).

What is the next step? 
We expect, to first order in $\hbar$, to find a transport 
equation whose solution furnishes us with the amplitude 
$A({\bf r})=e^{-S_1^i({\bf r},{\bf I})}$. 
We also have the phase $S_1^r({\bf r},{\bf I})$
to determine which, like $\Sigma_1^r({\bf r})$,
is not familiar from single-component WKB analysis. Once we have determined
these quantities our semiclassical wave function will be used to construct
the rule which quantises our classical dynamics.
%---------------------------------------------------------------------------
\subsection{The transport equation and other first order quantities}
\label{sec:SemiclassicalFirstOrderQuantities}

To derive the first order equations we take the expectation value of 
equation (\ref{eqn:UnitaryTransformedBdGequations}) and expand the result in 
powers of $\hbar$. One can then show (see Appendix \ref{sec:appendixA})
that there are two equations to first order in $\hbar$:
\begin{align}
	{\cal O}\left(\hbar\right)\qquad
	{\pmb \nabla}
	\cdot
	\left\{
	\left(
		\begin{array}{cc}
			\tilde{u}^*_{\bf I}  &
			\tilde{v}^*_{\bf I} 
		\end{array}
	\right)
	\left(
		\begin{array}{c}
			-\frac{{\bf P}^+}{2m}
			\tilde{u}_{\bf I}  \\
			+\frac{{\bf P}^-}{2m}
			\tilde{v}_{\bf I} 
		\end{array}
	\right)
	\right\} 
	& = 0,
\label{eqn:TransportEquation} \\
	{\cal O}\left(\hbar\right)\qquad
	2\; \mathrm{Im}
	\left(
		\begin{array}{cc}
			\tilde{u}^*_{\bf I}  &
			\tilde{v}^*_{\bf I} 
		\end{array}
	\right)
	\left(
		\begin{array}{c}
			-\frac{{\bf P}^+}{2m}
			\cdot
			{\pmb \nabla}
			\tilde{u}_{\bf I}  \\
			+\frac{{\bf P}^-}{2m}
			\cdot
			{\pmb \nabla}
			\tilde{v}_{\bf I} 
		\end{array}
	\right)
	& = 0,
\label{eqn:FirstOrderPhasesEquation}
\end{align}
where ${\bf P}^\pm={\bf p}_0({\bf r})\pm e{\bf A}({\bf r})$.
The first of these equations, (\ref{eqn:TransportEquation}),
is the transport equation and can be rewritten 
(see Appendix \ref{sec:appendixA}) as
\begin{equation}
	{\pmb \nabla}
	\cdot
	\left(
		e^{-2S^i_1({\bf r},{\bf I})}
		\left.
			\frac{\partial E_0^\alpha ({\bf p},{\bf r}) }
			     {\partial{\bf p}}
		\right|_{{\bf p}={\bf p}_0^\alpha({\bf r},{\bf I})}
	\right)
	= 0.
\label{eqn:AmplitudeTransportEquation}
\end{equation}
It expresses the fact that the product 
$(\mathrm{amplitude})^2 \times \mathrm{velocity}$, at each point, must
be conserved (there are no sources or sinks). Thus if the local velocity
of a quasiparticle increases the probability of finding it there decreases,
and vice versa. Solving equation 
(\ref{eqn:AmplitudeTransportEquation}) by the van Vleck ~\cite{vanvleck:28:0}
method gives
\begin{equation}
	e^{-S_1^i({\bf r},{\bf I})}
	=
	c\left|
		\det
		\frac{ 
			\partial^2 S_0^{\alpha,j}({\bf r},{\bf I}) 
		     }
		     {
			\partial {\bf r} \partial {\bf I}
		     }
	\right|^{1/2},
\label{eqn:VanVleckDet}
\end{equation}
where $c$ is a constant.

The second equation is more troublesome. 
It resembles the connection enforcing parallel transport in Berry's theory of 
geometric phases ~\cite{berry:84:0,robbins:97:0}. 
However, equation (\ref{eqn:FirstOrderPhasesEquation}) cannot be solved to
yield a geometric phase. Rather we find (see Appendix \ref{sec:appendixA})
\begin{equation}
	S_1^r({\bf r})
	=
	-\frac{1}{e}
	\int^t_{t_0}
	{\bf j}^{\alpha,j}({\bf r})
	\cdot
	\frac{ {\pmb \nabla}\phi({\bf r}) }{2}
	dt',
\label{eqn:FirstOrderPhase}
\end{equation}
where
\begin{equation}
	-\frac{1}{e}
	{\bf j}^{\alpha,j}({\bf r})
	=
	\frac{{\bf p}_0^{\alpha,j}({\bf r})}{m}
	+
	\left(
		(u^{\alpha,j}_{0,{\bf I}}({\bf r}))^2
		-
		(v^{\alpha,j}_{0,{\bf I}}({\bf r}))^2
	\right)
	\frac{e{\bf A}({\bf r}) }{m}.
\nonumber
\end{equation}
Equation (\ref{eqn:FirstOrderPhase}) is not a line integral. A 
physical interpretation to $S_1^r({\bf r})$ is given in Appendix
\ref{sec:appendixA}.
We should not be surprised that our theory contains both a topological
phase and another phase which cannot be expressed either geometrically or 
topologically. In a different context, 
it is known ~\cite{littlejohn:91:0,yabana:86:0,emmrich:96:0} 
that the asymptotics solutions
of matrix differential equations can contain such terms. 

Bringing all our results together, the multicomponent semiclassical 
wave functions for the Bogoliubov-de Gennes equations
with action $S_0^{\alpha,j}({\bf r},{\bf I})$ take the form
{\setlength{\multlinegap}{0pt}
\begin{multline}
	\left(
		\begin{array}{c}
			u_{{\bf I}}^\alpha ({\bf r}) \\
			v_{{\bf I}}^\alpha ({\bf r})
		\end{array}
	\right)
	=
	A^\alpha_j
	\left|
		\det
		\frac{ 
			\partial^2 S_0^{\alpha,j}({\bf r},{\bf I}) 
		     }
		     {
			\partial {\bf r} \partial {\bf I}
		     }
	\right|^{1/2}\!\!\!
	\left(
		\begin{array}{c}
			u^{\alpha,j}_{0,{\bf I}}({\bf r})
			e^{+i\phi({\bf r})/2} \\
			v^{\alpha,j}_{0,{\bf I}}({\bf r})
			e^{-i\phi({\bf r})/2}
		\end{array}
	\right)
	\times
\\
	\times 
	\exp
	\left(
%		\frac{i}{\hbar}
		i\hbar^{-1}
		S^{\alpha,j}_0({\bf r},{\bf I}) 
		+ i S^{\alpha,j}_1({\bf r},{\bf I})
	\right),
\label{eqn:WaveFunctionForOneD}
\end{multline}}
where $A_j^{\alpha}$ is a constant.
%--------------------------------------------------------------------------
\subsection{Torus quantisation}
\label{sec:TorusQuantisation}
To construct a quantisation rule we must consider the behaviour of 
(\ref{eqn:WaveFunctionForOneD}) along rays of the Hamiltonian system
(\ref{eqn:HamiltonsEquations}). Thus we must first discuss the properties
of the underlying classical dynamics.

It is well known ~\cite{goldstein:80:0} that for a 
classical mechanics with $N$ degrees of freedom to be integrable 
there must be $N$ constants of the motion. For such a system the 
dynamics takes place on an $N$-dimensional (Lagrangian) submanifold
in phase space, which has the topology of an $N$-torus ~\cite{arnold:89:0}.
To perform a canonical transformation from old coordinates $({\bf p},{\bf r})$
to new coordinates $({\bf I},{\pmb \varphi})$, where ${\pmb \varphi}$ are 
angle coordinates on the torus, one constructs the action function,
$S({\bf r},{\bf I})$, which is a solution of the Hamilton-Jacobi equation
for the given classical mechanics. Then the transformation is specified
by 
\begin{equation}
	{\bf p}=\frac{\partial S({\bf r},{\bf I})}
		     {\partial {\bf r}},
	\quad
	{\pmb \varphi}=\frac{\partial S({\bf r},{\bf I})}
		     {\partial {\bf I}}.
\nonumber 
\end{equation}
The action variables, ${\bf I}$, which are constants of the motion, are
given by 
\begin{equation}
	I_l
	=
	\frac{1}{2\pi}
	\oint_{\Gamma_l} {\bf p}\cdot d{\bf r},
\label{eqn:ActionVaribles}
\end{equation}
where $\Gamma_l$ is the $l$th irreducible loop on the $N$-torus. The energy
can then be expressed solely in terms of these constants, $E=E({\bf I})$,
and Hamiltons equations of motion can then be integrated explicitly.

It is also well known ~\cite{gutzwiller:90:0,ozorio:88:0}
that the semiclassical approximation to the spectrum of 
the Schr\"odinger equation with a Hamiltonian whose classical dynamics 
is integrable, proceeds via the semiclassical wave function (whose 
multicomponent generalisation we have given as 
(\ref{eqn:WaveFunctionForOneD})). The single-valuedness
of this wave function when followed around  each of the irreducible loops upon
the torus yields the $N$ quantisation conditions
\begin{equation}
	\oint_{\Gamma_l} {\bf p}\cdot d{\bf r}
	=2\pi I_l
	=2\pi \hbar \left( 
				n_l +\frac{m_l}{4}
		   \right),
\label{eqn:EBKconditions}
\end{equation}
where $n_l$ and $m_l$ are integers. $m_l$ is the
Maslov index ~\cite{maslov:81:0}, and accounts for a change of $\pi/2$
in the phase of the wave function each time a caustic is crossed. The Maslov
index of a closed curve is defined to be the number of times 
$\partial r/\partial p$ changes sign from negative to positive, minus the
number of times $\partial r/\partial p$ changes sign from positive to negative.
It is a topological invariant of a Lagrangian torus. 

The quantisation conditions (\ref{eqn:EBKconditions}) are the so called 
Einstein-Brillouin-Keller (EBK) quantisation rules, the generalisation of
the Bohr-Sommerfeld quantisation rules of `old' quantum mechanics to 
encompass non-separable problems.

What we require is the multicomponent generalisation of the EBK rules. Since 
the multicomponent theory was reduced to two Hamiltonian systems one might
think that the EBK rules apply to the tori for each Hamiltonian system.
We start by writing the most general semiclassical wave function, which
due to the multivaluedness of ${\bf p}^j_0({\bf r})$ is given by the 
superposition principle of quantum mechanics as
{\setlength{\multlinegap}{0pt}
\begin{multline}
	\left(
		\begin{array}{c}
			u_{{\bf I}}^\alpha ({\bf r}) \\
			v_{{\bf I}}^\alpha ({\bf r})
		\end{array}
	\right)
	=
	\sum_j
	A^\alpha_j
	\left|
		\det
		\frac{ 
			\partial^2 S_0^{\alpha,j}({\bf r},{\bf I}) 
		     }
		     {
			\partial {\bf r} \partial {\bf I}
		     }
	\right|^{1/2}\!\!\!
	\left(
		\begin{array}{c}
			u^{\alpha,j}_{0,{\bf I}}({\bf r})
			e^{+i\phi({\bf r})/2} \\
			v^{\alpha,j}_{0,{\bf I}}({\bf r})
			e^{-i\phi({\bf r})/2}
		\end{array}
	\right)\times
\\ 
	\times
	\exp
	\left(
		i\hbar^{-1}
		S^{\alpha,j}_0({\bf r},{\bf I}) 
		+ i S^{\alpha,j}_1({\bf r},{\bf I})
		+im\pi /2
	\right),
\label{eqn:WaveFunctionOnTorus}
\end{multline}}
where $j$ runs over all the branches, $S_0^{\alpha,j}$, which together
make up the torus. Firstly consider a situation where $S_1^{\alpha,j}=0$.
Such a situation occurs when ${\pmb \nabla}\phi=0$. Then it is clear that
single-valuedness of this wave function 
does yields a quantisation condition of the form (\ref{eqn:EBKconditions}).
Now generalise to the case where ${\pmb \nabla}\phi({\bf r})$ has global
curvature. Then the contribution to the change of phase around a closed
path arising from the topologically non-trivial $\phi({\bf r})$ is
$\pi m^\phi$ where
\begin{equation}
	m^\phi
	=
	\frac{1}{2\pi}
	\oint  
	{\pmb \nabla}\phi({\bf r})\cdot d{\bf r},
\label{eqn:StringTopologicalIndex}
\end{equation}
is an integer.
However the phase $S_1^{\alpha,j}({\bf r})$ poses a greater problem since
as discussed it is not locally path independent so that it prevents us from 
constructing an analog of (\ref{eqn:EBKconditions}) in the most general
circumstances which we will be interested in. This problem has been 
discussed in the context of multicomponent wave equations by Littlejohn
and Flynn ~\cite{littlejohn:91:0}. They proposed using Hamiltonians which
include first order terms in $\hbar$. For these new Hamiltonians the effect
of $S_1^{\alpha,j}$ is already included in the `zeroth order' theory.
Their results were limited to systems with no global degeneracies
and, as discussed by Emmrich and Weinstein ~\cite{emmrich:96:0}, integrability
of the underlying classical dynamics does not garantee an extension of the 
EBK rules for degenerate systems. 
Recently Bolte and Keppeler ~\cite{bolte:99:0}
presented a semiclassical theory for the Dirac equation based on semiclassical
trace formulae, rather than multicomponent WKB, at least in part to avoid these
difficulties (and also to handle non-integrable dynamics).
In our case, at the present time, this will not be necessary.
In the next section we will follow a similar proceedure to Littlejohn and
Flynn which however will differ in that the
Hamiltonians we construct will naturally be seen to 
contain terms up to $\hbar^2$. (This takes our work outside of the 
considerations of the above citations ~\cite{emmrich:96:0,littlejohn:91:0}.)
Later in this paper, when we study the 
single vortex, the inclusion of $\hbar^2$ terms in the Hamiltonian turns 
out to be essential in order to obtain the known (exact) wave function at 
the origin. (Note that there is no reason to object to 
$\hbar$-dependent terms in the `classical' Hamiltonians since it is an
effective (fictitous) classical mechanics that we are considering.)
%--------------------------------------------------------------------------
\section{An effective semiclassical theory}
\label{sec:EffectiveSemiclassicalTheoryForSuperconductors}
\subsection{The zeroth order theory}
We start by writing the spinor in terms of amplitudes and phases
\begin{equation}
	\left(
		\begin{array}{c}
			u_\lambda ({\bf r}) \\
			v_\lambda ({\bf r}) 
		\end{array}
	\right)
	=
	\left(
		\begin{array}{c}
			u_{0,\lambda} ({\bf r}) e^{+i\Sigma({\bf r})} \\
			v_{0,\lambda} ({\bf r}) e^{-i\Sigma({\bf r})}
		\end{array}
	\right)e^{iS({\bf r})}e^{-S_1^i({\bf r})},
\label{eqn:ResummedSpinor}
\end{equation}
but this time we {\it don't} expand $S$ and $\Sigma$ in $\hbar$. The amplitudes
$u_{0,\lambda} ({\bf r})$, $v_{0,\lambda} ({\bf r})$ and 
$e^{-S_1^i({\bf r})}$ are again determined by the zeroth and first 
order equations of the theory but these equations will be different
from those obtained by asymptotics in $\hbar$.
Substituting (\ref{eqn:ResummedSpinor}) into the BdG equations
and proceeding as before we obtain, instead of 
(\ref{eqn:UnitaryTransformedBdGequations}), the equations
\begin{multline}
	\left(
		\begin{array}{cc}
			E_\lambda-H(
				\hat{\bf p} +
				{\bf p}({\bf r};\hbar),
				{\bf r}
			) &
			-|\Delta({\bf r})| e^{i\phi({\bf r})} \\
			-|\Delta({\bf r})| e^{-i\phi({\bf r})} &
			E_\lambda+H^*(
				\hat{\bf p} +
				{\bf p}({\bf r};\hbar),
				{\bf r}
			)
		\end{array}
	\right)
	\left(
		\begin{array}{c}
			u_{0,\lambda} ({\bf r}) e^{+i\Sigma({\bf r})} \\
			v_{0,\lambda} ({\bf r}) e^{-i\Sigma({\bf r})}
		\end{array}
	\right)
\\
	\times
	e^{-S_1^i({\bf r})}
	=0,
\nonumber
\end{multline}
where ${\bf p}({\bf r};\hbar)=\hbar {\pmb \nabla}S$ is the $\hbar$-dependent
%-----------beginfootnote-------------------------------------------------
momentum\footnote{If we wish to recover our previous results we can expand:
${\bf p}({\bf r};\hbar)={\pmb \nabla}S_0+\hbar {\pmb \nabla}S_1^r+\cdots
={\bf p}_0({\bf r})+\hbar {\bf p}_1({\bf r})+\cdots$, and proceed with 
asymptotics in the small parameter $\hbar$.}.
%-------endfootnote-------------------------------------------------------
We are not yet ready to make any simplifying approximations since the spinor
still contains a phase. We use $U_\phi({\bf r})$ 
(\ref{eqn:SingularUnitaryMatrix}) to remove $\phi({\bf r})$ at every 
${\bf r}$-point and obtain
\begin{multline}
	\left(
		\begin{array}{cc}
			E_\lambda-H(
				\hat{\bf p} +
				{\bf p}({\bf r};\hbar) +
				\frac{\hbar}{2}{\pmb \nabla}\phi,
				{\bf r}
			) &
			-|\Delta({\bf r})|  \\
			-|\Delta({\bf r})|  &
			E_\lambda+H^*(
				\hat{\bf p} +
				{\bf p}({\bf r};\hbar)-
				\frac{\hbar}{2}{\pmb \nabla}\phi,
				{\bf r}
			)
		\end{array}
	\right)\times
\\
	\times
	\left(
		\begin{array}{c}
			u_{0,\lambda} ({\bf r}) 
			e^{+i\Sigma({\bf r})-i\phi({\bf r})/2} \\
			v_{0,\lambda} ({\bf r}) 
			e^{-i\Sigma({\bf r})+i\phi({\bf r})/2}
		\end{array}
	\right)e^{-S_1^i({\bf r})}
	=0.
\label{eqn:UnitaryTransformedBdGToAllOrders}
\end{multline}
For an appropriate zeroth order theory the spinor written in this form
is indeed real as we now demonstrate. 

To simplify the equations (\ref{eqn:UnitaryTransformedBdGToAllOrders})
we must drop the differential operators
whose action upon $u_{0,\lambda} ({\bf r})$ and
$v_{0,\lambda} ({\bf r})$ is small. Previously this was achieved by
discarding all terms containing $\hbar$ 
($\hat{{\bf p}}=-i\hbar {\pmb \nabla}$). However consistency would then 
require us to replace ${\bf p}({\bf r};\hbar)$ by ${\bf p}_0({\bf r})$
which we do not want. Rather than use $\hbar$ as our ordering parameter
we instead use $\hat{{\bf p}}$ itself. Thus our zeroth order theory is obtained
by dropping $\hat{{\bf p}}$ to give the analog of 
(\ref{eqn:RealZerothOrderBdGequations}) i.e.,
\begin{equation}
	\left(
		\begin{array}{cc}
			E_\lambda-H^e(
				{\bf p},
				{\bf r}
			) &
			-|\Delta({\bf r})|  \\
			-|\Delta({\bf r})|  &
			E_\lambda+H^h(
				{\bf p},
				{\bf r}
			)
		\end{array}
	\right)
	\left(
		\begin{array}{c}
			u_{0,\lambda} ({\bf r}) 
			e^{+i\Sigma({\bf r})-i\phi({\bf r})/2} \\
			v_{0,\lambda} ({\bf r}) 
			e^{-i\Sigma({\bf r})+i\phi({\bf r})/2}
		\end{array}
	\right)
	=0,
\nonumber
\end{equation}
where now ${\bf p}={\bf p}({\bf r};\hbar)$, $H^e({\bf p},{\bf r})$ 
is an electron Hamiltonian of the form
(\ref{eqn:HamiltonianOperator}) with the momentum operator $\hat{\bf p}$
replaced by ${\bf p}({\bf r};\hbar)$, {\it and} the vector potential 
${\bf A}({\bf r})$ replaced by an effective vector potential 
$ {\bf A}^{\mathrm{eff}}({\bf r}) = 
\frac{\hbar}{2e}{\pmb \nabla}\phi({\bf r})
+{\bf A}({\bf r})$.
The hole Hamiltonian,
$H^h({\bf p},{\bf r})$, is the same as $H^e({\bf p},{\bf r})$
except $e\rightarrow -e$. 

Since the resulting matrix Hamiltonian is real we have
\begin{equation}
	\Sigma ({\bf r})
	=
	\frac{\phi({\bf r})}{2},
\label{eqn:SigmaInTermsOfPhiToAllOrders}
\end{equation}
and again, as a consequence of (\ref{eqn:globalcurvature}), 
we see that $\Sigma$ can contain 
a topological contribution.

Our new theory yields the Hamilton-Jacobi equations 
\begin{equation}
E_{\bf I}=E^\alpha({\bf p},{\bf r}),
\end{equation}
where two new $\hbar$-dependent Hamiltonians, $E^{\alpha}({\bf p},{\bf r})$,
have replaced the Hamiltonians $E_0^\alpha({\bf p}_0,{\bf r})$ appearing in
our previous theory. 

Explicitly the new Hamiltonians are
\begin{equation}
	E^{\alpha}({\bf p},{\bf r})
	=
	{\bf p}\cdot{\bf v}_s({\bf r})
	+\alpha
	\sqrt{
		\left(
			\frac{p^2}{2m}+\frac{1}{2}mv_s^2({\bf r}) 
				+ V({\bf r}) -\epsilon_F
		\right)^2
		+ |\Delta({\bf r})|^2
	},
\label{eqn:hdependentClassicalHamiltonian}
\end{equation}
where $\alpha=\pm$ and 
$m{\bf v}_s=\frac{\hbar}{2}{\pmb \nabla}\phi+e{\bf A}({\bf r})$. 
The constancy of 
$E^{\alpha}({\bf p},{\bf r})$ defines implicitly the functions
${\bf p}^{\alpha}({\bf r};\hbar)=
\hbar \frac{\partial S^{\alpha}}{\partial {\bf r} }$,
and the solutions of the Hamilton-Jacobi equations,
when they exist, have the form
\begin{equation}
	S^\alpha({\bf r},{\bf I}) = \frac{1}{\hbar}
			\int^{{\bf r}(t)}_{{\bf r}_0(t_0)}
			{\bf p}^{\alpha}({\bf r},{\bf I};\hbar)
			\cdot
			d{\bf r},
\label{eqn:hdependentAction}
\end{equation}
where the integrals are taken along trajectories of the Hamiltonian
systems
\begin{equation}
	\dot{\bf r}^\alpha
	=
	\left(
		\frac{\partial E^\alpha ({\bf p},{\bf r}) }
			{ \partial {\bf p}}
	\right),
	\quad
	\dot{\bf p}^\alpha
	=
	-\left(
		\frac{\partial E^\alpha ({\bf p},{\bf r}) }
			{ \partial {\bf r}}
	\right).
\label{eqn:EffectiveHamiltonsEquations}
\end{equation}
The $\hbar$-dependent Hamiltonians (\ref{eqn:hdependentClassicalHamiltonian})
together with the Hamiltonian systems they define, and the $\hbar$-dependent
action (\ref{eqn:hdependentAction}) are a central result of this paper.
$S^\alpha({\bf r},{\bf I})$ is clearly a locally path independent quantity
since ${\bf p}^\alpha({\bf r},{\bf I};\hbar)$ is a solution of Hamilton's
equations and thus lies in a Lagrangian torus in phase space. If we can prove
that no other phases appear in our theory the action can immediately be
used to construct a quantisation condition.

You will notice that the 
Hamiltonians (\ref{eqn:hdependentClassicalHamiltonian})
include a term in $v_s^2({\bf r})$, that is a term of order $\hbar^2$ as 
alluded to in the preceeding section.

In concluding the zeroth order theory we see that
the normalised amplitudes, representing the internal structure of our
`particle' as it is transported along the trajectory, become
\begin{align}
	u_{0,{\bf I}}^{\alpha,j} ({\bf r})
	&=
	\sqrt{
		\frac{1}{2}
		\left(\textstyle
			1+\beta
			  \frac{	
				\sqrt{ 
					\left( 
						E^\alpha-{\bf p}^j\cdot{\bf v}_s
					\right)^2
					-
					\left| \Delta({\bf r}) \right|^2
				     }
			      }
			      {
				E^\alpha-{\bf p}^j\cdot{\bf v}_s
			      }
		\right)
	     },
\nonumber \\
	v_{0,{\bf I}}^{\alpha,j} ({\bf r})
	&=
	\sqrt{
		\frac{1}{2}
		\left(\textstyle
			1-\beta
			  \frac{
				\sqrt{ 
					\left( 
						E^\alpha-{\bf p}^j\cdot{\bf v}_s
					\right)^2
					-
					\left| \Delta({\bf r}) \right|^2
				     }
			      }
			      {
				E^\alpha-{\bf p}^j\cdot{\bf v}_s
			      }
		\right)
	     }.
\label{eqn:NormalisedAmplitudesEffective}
\end{align}
Thus the amplitudes too have become $\hbar$-dependent.
%-------------------------------------------------------------------------
\subsection{First order theory}
\label{sec:EffectiveSemiclassicalFirstOrderTheory}
To derive the new first order equation we take the expectation value of
equation (\ref{eqn:UnitaryTransformedBdGToAllOrders}).
By defining
\begin{equation}
	G=
	\left(
		\begin{array}{c}
			u_{0,\lambda} ({\bf r}) 
			e^{+i\Sigma({\bf r})-i\phi({\bf r})/2} \\
			v_{0,\lambda} ({\bf r}) 
			e^{-i\Sigma({\bf r})+i\phi({\bf r})/2}
		\end{array}
	\right)e^{-S_1^i({\bf r})},
\label{eqn:Gspinor}
\end{equation}
and introducing $\hat{D}(\hbar)$ for the $\hbar$-dependent 
matrix differential operator:
\begin{equation}
	\hat{D}(\hbar)
	=
	\left(
		\begin{array}{cc}
			E_\lambda-H(
				\hat{\bf p} +
				{\bf p}({\bf r};\hbar) +
				\frac{\hbar}{2}{\pmb \nabla}\phi,
				{\bf r}
			) &
			\!\!\!\!\!\!\!\!\!\!\!\!\!-|\Delta({\bf r})|  \\
			-|\Delta({\bf r})|  &
			\!\!\!\!\!\!\!\!\!\!\!\!\!E_\lambda+H^*(
				\hat{\bf p} +
				{\bf p}({\bf r};\hbar)-
				\frac{\hbar}{2}{\pmb \nabla}\phi,
				{\bf r}
			)
		\end{array}
	\right),
\nonumber
\end{equation}
we can write this expectation as
\begin{equation}
	G^\dagger \hat{D}(\hbar)G
	=0.
\nonumber 
\end{equation}
Expanding this equation upto and including first order in $\hat{{\bf p}}$ 
we find
\begin{equation}
	0=G_0^\dagger D_0(\hbar) G_0 +G_1^\dagger D_0(\hbar) G_0  
		+ G_0^\dagger D_0(\hbar) G_1
		 +G_0^\dagger \hat{D}_1(\hbar) G_0,
\label{eqn:FirstOrderhbarMatrices}
\end{equation}
where $D_0(\hbar)$ is the zeroth order $\hbar$-dependent Hamiltonian matrix
\begin{equation}
	D_0(\hbar)
	=
	\left(
		\begin{array}{cc}
			E_{\bf I}-H^e(
				{\bf p},
				{\bf r}
			) &
			-|\Delta({\bf r})|  \\
			-|\Delta({\bf r})|  &
			E_{\bf I}+H^h(
				{\bf p},
				{\bf r}
			)
		\end{array}
	\right),
\label{eqn:ZerothOrderhDependentMatrix}
\end{equation}
$G_0$ is given by equation (\ref{eqn:Gspinor}) together with the zeroth order
condition (\ref{eqn:SigmaInTermsOfPhiToAllOrders}) i.e.
\begin{equation}
	G_0 =
	\left(
		\begin{array}{c}
			u_{0,{\bf I}} ({\bf r}) 
			 \\
			v_{0,{\bf I}} ({\bf r}) 
		\end{array}
	\right)e^{-S_1^i({\bf r})},
\label{eqn:G_0spinor}
\end{equation}
which does not contain any phases, $G_1$ allows for corrections beyond those
considered in $G_0$ (in analogy with the semiclassical theory (Appendix
\ref{sec:appendixA})), and $\hat{D}_1(\hbar)$ is 
\begin{equation}
	\hat{D}_1=
	\left(
		\begin{array}{c}
			-\frac{1}{2m}
	 		\left(
				\frac{\hbar}{i}{\pmb \nabla}
				\cdot {\bf P}^+
				+
				{\bf P}^+
				\cdot \frac{\hbar}{i}{\pmb \nabla}
			\right) \qquad \qquad 0 \qquad \quad  \\
			\qquad \quad 0 \qquad \qquad
			+\frac{1}{2m}
	 		\left(
				\frac{\hbar}{i}{\pmb \nabla}
				\cdot {\bf P}^-
				+
				{\bf P}^-
				\cdot \frac{\hbar}{i}{\pmb \nabla}
			\right)
		\end{array}
	\right),
\nonumber
\end{equation}
where ${\bf P}^\pm(\hbar) = {\bf p}({\bf r};\hbar) \pm m{\bf v}_s({\bf r})$.
Our analysis now parallels our previous semiclassical discussion 
(Appendix \ref{sec:appendixA}):
the first term in equation (\ref{eqn:FirstOrderhbarMatrices}) is the zeroth
order equation whilst the second and third terms can both be shown to depend
upon $D_0(\hbar) G_0=0$, so that their vanishing yields no new information.
We are left with the first order equation
\begin{equation}
	{\cal{O}}({\hat{\bf p}})
	\qquad
	0=G_0^\dagger \hat{D}_1(\hbar) G_0.
\nonumber
\end{equation}
Defining the vector matrix ${\bf M}$ by 
\begin{equation}
	{\bf M}
	=
	\left(
		\begin{array}{cc}
			-\frac{{\bf P}^+(\hbar)}{2m} & 0 \\
			0& +\frac{{\bf P}^-(\hbar)}{2m}
		\end{array}
	\right),
\label{eqn:EffectiveVectorMatrix}
\end{equation}
this can be rewritten as
\begin{equation}
	0=
	{\pmb \nabla}
	\cdot
	\left\{
		G_0^\dagger
		{\bf M}
		G_0
	\right\}
	+
	\;2i \;\text{Im}
		\left[
		G_0^\dagger{\bf M}
		\cdot
		{\pmb \nabla}G_0
	\right].
\end{equation}
The first term is purely real, whilst the second is purely imaginary.
In our previous semiclassical theory, where we had the spinor $F_0$ in place
of $G_0$, the first term led to the transport equation 
(see (\ref{eqn:TransportEquation}) and
(\ref{eqn:AmplitudeTransportEquation})) whilst
the second led to an equation for the first order phase $S_1^r({\bf r})$,
(see (\ref{eqn:FirstOrderPhasesEquation}) and (\ref{eqn:FirstOrderPhase})). 
This was because $F_0$ was complex. In the present
effective semiclassical theory $G_0$ is real (\ref{eqn:G_0spinor}). 
Consequently the second term
in the above is trivially zero. We are left with a new transport equation
which can be written as
\begin{equation}
	{\pmb \nabla}
	\cdot
	\left(
		e^{-2S^i_1({\bf r},{\bf I})}
		\left.
			\frac{\partial E^\alpha ({\bf p},{\bf r}) }
			     {\partial{\bf p}}
		\right|_{{\bf p}={\bf p}^\alpha({\bf r},{\bf I};\hbar)}
	\right)
	= 0,
\label{eqn:hDependentAmplitudeTransportEquation}
\end{equation}
in terms of our new $\hbar$-dependent Hamiltonians 
(\ref{eqn:hdependentClassicalHamiltonian}). Its solution yields the new
$\hbar$-dependent determinant
\begin{equation}
	e^{-S_1^i({\bf r},{\bf I})}
	=c
	\left|
		\det
		\frac{ 
			\partial^2 S^{\alpha,j}({\bf r},{\bf I};\hbar) 
		     }
		     {
			\partial {\bf r} \partial {\bf I}
		     }
	\right|^{1/2}.
\label{eqn:hDependentVanVleckDet}
\end{equation}

Actually one must be more careful than we have been here in deriving this
effective theory. Whilst one would very much
like there to be no first order phase corrections ($S_1^r$, $\Sigma_1^r$)
we simply cannot demand that this is the case. As soon as we 
approximate to obtain a zeroth order theory $\Sigma$ and $S$ are only 
determined
to this order and we must allow for the possibility of corrections,
$S_1^r({\bf r})$, $\Sigma_1^r({\bf r})$.
If we do this $G$, equation (\ref{eqn:Gspinor}), is replaced by

\begin{equation}
	G'=
	\left(
		\begin{array}{c}
			u_{0,{\bf I}} ({\bf r}) 
			e^{+i\Sigma({\bf r})+i\Sigma^r_1({\bf r})
			   -i\phi({\bf r})/2} \\
			v_{0,{\bf I}} ({\bf r}) 
			e^{-i\Sigma({\bf r})-i\Sigma^r_1({\bf r})
			   +i\phi({\bf r})/2}
		\end{array}
	\right) e^{+iS_1^r({\bf r})} e^{-S_1^i({\bf r})},
\label{eqn:G'spinor}
\end{equation}
and $G_0$ by
\begin{equation}
	G_0' =
	\left(
		\begin{array}{c}
			u_{0,{\bf I}} ({\bf r})e^{+i\Sigma_1^r({\bf r})} 
			 \\
			v_{0,{\bf I}} ({\bf r}) e^{-i\Sigma_1^r({\bf r})} 
		\end{array}
	\right) e^{+iS_1^r({\bf r})} e^{-S_1^i({\bf r})}.
\label{eqn:G_0'spinor}
\end{equation}
Then $\text{Im}  \left[ G_0^\dagger{\bf M}\cdot{\pmb \nabla}G_0 \right]$
is no longer trivially zero and yields
\begin{equation}
	S_1^r({\bf r})
	=
	-\frac{1}{e}
	\int^t_{t_0}
	{\bf j}^{\alpha,j}({\bf r})
	\cdot
	{\pmb \nabla}\Sigma_1^r({\bf r})
	dt',
\label{eqn:FirstOrderPhase'}
\end{equation}
where
\begin{equation}
	-\frac{1}{e}
	{\bf j}^{\alpha,j}({\bf r})
	=
	\frac{{\bf p}^{\alpha,j}({\bf r};\hbar)}{m}
	+
	\left(
		(u^{\alpha,j}_{0,{\bf I}}({\bf r}))^2
		-
		(v^{\alpha,j}_{0,{\bf I}}({\bf r}))^2
	\right)
	\frac{e{\bf A}^{\text{eff}}({\bf r}) }{m}.
\nonumber
\end{equation}
But there is a crucial difference between (\ref{eqn:FirstOrderPhase'}) and
(\ref{eqn:FirstOrderPhase}), namely it depends upon ${\pmb \nabla}\Sigma_1^r$
rather than ${\pmb \nabla}\phi$. What can we say about 
${\pmb \nabla}\Sigma_1^r$? Appealing to self-consistency we have
\begin{equation}
	u_\lambda({\bf r}) v^*_\lambda({\bf r}) 
	\propto 
	e^{i2\Sigma({\bf r})+i2\Sigma_1^r({\bf r})}
	=
	e^{i\phi({\bf r})+i2\Sigma_1^r({\bf r})}. 
\nonumber
\end{equation}
Thus $\Sigma_1^r=m\pi$, $m$ an integer, 
so ${\pmb \nabla}\Sigma_1^r=0$, and hence $S_1^r=0$. Furthermore using the
same arguments as followed equation (\ref{eqn:SigmaInTermsOfPhi}), the need for
$\Sigma_1^r$ can be eliminated. Since $S_1^r=0$, and $\Sigma_1^r=0$ our
form for the spinor, equation (\ref{eqn:ResummedSpinor}), and all that follows
up to (\ref{eqn:hDependentAmplitudeTransportEquation}) is indeed correct.
We have succeeded in constructing a theory where there are no first order
phase corrections and have thus removed the obstacle to the derivation of a
generalised Bohr-Sommerfeld or EBK quantisation rule.

The general wave function for our effective semiclassical theory now
takes the form
\begin{multline}
	\left(
		\begin{array}{c}
			u_{{\bf I}}^\alpha ({\bf r}) \\
			v_{{\bf I}}^\alpha ({\bf r})
		\end{array}
	\right)
	=
	\sum_j
	A^\alpha_j
	\left|
		\det
		\frac{ 
			\partial^2 S^{\alpha,j}({\bf r},{\bf I};\hbar) 
		     }
		     {
			\partial {\bf r} \partial {\bf I}
		     }
	\right|^{1/2}\!\!\!
	\left(
		\begin{array}{c}
			u^{\alpha,j}_{0,{\bf I},\hbar}({\bf r})
			e^{+i\phi({\bf r})/2} \\
			v^{\alpha,j}_{0,{\bf I},\hbar}({\bf r})
			e^{-i\phi({\bf r})/2}
		\end{array}
	\right)\times
\\
	\times
	\exp
	\left(
		i\hbar^{-1}
		S^{\alpha,j}({\bf r},{\bf I};\hbar)
		+im\pi/2 
	\right),
\label{eqn:WaveFunctionEffective}
\end{multline}
and the single-valuedness of this after returning from circuits around
each of the irreducible loops, $\Gamma_l$ on the 3-torus yields the 
general quantisation conditions
\begin{equation}
	\oint_{\Gamma_l} {\bf p}^\alpha ({\bf r};\hbar)\! \cdot d{\bf r}
	=2\pi \hbar \left( 
				n_l^\pm 
				+ \frac{m_l}{4}
				\mp \frac{m_l^\phi}{2}
		   \right),
\label{eqn:MulticomponentConditions}
\end{equation}
where as before
\begin{equation}
	m_l^\phi
	=
	\frac{1}{2\pi}
	\oint_{\Gamma_l}  
	{\pmb \nabla}\phi({\bf r})\cdot d{\bf r},
\label{eqn:StringTopologicalIndexAgain}
\end{equation}
takes integer values. For a given $\Gamma_l$ the line integral in
equation (\ref{eqn:MulticomponentConditions}) is fixed so we must have
$n^+_l=n^-_l + m_l^\phi$, i.e. we need only one of $n^\pm_l$. Choosing
$n^+_l=n_l$ equation (\ref{eqn:MulticomponentConditions}) becomes
\begin{equation}
	\oint_{\Gamma_l} {\bf p}^\alpha ({\bf r};\hbar)\! \cdot d{\bf r}
	=2\pi \hbar \left( 
				n_l
				+ \frac{m_l}{4}
				- \frac{m_l^\phi}{2}
		   \right),
\label{eqn:MulticomponentEBKcondition}
\end{equation}
for $n_l$, $m_l$, and $m_l^\phi$ integers. This is our generalisation of the
EBK quantisation condition to apply to the superconducting case.
It includes two topological integers. The first is the familiar Maslov index, 
whilst the second arises from the vortex singularities,
which, if present,  shift the quantum numbers by half-integers. The action
integral itself, as discussed, is defined upon an $\hbar$-dependent 
Lagrangian submanifold
in phase space. The trajectories of the Hamiltonian system which wind around
this manifold are specified by the $\hbar$-dependent Hamiltonians, 
$E^\alpha({\bf p},{\bf r})$, equations 
(\ref{eqn:hdependentClassicalHamiltonian}).
The question then arises as to the interpretation of a theory depending
upon these Hamiltonians. In particular how should
the appearance of $\hbar {\pmb \nabla} \phi({\bf r})/2$ be understood?
%-------------------------------------------------------------------------
\subsection{Interpretation: A Semiclassical theory in the presence of lines
of phase singularities}
Consider the BdG equations written in the (exact) form
\begin{multline}
	\left(
		\begin{array}{lr}
			E_\lambda
			-\frac{1}{2m}
			\left(
				\hat{\bf p}
				 +
				\frac{\hbar}{2} 
				{\pmb \nabla} \phi({\bf r})
				+e{\bf A}({\bf r})
			\right)^2
			-V({\bf r}) +\epsilon_F	
			&
			-|\Delta({\bf r})| \qquad \\
			\qquad -|\Delta({\bf r})|  &
			\!\!\!\!\!\!\!\!\!\!\!\!
			\!\!\!\!\!\!\!\!\!\!\!\!
			\!\!\!\!\!\!\!\!\!\!\!\!
			\!\!\!\!\!\!\!\!\!\!\!\!			
			\!\!\!\!\!\!\!\!\!\!\!\!
			\!\!\!\!\!\!\!\!\!\!\!\!
			E_\lambda+\frac{1}{2m}
			\left(
				\hat{\bf p}
				 -
				\frac{\hbar}{2} 
				{\pmb \nabla} \phi({\bf r})
				-e{\bf A}({\bf r})
			\right)^2
			+V({\bf r}) -\epsilon_F	
		\end{array}
	\right)
\\ 
	\times
	\left(
		\begin{array}{c}
			u_{\lambda} ({\bf r}) 
			e^{-i\phi({\bf r})/2} \\
			v_{\lambda} ({\bf r}) 
			e^{+i\phi({\bf r})/2}
		\end{array}
	\right)
	=
	0.
\label{eqn:BdGWithQuantisedString}
\end{multline}
The term $\hbar{\pmb \nabla} \phi({\bf r})/2$ enters into the particle and
hole Hamiltonians as an effective vector potential
\begin{equation}
	{\bf A}^{\text{PS}}({\bf r})
	=
	\frac{\hbar}{2e}
	{\pmb \nabla} \phi({\bf r}).
\label{eqn:QuantisedStringVectorPotential}
\end{equation}
There is a flux associated with this vector potential which we can find by
integrating (\ref{eqn:QuantisedStringVectorPotential}) along a path 
enclosing a vortex singluarity:
\begin{align}
	\oint_c
	{\pmb \nabla} \phi({\bf r})\cdot d{\bf r}
	&=
	\frac{2e}{\hbar}
	\oint_c
	{\bf A}^{\text{PS}}({\bf r}) \cdot d{\bf r},
\nonumber \\
	2\pi
	&=
	\frac{2e}{\hbar}
	\Phi_0.
\label{eqn:QuantisedStringFlux}
\end{align}
The left hand side follows from the single-valuedness of $\Delta({\bf r})$.
On the right hand side we have introduced $\Phi_0$ for the flux. It takes
the value
\begin{equation}
	\Phi_0
	=
	\frac{\hbar \pi}{e},
\label{eqn:SuperconductingFluxQuantum}
\end{equation}
i.e it is equal to one superconducting flux quantum. Where is this flux?
We can contract the curve, $c$, until only the line of phase singularities
is left inside. Equation (\ref{eqn:QuantisedStringFlux}) remains true. Thus
each line of phase singluarities, defined by $|\Delta({\bf r})|=0$, carries
a flux, $\Phi_0$, along it. A similar object was studied by Dirac
 ~\cite{dirac:31:0}, and is refered to as a Dirac string. However our string
has a physical reality, lying along the node of the order parameter unlike
a Dirac string which need not lie along a node. 
Hence we will stick with the `line of phase singularities' terminology. 
${\bf A}^{\text{PS}}({\bf r})$ is then the vector potential
associated with such a line of singularities. We can then interpret
equation (\ref{eqn:BdGWithQuantisedString}) as describing
superconducting quasiparticles in the presence of lines of phase singularities
of the pairing potential, each carrying
a flux, $\Phi_0$. Replacing $\hat{\bf p}$ by ${\bf p}$ and diagonalising we
obtain the Hamilton-Jacobi equations $E_{\bf I}=E^\alpha({\bf p},{\bf r})$, 
with $E^\alpha({\bf p},{\bf r})$ given by 
equation (\ref{eqn:hdependentClassicalHamiltonian}).
Thus we have constructed a (fictitous) classical mechanics describing 
quasiparticle excitations in superconductors propagating in the presence
of lines of phase singularities carrying flux $\Phi_0=h/2e$.

When the classical mechanics is integrable (a single s-wave vortex is such an 
example) then the semiclassical wave function takes the form 
(\ref{eqn:WaveFunctionEffective}) and we can apply the quantisation conditions
(\ref{eqn:MulticomponentEBKcondition}) to obtain the semiclassical excitation
spectrum. If however the classical dynamics is non-integrable then the 
solution cannot have the form (\ref{eqn:WaveFunctionEffective}). We return
to this point at the end of the paper.

In the interest of clarity concerning the above discussion we would like
to make the following remark.
If for a moment we consider a single vortex whose axis lies along the 
$z$-axis, then by symmetry
\begin{equation}
	\Delta({\bf r})
	=
	|\Delta(r)| e^{i\theta},
\nonumber
\end{equation}
in polar coordinates. The vector potential ${\bf A}^{\text{PS}}({\bf r})$
is then explicitly
\begin{equation}
	{\bf A}^{\text{PS}}({\bf r})
	=
	\frac{\Phi_0}{2\pi r}
	\hat{\bf \theta}.
\label{eqn:QuantisedStringA}
\end{equation}
Now the vector potential associated with an Aharanov-Bohm (AB) flux tube
is
\begin{equation}
	{\bf A}^{\text{AB}}({\bf r})
	=
	\frac{\Phi^{\text{AB}}}{2\pi r}
	\hat{\bf \theta},
\label{eqn:AharanovBohmA}
\end{equation}
for $r>r_c$, the solenoid radius. If we consider the limit $r_c \rightarrow 0$ 
as representing an idealised AB flux tube, or AB flux line then equations 
(\ref{eqn:QuantisedStringA}) and (\ref{eqn:AharanovBohmA}) become identical
for the choice $\Phi^{\text{AB}}=\Phi_0$. 
For this reason a vortex has been 
described in the literature ~\cite{vafek:00:0} 
as an effective magnetic Aharonov-Bohm half-flux. However we would like to
point out here that there is a crucial difference between an AB flux line
and a superconducting vortex, namely
that the single-valuedness of the AB wave function does {\it not} quantise the
flux, $\Phi^{\text{AB}}$, which remains classical ~\cite{berry:80:0}, whilst
single-valuedness of the order parameter, $\Delta({\bf r})$, {\it does}
quantise the flux, $\Phi_0$. For this reason we prefer the line of phase
singularities terminology.

We have one last point to make. In our original discussion of the general
form for the wave function we remarked (see the paragraph following 
equation (\ref{eqn:ComplexSpinorToOrderh})) that if $\phi({\bf r})$ is not
expanded in $\hbar$, which we claim to be the correct procedure, then there
is no fast component to the spinor i.e., $\Sigma_0^r({\bf r})=0$. Conversely
had we insisted upon expanding $\phi({\bf r})=\hbar^{-1}\phi_0({\bf r})+\cdots$
the zeroth order theory would have included the term 
$\hbar {\pmb \nabla}\phi|_{\hbar=0}={\pmb \nabla}\phi_0$. In the light of 
our effective semiclassical theory we now see that even though $\phi({\bf r})$
is not to be expanded, i.e., $\hbar{\pmb \nabla}\phi({\bf r})\sim \hbar$, the
most useful form for the theory has $\hbar{\pmb \nabla}\phi$ appearing in the
``zeroth order'' theory Hamiltonians.

This concludes our formalism for seeking semiclassical solutions to
superconducting problems described by the 
BdG equations (\ref{eqn:BdGequations}).
The general procedure is then as follows: (i) Choose ${\bf A}({\bf r})$ and 
$\Delta({\bf r})$ appropriate to the situation of interest. One can then 
immediately write down the effective Hamiltonians in explicit form. (ii)
Investigate the orbits of the Hamiltonian system and identify those of
interest. (iii) Compute the relevant Maslov indicies and any topological 
indicies arising from vortex singluarities. 
(iv) Deploy the generalised EBK quantisation condition
(\ref{eqn:MulticomponentEBKcondition}) to obtain the semiclassical spectrum.
(Steps (iii) and (iv) can be interchanged if analytic expressions are being
sort.) Depending upon the situation it may also be important to: (v)
investigate tunnelling between classical orbits and its effect upon the 
semiclassical spectrum, for example the lifting of degeneracies to yield
`avoided crossings'. (vi) One may also deploy the wave function to study 
properties other than the energy spectrum
(for example, quasiparticle current flow).

In the next two sections we will apply this theory to two well known problems.
The first we will
consider is a superconductor-normal metal -superconductor (SNS) junction. 
For this system we will focus upon understanding the
various branches of $p_y^\beta(y)$ which make up a trajectory in a
superconducting system, and upon the type of quasiparticles they represent.
We will then quantise the orbit and obtain the spectrum. Furthermore
we will compute the Maslov index, though via the more `traditional' route
of asymptotic analysis, and show that it is what one would expect 
from following Maslov's topological prescription. Since for the SNS junction
we take $\hbar{\pmb \nabla}\phi=0$ the two theories presented in sections
\ref{sec:SemiclassicalTheoryForSuperconductors} and 
\ref{sec:EffectiveSemiclassicalTheoryForSuperconductors} are identical. 
By contrast the second problem we
apply our theory to, the single vortex, has $\hbar{\pmb \nabla}\phi \neq 0$.
The significance of the topological phase and $\hbar$-dependent terms in
the theory will then become apparent.
%--------------------------------------------------------------------------
\section{The Superconductor-Normal metal-Superconductor Junction}
\label{sec:SNS_junction}
\begin{figure}[htbp]
	\begin{center}
		\epsfig{file=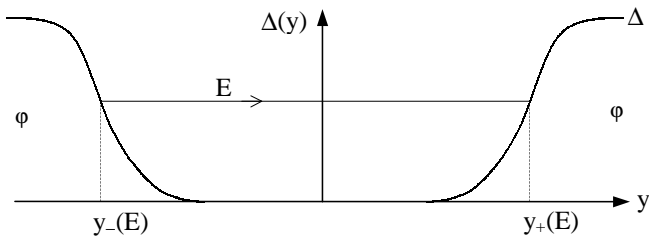,width=86mm,clip=}
		\caption{Illustration of the profile of 
			$|\Delta(y)|$ with classical turning
			points, $y_\pm(E)$, indicated.}
		\label{fig:SNSsmoothDeltaProfile}
	\end{center}
\end{figure}

A superconductor-normal metal-superconductor junction (SNS) consists of
a taking a superconducting wire and replacing a segment with a piece of
normal metal. The inclusion of the normal layer makes the order parameter,
$\Delta({\bf r})$, inhomogeneous along the wire. By assuming the wire is
sufficiently thick to neglect finite size effects the profile of 
$\Delta({\bf r})$ will only vary along the length of the wire. 
We choose the $y$-axis to lie along this direction and take $y=0$ to be
at the middle of the normal layer. We shall allow $|\Delta(y)|$ to have a 
smoothly varying profile at 
each interface. This contrasts with the more usual approach in 
the literature ~\cite{kulik:70:0,bardeen:72:0} of taking the profile of
$|\Delta|$ to be a step function at each interface.
The `width' of the normal layer, as far as an excitation is concerned, is
decided by the turning points of the classical trajectory and is thus
energy dependent. These turning points will be designated by $y_-(E)$ and
$y_+(E)$ ($y_-< y_+$), see FIG.~\ref{fig:SNSsmoothDeltaProfile}.
%------------------Figure to insert here----------------------------
%-------------------------------------------------------------------
The phases of the order parameter are taken to be constant and both equal
to $\phi$, i.e there is no phase gradient across the junction. Then,
taking the vector potential ${\bf A}({\bf r})$ to be zero, the classical
Hamiltonians, $E^\alpha({\bf p},{\bf r})$, 
equations (\ref{eqn:hdependentClassicalHamiltonian}),
are found to be
\begin{equation}
	E^{\alpha}({\bf p},{\bf r})
	=
	\alpha
	\sqrt{
		\left(
			\frac{p_x^2}{2m}+\frac{p_y^2}{2m}+\frac{p_z^2}{2m}
			-\epsilon_F
		\right)^2
		+ |\Delta(y)|^2
	},
\label{eqn:ClassicalHamiltonianForSNS}
\end{equation}
where $\alpha=\pm$ distinguishes the two Hamiltonians. However in keeping with
the definition of $E_{\bf I}$ as an excitation energy ($E_{\bf I}\geq0$)
we can discard the $\alpha=-1$ Hamiltonian, and consider only 
$E^+({\bf p},{\bf r})$. 
%--------------------------------------------------------------------------
\subsection{The $y$ behaviour of the excitation orbits}
Setting $E^+({\bf p},{\bf r})=E$, 
equation (\ref{eqn:ClassicalHamiltonianForSNS}) can 
be inverted giving $p_y^\beta(y)$ ($\beta=\pm$) as
\begin{equation}
	p_y^\beta(y)
	=
	\sqrt{ 
		p_F^2 - p_x^2-p_z^2
		+ \beta 2m
		\sqrt{
			E^2 - |\Delta(y)|^2
		}
	}.
\label{eqn:MomentumBranchesForSNS1}
\end{equation}
Thus $E^+({\bf p},{\bf r})=E$ defines four momentum branches (including
$-p_y^\pm(y)$) as a function of position, 
from which the orbit of the excitation is to be constructed. 

Now $p_y^+$ represents a quasiparticle, and $p_y^-$ a quasihole
(see Appendix \ref{app:BetajDependence}). This is clear from considering 
(\ref{eqn:MomentumBranchesForSNS1}) deep inside the normal layer where
$|\Delta(y)|\rightarrow0$:
\begin{equation}
	\lim_{|\Delta|\rightarrow0}
	p_y^\pm(y)
	=
	\sqrt{ 		
		p_F^2 \pm 2mE - 
		p_x^2-p_z^2
	}.
\end{equation}
When $E$ is non-zero 
$p_y^+$ lies outside the Fermi sea - a particle-like excitation - whilst
$p_y^-$ lies inside, so describes a hole-like excitation.

Throughout the normal region, for which
$|\Delta(y)|=0$, the momentum branches $p_y^\pm(y)$ are
independent of $y$, i.e are straight lines in the $p_y-y$ phase plane.

\begin{figure}
	\begin{center}
		\epsfig{file=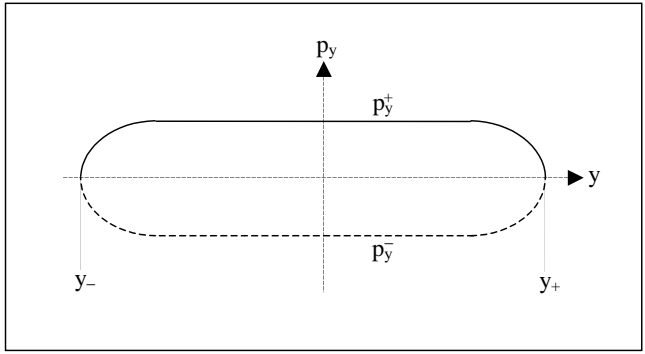,width=86mm,clip=}
		\caption{The form of $p_y^\pm(y)$ as $|\Delta(y)|
			\rightarrow E$.
			The $p_y^\pm$ are constant and positive
			 deep in the normal layer
			$y_- \ll y \ll y_+$. 
			For $|\Delta(y)| \rightarrow E$ 
			($y \rightarrow y_\pm$) $p_y^+$ decreases whilst
			$p_y^-$ increases.}
		\label{fig:OneAndreevOrbit}
	\end{center}
\end{figure}

Let us write $p_y^\pm(y)$ in (\ref{eqn:MomentumBranchesForSNS1}) as
\begin{equation}
	p_y^\beta(y)
	=
	\sqrt{ 
		p_F^2 - p_x^2-p_z^2
		+ \beta \epsilon(y)
	},
\label{eqn:MomentumBranchesForSNS2}
\end{equation}
where $\epsilon(y)=2m\sqrt{E^2 - |\Delta(y)|^2}$.
Approaching the interface $|\Delta(y)|\rightarrow E$, 
so $\epsilon(y)\rightarrow0^+$. Then from 
(\ref{eqn:MomentumBranchesForSNS2}) $p_y^+$ decreases and $p_y^-$
increases (FIG.~\ref{fig:OneAndreevOrbit}). The length scale over
which $\epsilon(y)$ changes is given by the coherence length, $\xi$.
%----------------------------------------------------------------------------
%----------------------------------------------------------------------------
At the interfaces $\epsilon(y_\pm)=0$ and
$p_y^\pm(y)$ are
\begin{equation}
	\lim_{|\Delta|\rightarrow E}
	p_y^\pm(y) =
	\sqrt{ 
		p_F^2 -p_x^2-p_z^2
	}.
\label{eqn:MomentumBranchesAtInterface}
\end{equation}
Thus at the points $y_\pm$, the particle and hole momenta are equal.
Notice however that  
%-----------footnote-------------------------------------------------
$p_y^\pm(y_\pm)\neq0$.\footnote{From equation
(\ref{eqn:MomentumBranchesAtInterface}),
$p_y^\pm(y_\pm)\neq0$ unless $p_F^2 = p_x^2+p_z^2$,
i.e. unless all the kinetic energy is parallel to the interface - a situation
which we shall not consider here. However, see for example 
$\check{\text{S}}$ipr and 
Gy\"orffy ~\cite{sipr:97:0} for details of how the physics changes radically in this
limit.}
%-----------endfootnote------------------------------------------------
This behaviour at the coelesence of the
momentum branches contrasts with typical classical orbits
where the turning points are characterised by $p_y(y_i)=0.$

\begin{figure}
	\begin{center}
		\epsfig{file=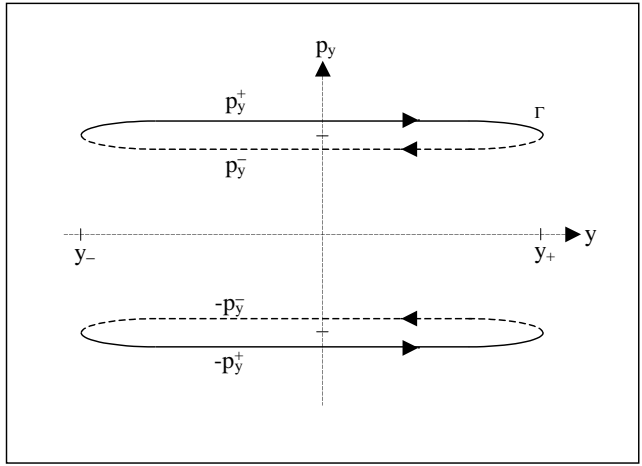,width=86mm,clip=}
		\caption{The branches of $p_y(y)$ with the 
			corresponding velocity directions indicated.}
		\label{fig:AllAndreevOrbitsSNS}
	\end{center}
\end{figure}

To understand  what happens when the two momenta coalesce we 
consider the velocity associated with an excitation moving along each of
the trajectories $p_y^\pm$. For this we use Hamilton's equations
(\ref{eqn:HamiltonsEquations}),
together with the Hamiltonian (\ref{eqn:ClassicalHamiltonianForSNS}) for 
this problem. We have
\begin{align}
	\dot{\bf r}
	&=
	\left(
		\frac{\partial E ({\bf p},{\bf r}) }
			{\partial {\bf p}}
	\right),
\nonumber \\
	&=
	\frac{\partial }{\partial {\bf p}}
	\sqrt{
		\left(
			\frac{{\bf p}^2}{2m}
			-\epsilon_F
		\right)^2
		+ |\Delta(y)|^2
	},
\nonumber \\
	&=
	\frac{
		\left(
			\frac{{\bf p}^2}{2m}
			-\epsilon_F
		\right)
	     }
	     {
		\sqrt{
			\left(
				\frac{{\bf p}^2}{2m}
				-\epsilon_F
			\right)^2
			+ |\Delta(y)|^2
		     }
	     }
	\frac{{\bf p}}{m},
\nonumber
\end{align}
all evaluated for ${\bf p}={\bf p}^\pm({\bf r})$ and ${\bf r}={\bf r}^\pm(t)$.
By using equation (\ref{eqn:MomentumBranchesForSNS1}) the velocity is
\begin{equation}
	{\bf v}^\pm
	= \pm \frac{
			\sqrt{
				E^2 - |\Delta(y)|^2
			     }
		     }{E}
	\frac{
		\left(
			p_x,p_y^\pm(y),p_z
		\right)
	     }
	     {m}.
\label{eqn:GroupVelocitySNS}
\end{equation}
In the limit $|\Delta(y)|\rightarrow 0$,
${\bf v}^\pm\rightarrow \pm \left( p_x,p_y^\pm(y),p_z \right)$. Let
$p_y^\pm>0$ as in FIG.~\ref{fig:OneAndreevOrbit}, then for a particle the
velocity ${\bf v}^+$ is directed to the right, whilst
the hole velocity is directed to the left. This situation is reversed if the
negative branches of $p_y^\pm$ are considered. 
FIG.~\ref{fig:AllAndreevOrbitsSNS} shows all the branches of $p_y(y)$ 
together with arrows indicating the velocity of the excitation.
%----------------------------------------------------------------------
%---------------------------------------------------------------------
Consider a particle travelling along
$p_y^+>0$ approaching $y_+$. The velocity ${\bf v}^+>0$ decreases until it
is zero at $y_+$. At this point $p_y^+(y_+)=p_y^-(y_+)$ and the particle
converts to a hole. It then moves away from the interface with 
${\bf v}^-<0$. Thus even though $p^\pm_y(y)\neq 0$, $y=y_\pm$ is a
turning point of the orbit.
From (\ref{eqn:GroupVelocitySNS}) we see the quantity responsible for the 
change in sign of the velocity is the scalar function 
$\pm\sqrt{E^2 - |\Delta(y)|^2}/E$ which multiplies {\it all three}
components of the momentum ${\bf p}$. Thus at reflection all velocity
components reverse their direction despite both $p_x$ and $p_z$ being
constants of the 
%-----------------------footnote----------------------------------------
motion.\footnote{It is clear that $p_x$ and $p_z$ are constants of the 
motion because both $x$ and $z$ are cyclic coordinates so that 
$\dot{p_x}=0$ and $\dot{p_z}=0$.}
%----------------------endfootnote--------------------------------------
Such a process is called Andreev reflection ~\cite{andreev:64:0},
and we have demonstrated that our `classical' Hamiltonians describe this
remarkable property of quasiparticle excitations in a superconductor.

\begin{figure}
	\begin{center}
		\epsfig{file=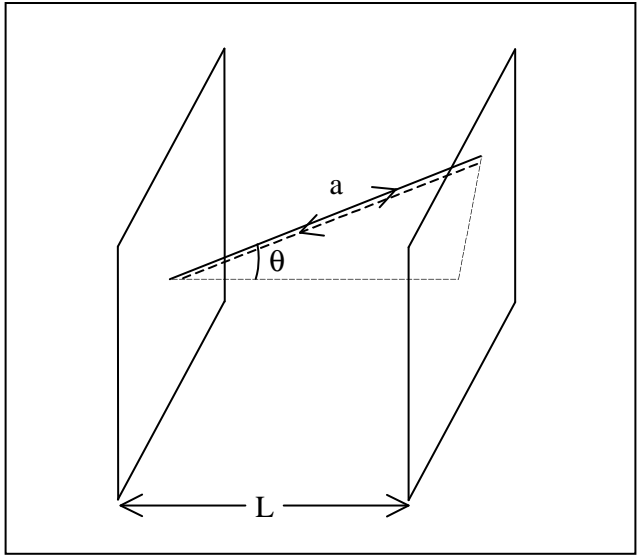,width=86mm,clip=}
		\caption{Illustration of the unique nature of Andreev
			scattering - retroreflection.}
		\label{fig:RealSpaceAndreevReflectionSNS}
	\end{center}
\end{figure}

Once the hole has travelled across the normal layer the reverse 
process happens, the hole turns into a particle, and the excitation completes a
periodic orbit. Thus we have bound excitations in the normal layer whose
orbit consists of both quasiparticle and quasihole segments. 
We call $y_\pm$ 
(defined implicitly by $\epsilon(y_\pm)=0$) Andreev turning points to
distinguish them from  normal turning points, $y_i$, for which
$p_y^\pm(y_i)=0$.
FIG.~\ref{fig:RealSpaceAndreevReflectionSNS} conveys this remarkable
feature of the reflection in real space.
%---------------------figure----------------------------------------------
%---------------------------------------------------------------------

Now that we have identified the classical orbits we can turn our attention
to the semiclassical spectrum.
%------------------------------------------------------------------------
\subsection{Calculation of semiclassical spectrum}
We deploy our generalised quantisation rule
\begin{equation}
	\oint_{\Gamma} {\bf p}^\alpha ({\bf r};\hbar)\! \cdot d{\bf r}
	=2\pi \hbar \left( 
				n
				+ \frac{m}{4}
				- \frac{m^\phi}{2}
		   \right).
\nonumber
\end{equation}
Since $p_x$ and $p_z$ are constant along the orbit, which retraces itself,
$\oint p_x dx=\oint p_z dz =0$. Furthermore $m^\phi=0$ since there are no
vortices. Our quantisation condition becomes
\begin{equation}
	\oint_{\Gamma} p_y^\beta (y;\hbar) dy
	=2\pi \hbar \left( 
				n
				+ \frac{m}{4}
		   \right),
\label{eqn:BohrSommerfeldRuleSNS}
\end{equation}
with $\beta=\pm$ along the appropriate segments of the trajectory, $\Gamma$,
drawn in FIG.~\ref{fig:AllAndreevOrbitsSNS}. If we calculate the right hand
side of (\ref{eqn:BohrSommerfeldRuleSNS}), which depends upon $E$ through
$p_y^\beta$, we can find the semiclassical
spectrum of bound Andreev excitations in the SNS system. This we now
do.

For our example we have
\begin{equation}
	\oint_{\Gamma} p_y^\beta(y) \ dy
	=
	\int^{y_+}_{y-} p^+_y(y) \ dy
	-
	\int^{y_+}_{y-} p^-_y(y) \ dy.
\label{eqn:ParticleHoleBohrRule}
\end{equation}
At this stage we approximate the momentum branches $p^\pm_y(y)$ by
replacing $|\Delta(y)|$ by a step $|\Delta|$. There is no need to 
do
%-----------footnote---------------------------------------------
this\footnote{In appendix \ref{sec:ComplexMethodSNS}, when the Maslov
index $m_l$ is calculated, it is done so for a general
smoothly varying $|\Delta(y)|$, see equation (\ref{eqn:SmoothDeltaExpansion}).}
%-----------endfootnote--------------------------------------------
other than to facilitate a comparison of our results with
Andreev ~\cite{andreev:66:0}. 
With this
approximation the momentum branches $p_y^\pm$ in the normal layer are
given by
\begin{align}
	p_y^\pm
	&=
	\sqrt{
			p_F^2 - p_\bot^2 \pm 2mE
	     }, \nonumber \\
	&=
	\sqrt{
		\left(
			p_F^2 - p_\bot^2 
		\right)
	     }
	\left(
		1\pm \frac{2mE}{p_F^2 - p^2_\bot}
	\right)^{1/2},
\nonumber
\end{align}
where $p_\bot^2=p_x^2+p_z^2$. The quantity
$2mE/(p_F^2 - p^2_\bot) \ll 1$ for almost all values of $p_\bot$
except $p_\bot \rightarrow p_F$. The latter limit implies that the 
excitation hits the interface at glancing incidence which
we do not consider here. 
Let us therefore expand $p_y^\pm$ in powers of $E$. 
Upto and including first order in $E$ the $p_y^\pm$ become
\begin{equation}
	p_y^\pm = p_F|\cos \theta|
		\pm
		\frac{1}{2}
		\frac{2mE}{p_F|\cos \theta|}.
\label{eqn:PyToOrderEOneAngularCoords}
\end{equation}
Here we have used $p_\bot=p_F\sin\theta$ (valid to this order). 
(The modulus has been inserted to remind us that 
$-\pi/2 \leq \theta \leq \pi/2$.)
Returning to (\ref{eqn:ParticleHoleBohrRule}) and using 
(\ref{eqn:PyToOrderEOneAngularCoords}) we have
\begin{equation}
	\int^{y_+}_{y-} dy \ (p^+_y(y) - p^-_y(y))
	=
	\frac{2E}{v_F|\cos \theta|} L,
\label{eqn:BohrSommerfeldAreaSNS}
\end{equation}
where $L=y_+-y_-$ is the width of the normal layer which is the same for all
bound excitations in the step $|\Delta|$ approximation.
Quantising (\ref{eqn:BohrSommerfeldAreaSNS}) directly using 
(\ref{eqn:BohrSommerfeldRuleSNS}) we find ~\cite{andreev:66:0}:
\begin{equation}
	E_n(\theta)
	=
	\pi\hbar v_F
	\left( 
		n + \frac{m}{4}
	\right) 
	\frac{|\cos \theta|}{L}.
\label{eqn:AndreevSpectrum}
\end{equation}
This is the well known spectrum of an excitation trapped in the 
normal layer of an SNS junction (usually obtained by wave function
matching). Since the orbits we have quantised are topologically circles
the Maslov index is $m=2$ (see section \ref{sec:TorusQuantisation}).
We will however calculate it explicitly below.
A number of text books discuss the spectrum, (\ref{eqn:AndreevSpectrum}),
(see for example Abrikosov ~\cite{abrikosov:88:0}) so we will not dwell on
it further. We do, however, want to draw attention to the following points.
Our use of Hamiltons equations, together with a smoothly varying $|\Delta(y)|$,
gave a simple and clear picture of the `particle' to `hole' conversion at
the interface. It is an attractive feature of this theory that our
classical Hamiltonian system describes Andreev retroreflection, rather than
having it arise from wave function matching. The generalised quantisation rule
correctly reproduces the Andreev spectrum. In his book ~\cite{abrikosov:88:0}
Abrikosov invokes Bohr's quantisation rule to obtain this spectrum. However
the justification for using such a quantisation rule is non-trivial as we
have demonstrated.

We now conclude our discussion of the semiclassical theory in the context of
the SNS junction by utilizing the wave function to determine the Maslov index.
%----------------------------------------------------------------------------
\subsection[The semiclassical wave functions for excitations]{The semiclassical wave functions for excitations in SNS junctions}
\label{sec:whereWaveFunctionForOneDSNS2is}
The wave function for this example consists of plane waves in
the $x$ and $z$ directions 
(since $p_x$ and $p_z$ are constants of the motion) together with the one
dimensional form of our semiclassical wave function. 
(Since ${\pmb \nabla}\phi=0$ either of the semiclassical wave functions,
(\ref{eqn:WaveFunctionForOneD}) or (\ref{eqn:WaveFunctionEffective}) in their
one dimensional form can be used because they are identical in this case.)
Thus we have
\begin{multline}
	\left(
		\begin{array}{c}
			u_{I,p_x,p_z} ({\bf r}) \\
			v_{I,p_x,p_z} ({\bf r})
		\end{array}
	\right)
	=
	\sum_{\beta,j}
	A_{\beta,j}
	\left|
		\frac{ 
			\partial^2 S^{\beta,j}(y,I) 
		     }
		     {
			\partial y \partial I
		     }
	\right|^{1/2}\!\!\!
	\left(
		\begin{array}{c}
			u^{\beta,j}_{0,I}(y)
			e^{+i\phi/2} \\
			v^{\beta,j}_{0,I}(y)
			e^{-i\phi/2}
		\end{array}
	\right)\times
\\
	\times
	e^{ip_x x/\hbar+ip_z z/\hbar}
	\exp
	\left(
		i\hbar^{-1}
		S^{\beta,j}(y,I)
		+im\pi/2
	\right)
	,
\label{eqn:WaveFunctionForOneDSNS2}
\end{multline}
where now $\beta$ labels the two distinct momentum branches $p_y^+$ and
$p_y^-$ whilst $j$ specifies the sign of those branches. It is useful to 
separate the four branches $p_y^+,p_y^-,-p_y^+,-p_y^-$, in this way because
both the spinor amplitudes $u^{\beta,j}_{0,I}(y)$ and
$v^{\beta,j}_{0,I}(y)$, and the amplitude factor
$|\partial^2 S^{\beta,j}/\partial y \partial I |^{1/2}$
will turn out to be insensitive to the sign of the momentum. 
The amplitude factor can be rewritten as
\begin{equation}
	\left|
		\frac{ 
			\partial^2 S^{\beta,j}(y,I) 
		     }
		     {
			\partial y \partial I
		     }
	\right|^{1/2}
	=
	\omega^{1/2}
	\left|
		\frac{ 
			\partial E
		     }
		     {
			\partial p_y
		     }
	\right|^{-1/2},
\end{equation}
where $\omega(I)=\partial E/\partial I$ is a constant which we will 
absorb into the amplitudes $A_{\beta,j}$.  
We then have
$S^{\beta,j}_0(y,I)$, $u^{\beta,j}_{0,I}(y)$, $v^{\beta,j}_{0,I}(y)$ 
and $|\partial E/\partial p_y|^{-1/2}$ to investigate.
%---------------------subsubsection------------------------------------
\subsubsection{Spinor amplitudes}
The spinor amplitudes are given by (\ref{eqn:NormalisedAmplitudesEffective})
with ${\bf v}_s({\bf r})=0$ and $|\Delta({\bf r})|=|\Delta(y)|$ i.e
\begin{equation}
	u^\beta_{0,I}(y)
	=
	\sqrt{
		\frac{1}{2}
		\left(
			1+\frac{
				\beta \sqrt{ E^2-|\Delta(y)|^2 } 
			       }
			       { E }
		\right)
	     } .
\label{eqn:NormaliseUinE}
\end{equation}
and
\begin{equation}
	v^\beta_{0,I}(y)
	=
	\sqrt{
		\frac{1}{2}
		\left(
			1-\frac{
				\beta \sqrt{ E^2-|\Delta(y)|^2 } 
			       }
			       { E }
		\right)
	     } \ .
\label{eqn:NormaliseVinE}
\end{equation}
Equations (\ref{eqn:NormaliseUinE}) and (\ref{eqn:NormaliseVinE}) show that
the semiclassical expressions for the spinor amplitudes reproduce the 
familiar form for the SNS junction problem 
(see for example ~\cite{blonder:82:0}).
Briefly consider the two limiting cases $|\Delta(y)|\rightarrow 0$, and
$|\Delta(y)|\rightarrow E$. In the first instance, deep inside the normal
layer, the spinor amplitudes corresponding to $p^+_y$ and $p^-_y$
are easily seen to be
\begin{align}
	\lim_{|\Delta(y)|\rightarrow 0}
	&\left(
		\begin{array}{c}
			u_{0,I}^+ (y) \\
			v_{0,I}^+ (y) 
		\end{array}
	\right)
	=
	\left(
		\begin{array}{c}
			1 \\
			0
		\end{array}
	\right) , \qquad (\mathrm{particle}),
\nonumber \\
	\lim_{|\Delta(y)|\rightarrow 0}
	&\left(
		\begin{array}{c}
			u_{0,I}^- (y) \\
			v_{0,I}^- (y) 
		\end{array}
	\right)
	=
	\left(
		\begin{array}{c}
			0 \\
			1
		\end{array}
	\right) , \qquad (\mathrm{hole}),
\nonumber
\end{align}
so that the spinor amplitude for the  particle-like excitation, corresponding
to the momentum branch $p_y^+(y)$ is 
\begin{math}
	\left( \begin{smallmatrix}
		1 \\
		0
		\end{smallmatrix}
	\right),
\end{math}
and for $p_y^-(y)$, it is
\begin{math}
	\left( \begin{smallmatrix}
		0 \\
		1
		\end{smallmatrix}
	\right).
\end{math}
Notice then that particle- and hole-like excitations decouple in the normal
layer, as is to be expected.
An excitation is either particle or hole, not a mixture. In the second limit,
$|\Delta(y)|\rightarrow E$, we have
\begin{align}
	\lim_{|\Delta(y)|\rightarrow E}
	&\left(
		\begin{array}{c}
			u_{0,I}^+ (y) \\
			v_{0,I}^+ (y) 
		\end{array}
	\right)
	=
	\left(
		\begin{array}{c}
			\frac{1}{\surd 2} \\
			\frac{1}{\surd 2}
		\end{array}
	\right) ,
\nonumber \\
	\lim_{|\Delta(y)|\rightarrow E}
	&\left(
		\begin{array}{c}
			u_{0,I}^- (y) \\
			v_{0,I}^- (y) 
		\end{array}
	\right)
	=
	\left(
		\begin{array}{c}
			\frac{1}{\surd 2}\\
			\frac{1}{\surd 2}
		\end{array}
	\right) .
\nonumber
\end{align}
Thus at the turning point an excitation is an equal mixture of particle and
hole. If we assign an effective charge 
$e^*(y)=e\left( u_{0,I}^2 (y)-v_{0,I}^2(y) \right)$ to
a given spinor then a particle-like excitation has $e^*(y)>0$, a hole-like
excitation has $e^*(y)<0$ and at the turning point $e^*(y)=0$.
(Recall that charge is not a good quantum number in a superconductor.)
As a final comment, notice that for $|\Delta(y)|>E$, 
$u_{0,I}^\beta (y)$ and $v_{0,I}^\beta (y)$ become
complex conjugates. This coincides with the momenta 
$p_y^\beta(y)$ becoming complex.
%------------------------------------------------------------------------
\subsubsection{The velocity dependent amplitude}
The wave function (\ref{eqn:WaveFunctionForOneDSNS2}) contains not only the
position dependent particle and hole
amplitudes $u^\beta_{0,I}(y)$ and 
$v^\beta_{0,I}(y)$, but also the overall amplitude
$\left( \partial E /\partial p\right)^{-1/2}$.
We have
\begin{equation}
	\left(
		 \frac{\partial E({\bf p},y)}
		      { \partial p_y} 
	\right)^{-1/2}_{p_y=p_y^\beta(y)}
	=
	\frac{
		E^{1/2}
	     }
	     {
		\sqrt[\leftroot{8} 4]{
				E^2 - |\Delta(y)|^2
			}
	     }
	\frac{m^{1/2}}
	     {
		\sqrt{ p_y^\beta(y) }
	     }.
\label{eqn:OverallAmplitudeSNS}
\end{equation}
Here the $\surd p_y^\beta(y)$, familiar from semiclassical theory for one
component systems, is supplimented by another term. For the SNS junction
we again consider the two limits $|\Delta(y)|\rightarrow 0$ and 
$|\Delta(y)|\rightarrow E$. In the first instance we have
\begin{equation}
	\lim_{|\Delta(y)|\rightarrow 0}
	\left(
		 \frac{\partial E({\bf p},y)}
		      { \partial p_y} 
	\right)^{-1/2}_{p_y=p_y^\beta(y)}
	=
	\frac{m^{1/2}}
	     {
		\sqrt{ p_y^\beta(y) }
	     },
\nonumber
\end{equation}
so that a particle-like excitation has an amplitude inversely proportional to
the square  root of the particle momentum, and the hole-like excitation 
similarly depends upon the hole momentum. In the second limit as the excitation
approaches the interface 
$\left( \partial E /\partial p\right)^{-1/2}$
diverges, despite $p_y^\beta(y)\neq 0$, due to the 
\begin{math}
	\begin{smallmatrix}
		\left(  E^2-|\Delta(y)|^2 \right)^{-1/4} \\
	\end{smallmatrix}
\end{math}
dependence. This divergence signals the breakdown of the semiclassical
approximation for the wave function when we are too close to the turning
points (i.e., caustics) at which the velocity goes to zero.
%------------------------------------------------------------------------------
\subsubsection{The action $S_0^{\beta,j}$}
The action appearing in (\ref{eqn:WaveFunctionForOneDSNS2}) is given by the
integral of the appropriate momentum branch. We have
\begin{equation}
	S_0^{\beta,j}(y)
	=
	j\int_{y_0}^y dy' \ p_y^\beta(y'),
\label{eqn:ActionSNS} 
\end{equation}
where $j=\pm$ gives the sign of the 
momenta, and $\beta$ as usual distinguishes the branches $p^+_y$
and $p^-_y$. The lower limit $y_0$ is an arbitrary constant called the phase
reference point.

Although the amplitudes $u^\beta_{0,I}(y)$, 
$v^\beta_{0,I}(y)$, and
$\left( \partial E /\partial p\right)^{-1/2}$
acquire their normal state forms as $|\Delta(y)|\rightarrow 0$, the action
(\ref{eqn:ActionSNS}), being an integrated quantity, retains a `memory' of
the interface region encountered by the quasiparticle.

If we consider the form of $S_0^{\beta,j}(y)$ inside the superconductor
($|\Delta(y)|>E$) then the momenta, $p^\beta_y(y)$, become complex and
the action correspondingly has both a real and imaginary part:
\begin{equation}
	S_0^{\beta,j}(y)
	=
	j\int_{y_0}^y dy' \ p_y^r(y')
	+
	\mathrm{i} (j\beta) \int_{y_0}^y dy'\ p_y^i(y'),
\label{eqn:ComplexActionSNS} 
\end{equation}
where $p_y^r$ and $p_y^i$ are the real and imaginary parts of the 
momenta. 
%----------------------------------------------------------------------
\subsection{Derivation of the  quantisation condition and
determination of the Maslov index}
\begin{figure}
	\begin{center}
		\epsfig{file=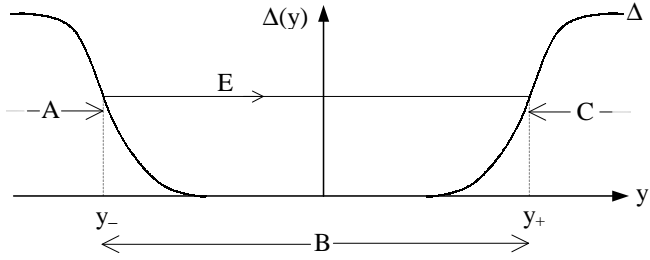,width=86mm,clip=}
		\caption[Classically forbidden and allowed regions.]
			{The classically forbidden regions, A, C, and 
			the classically allowed region B for an excitation
			confined by $\Delta(y)$.}
		\label{fig:SNSProfileABC}
	\end{center}
\end{figure}

We now construct a specific solution. In all that follows we will concentrate
upon the $y$-dependence of the wave functions (i.e., we will drop the
plane wave factors for the $x$ and $z$ directions since they play no r\^ole
in this derivation).
As a consequence of (\ref{eqn:ComplexActionSNS})
the wave function in the superconducting region is comprised of only those
solutions which satisfy the physical boundary conditions, decaying as
$y \rightarrow \pm\infty$. Let us label the three distinct regions of the
SNS junction A, B, C FIG.~\ref{fig:SNSProfileABC} where B is the `classical'
region $y_-<y<y_+$. 
%------------figure-------------------------------------------------------
%------------endfigure----------------------------------------------------
Then in region A the wave function will consist of solutions which decay as
$y\rightarrow -\infty$. These contain the terms
\begin{equation}
	\exp \left(
			\pm
			\frac{i}{\hbar}
			\int_y^{y_-} dy' \ p_y^r(y')
	     \right)
	\exp \left(
			-\frac{1}{\hbar}
			\int_y^{y_-} dy' \ p_y^i(y')
	     \right),\ \ 
	\text{(region A)}
\label{eqn:WavesDecayingToMinusInfty}
\end{equation}
where $p_y^i (y)\geq 0$ and the phase reference point has been taken to 
be $y_-$. Similarly in region C solutions will contain the terms
\begin{equation}
	\exp \left(
			\pm
			\frac{i}{\hbar}
			\int_{y_+}^{y} dy' \ p_y^r(y')
	     \right)
	\exp \left(
			-\frac{1}{\hbar}
			\int_{y_+}^{y} dy' \ p_y^i(y')
	     \right),\ \
	\text{(region C)}
\label{eqn:WavesDecayingToPlusInfty}
\end{equation}
where now the phase reference point is $y_+$. Finally, in region B, 
the eigenfunction will consist of a
superposition of all four solutions corresponding to the four actions
$S_0^{\beta,j}(y)$ (\ref{eqn:ActionSNS}), one for each of the
classical momentum branches displayed previously in figure 
\ref{fig:AllAndreevOrbitsSNS}. 

Let us write the appropriate superposition in 
each region as $\Psi_A^-$ (phase reference $y_-$), $\Psi_B^-$, $\Psi_B$
(phase reference $y_+$), and $\Psi_C$. 
Our asymptotic solutions are only valid away from the turning points at 
$y_\pm$. 
We match $\Psi_C \rightarrow \Psi_B$ and $\Psi_B^- \rightarrow \Psi_A^-$ by
analytically continuing our solutions into the complex coordinate 
plane in order to circumvent $y_\pm$. The interested reader is refered to 
Appendix \ref{sec:ComplexMethodSNS} for the technical details. 
Here we summarise the results.
%\begin{widetext}
Starting from a complex evanescent solution satisfying the boundary
conditions at $y=+\infty$
\begin{align}
	\Psi_C(y)
	&=
		\frac{ A }
		     {
			\sqrt{
				\left(
					\frac{\partial E }
			    		{\partial p_y}
				\right)
			     }
		     }_{\!\!p_y^+(y)} \!\!\!\!\!\!\!\!\!\!\!
		\left(  \!\!
			\begin{array}{c}
				u^+_{0,I}(y)e^{+i\phi/2} \\
				v^+_{0,I}(y)e^{-i\phi/2}		
			\end{array}
			\!
		\right)
		e^{\textstyle
			+\frac{i}{\hbar}
			\int_{y_+}^y p_y^{r}(y')dy'
		  }
	e^{\textstyle
		-\frac{1}{\hbar} 	
		\left|
			\int^y_{y_+} p_y^{i}(y')dy'
		\right|
	  }
\nonumber \\
	&+
		\frac{ B }
		      {
			\sqrt{
				\left(
					\frac{\partial E }
		    			{\partial p_y}
				\right)
			     }
		      }_{\!\!p_y^-(y)} \!\!\!\!\!\!\!\!\!\!\!
		\left(  \!\!
			\begin{array}{c}
				u^-_{0,I}(y) e^{+i\phi/2}\\
				v^-_{0,I}(y) e^{-i\phi/2}	
			\end{array}
			\!
		\right)
		 e^{\textstyle
			-\frac{i}{\hbar}
			\int_{y_+}^y p_y^{r}(y')dy'
		   } 
	e^{\textstyle
		-\frac{1}{\hbar} 	
		\left|
			\int^y_{y_+} p_y^{i}(y')dy'
		\right|
	  },
\label{eqn:TheExplicitDecayingSolution}
\end{align}
we follow a contour in the upper half of the complex plane around
$y_+$ to obtain
\begin{align}
	\Psi_B(y)
	&=
	\frac{ 1 }
	     {
		\sqrt{
			\left(
				\frac{\partial E }
		    		     {\partial p_y}
			\right)
		     }
	     }_{\!\!p_y^+(y)} \!\!\!\!\!\!\!\!\!\!\!
	\left( 
		\begin{array}{c}
			u^+_{0,I}(y)e^{+i\phi/2} \\
			v^+_{0,I}(y)e^{-i\phi/2}		
		\end{array}
	\right)
	\times
\nonumber \\
	&\qquad \qquad \qquad
	\times
	\left(
		 A e^{\textstyle
			-\frac{i}{\hbar}
			\int^{y_+}_y  p_y^+(y')dy'
		    }
		 -iB
		e^{\textstyle 
			+\frac{i}{\hbar}
			\int^{y_+}_y  p_y^+(y')dy'
		  }
	\right)
\nonumber \\
	&+
	 \frac{ 1 }
	      {
		\sqrt{
			\left(
				\frac{\partial E }
	    			     {\partial p_y}
			\right)
		     }
	      }_{\!\!p_y^-(y)} \!\!\!\!\!\!\!\!\!\!\!
	\left(
		\begin{array}{c}
			u^-_{0,I}(y) e^{+i\phi/2}\\
			v^-_{0,I}(y) e^{-i\phi/2}	
		\end{array}
	\right) \times
\nonumber \\
	&\qquad \qquad \qquad
	\times
	\left(
	 	B e^{\textstyle
			+\frac{i}{\hbar}
			\int^{y_+}_y p_y^-(y')dy'
		    }
		-iA
		 e^{\textstyle
			-\frac{i}{\hbar}
			\int^{y_+}_y p_y^-(y')dy'
		    }
	\right),
\label{eqn:TheExplicitOscillatorySolution}
\end{align}
for $y_- < y < y_+$, and continuing around $y_-$ we find 
\begin{align}
	\Psi_A^-(y)
	&=
	\frac{ B }
	     {
		\sqrt{
			\left(
				\frac{\partial E }
		    		     {\partial p_y}
			\right)
		     }
	     }_{\!\!p_y^-(y)} \!\!\!\!\!\!\!\!\!\!\!
	\left(
		\begin{array}{c}
			u^-_{0,I}(y)e^{+i\phi/2} \\
			v^-_{0,I}(y)e^{-i\phi/2}		
		\end{array}
	\right) \!\!
	\left( \!
		e^{\textstyle
			\frac{i}{\hbar}
			\int^{y_+}_{y_-} p_y^+(y')dy'
		  }
		\!\!\!\!+
		e^{\textstyle
			\frac{i}{\hbar}
			\int^{y_+}_{y_-} p_y^-(y')dy'
		  }\!
	\right)
\nonumber \\
	&\quad \times
	e^{\textstyle
			\frac{i}{\hbar}
			\int^{y_-}_y p_y^{r}(y')dy'
		   }
	e^{\textstyle
		+\frac{1}{\hbar} 	
		\left|
		\int_y^{y_-} p_y^{i}(y')dy'
		\right|
	  }
\nonumber \\
	&-\frac{ iA }
	     {
		\sqrt{
			\left(
				\frac{\partial E }
		    		     {\partial p_y}
			\right)
		     }
	     }_{\!\!p_y^-(y)} \!\!\!\!\!\!\!\!\!\!\!
	\left(
		\begin{array}{c}
			u^-_{0,I}(y)e^{+i\phi/2} \\
			v^-_{0,I}(y)e^{-i\phi/2}		
		\end{array}
	\right)
	e^{\textstyle
		-\frac{i}{\hbar}
		\int^{y_+}_{y_-} p_y^-(y')dy'
	  }
	e^{\textstyle
		-\frac{i}{\hbar}
		\int^{y_-}_y p_y^{r}(y')dy'
	   }
\nonumber \\
	&\quad \times
	e^{\textstyle
		-\frac{1}{\hbar} 	
		\left|
		\int_y^{y_-} p_y^{i}(y')dy'
		\right|
	  }
\nonumber \\
	&-\frac{ iB }
	     {
		\sqrt{
			\left(
				\frac{\partial E }
		    		     {\partial p_y}
			\right)
		     }
	     }_{\!\!p_y^+(y)} \!\!\!\!\!\!\!\!\!\!\!
	\left(
		\begin{array}{c}
			u^+_{0,I}(y)e^{+i\phi/2} \\
			v^+_{0,I}(y)e^{-i\phi/2}		
		\end{array}
	\right)
	e^{\textstyle
		+\frac{i}{\hbar}
		\int^{y_+}_{y_-} p_y^+(y')dy'
	  }
	e^{\textstyle
		+\frac{i}{\hbar}
		\int^{y_-}_y p_y^{r}(y')dy'
	   }
\nonumber \\
	&\quad \times
	e^{\textstyle
		-\frac{1}{\hbar} 	
		\left|
		\int_y^{y_-} p_y^{i}(y')dy'
		\right|
	  } 
\nonumber \\
	&+\frac{ A }
	     {
		\sqrt{
			\left(
				\frac{\partial E }
		    		     {\partial p_y}
			\right)
		     }
	     }_{\!\!p_y^+(y)} \!\!\!\!\!\!\!\!\!\!\!
	\left(
		\!\!\!
		\begin{array}{c}
			u^+_{0,I}(y)e^{+i\phi/2} \\
			v^+_{0,I}(y)e^{-i\phi/2}		
		\end{array}
		\!\!\!
	\right) \!\!
	\left( \!
		e^{\! \textstyle
			-\frac{i}{\hbar}
			\int^{y_+}_{y_-} p_y^+(y')dy'
		  }
		\!\!\!\!+
		e^{\! \textstyle
			-\frac{i}{\hbar}
			\int^{y_+}_{y_-} p_y^-(y')dy'
		  }\!
	\right)
\nonumber \\
	&\quad \times
	e^{\textstyle
		-\frac{i}{\hbar}
		\int^{y_-}_y p_y^{r}(y')dy'
	   }
	e^{\textstyle
		+\frac{1}{\hbar} 	
		\left|
		\int_y^{y_-} p_y^{i}(y')dy'
		\right|
	  },
\label{eqn:TheExplicitDecayingAndGrowingSolution}
\end{align}
for $y<y_-$. (The spinor and square root prefactors in $\Psi_A^-$ and 
$\Psi_C$ which are complex can be found in appendix 
\ref{sec:ComplexMethodSNS}, equation (\ref{eqn:ComplexPrefactors}).)
To satisfy the boundary condition at $y=-\infty$ that our
solution vanishes, the coefficients of the first and fourth terms 
of $\Psi_A^-(y)$ must be zero, i.e. 
\begin{equation}
	e^{
			\frac{i}{\hbar}
			\int^{y_+}_{y-} p^+_y(y') \ dy'
			-\frac{i}{\hbar}
			\int^{y_+}_{y-} p^-_y(y') \ dy'
	  }
	=
	e^{i\pi(2n+1)
	  }.
\nonumber
\end{equation}
Our quantisation condition is then
\begin{equation}
	\frac{1}{\hbar}	
	\int^{y_+}_{y-} p^+_y(y') - p^-_y(y')\ dy'
	=
	2\pi \left(n + \frac{1}{2} \right),
\label{eqn:ProvedBohrSommerfeldRuleSNS}
\end{equation}
which is precisely the quantisation rule used earlier
(equation (\ref{eqn:BohrSommerfeldRuleSNS})) to derive the spectrum of
Andreev quasiparticles. We have also proved that the Maslov index is $m=2$
(valid for quasiparticles bound by a smoothly varying order parameter, 
$|\Delta(y)|$).
%--------------------------------------------------------------------------
%--------------------------------------------------------------------------
\section{The Single Vortex}
\label{sec:theSingleVortex}
The single vortex is an ideal problem to demonstrate how our semiclassical
theory works since, as will be seen below, both the topological phase and
the $\hbar$-dependence of the Hamiltonian play important r\^oles. We will
also (in an accompanying appendix) be able to make a direct comparison
between the two semiclassical theories.

\begin{figure}[tbp]
	\begin{center}
		\epsfig{figure=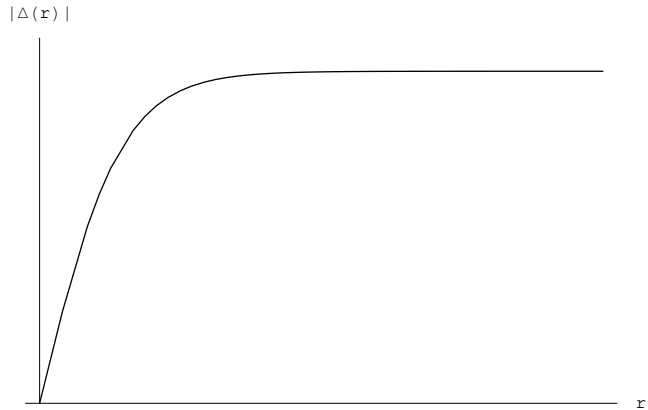,clip=,width=86mm}
		\caption{Profile of radial dependence of $\Delta({\bf r})$}
		\label{fig:deltaprofile}
	\end{center}
\end{figure}

The order parameter for a single vortex in a cylindrical
coordinate system takes the form
\begin{equation}
	\Delta({\bf r})
	=
	|\Delta(r)|e^{-i\theta},
\nonumber
\end{equation}
where the radial profile $|\Delta(r)|$ is shown in FIG.~\ref{fig:deltaprofile}.
%-------------------------------------------------------------------------
%--------------------------------------------------------------------------
Since ~\cite{degennes:89:0}
\begin{equation}
	\frac{e{\bf A}({\bf r})}{\hbar {\pmb \nabla}\phi}
	\sim
	\frac{B}{B_{c2}}
	\ll 1,
\nonumber
\end{equation}
for $r\lesssim \xi$, in this regime
\begin{equation}
	m{\bf v}_s({\bf r})
	=
	\frac{\hbar {\pmb \nabla}\phi}{2}
	=-\frac{\hbar}{2r}\hat{\bf \theta},
\nonumber
\end{equation}
i.e. the r\^ole of the superfluid is {\it entirely} represented through this
$\hbar$-dependent term. 
Notice, since ${\bf A}({\bf r})\cong 0$ the classical dynamics governed by
$E_0^\alpha({\bf p}_0,{\bf r})$ is not influenced by the superfluid flow,
whose r\^ole appears in a non-zero 
phase $S_1^{\alpha,j}({\bf r})$. However the classical dynamics governed by
$E^\alpha({\bf p},{\bf r})$ does include the influence of ${\bf v}_s({\bf r})$.
It is the latter which we investigate here.

As already dicussed, $\hbar {\pmb \nabla}\phi/2e$ can be interpreted as
the vector potential associated with a line of phase singularities
which in the present case runs along the $z$-axis carrying a flux, 
$\Phi_0$, from positive
to negative $z$. From this vantage point we now investigate the behaviour of
quasiparticles in the presence of a vortex.

We follow Caroli, de Gennes and Matricon ~\cite{caroli:64:0} by seeking a
solution to the positive angular momentum branch of the spectrum, i.e.
we take $p_\theta \geq 0$. Since then only 
$E^+({\bf p},{\bf r}) \geq 0$
our Hamiltonian describing the quasiparticle-hole excitations is
\begin{equation}
	E^+(p_r,p_\theta,p_z,r)
	=
	-\frac{p_\theta \hbar}{2m r^2}
	+
	\sqrt{
		\left(
			\frac{p_r^2}{2m}
			+\frac{p_\theta^2 +\hbar^2/4}{2m r^2}
			+ \frac{p_z^2}{2m}
			-\epsilon_F
		\right)^2
		+
		|\Delta(r)|^2
	     }.
\label{eqn:SingleVortexHamiltonian}
\end{equation}
Since it does not depend upon $\theta$ or $z$, $p_\theta$ and $p_z$ are
cyclic variables (constants of the motion). $p_z$ is a continuous variable
but $p_\theta$ must be quantised.
%-------------------------------------------------------------------------
\subsection{$p_\theta$ quantisation}
We apply the generalised EBK quantisation rule 
(\ref{eqn:MulticomponentEBKcondition}) to a path encircling the origin.
For this path the Maslov index is zero, due to the fact that there 
are no turning points, but the topological phase associated
with the vortex (\ref{eqn:StringTopologicalIndexAgain}) is important:
\begin{equation}
	\frac{1}{\hbar}
	\int_0^{2\pi}
	p_\theta d\theta
	=
	2\pi \left(
			\nu
			-
			\frac{m^\phi}{2}
	    \right),
\nonumber
\end{equation}
where $\nu$ and $m^\phi$ are integers.
Writing $p_\theta=\hbar \mu$, $\mu$ an angular momentum quantum number, and
using $m^\phi=\frac{1}{2\pi}\oint {\pmb \nabla}\phi \cdot d{\bf r}=-1$, 
we find
\begin{equation}
	\mu = \nu +\frac{1}{2}, \qquad \nu=0,1,2,\dots
\nonumber
\end{equation}
The topological phase forces $\mu$ to be {\it half integer}. This agrees with 
the classic work of Caroli, de Gennes and Matricon ~\cite{caroli:64:0} who
obtained the result in a very different way, namely by matching wave 
functions. 

From 
our present vantage point we see that the effect of the half-flux flowing 
along the $z$-axis string is to give both particles and holes
an angular momentum `kick'.  

%----------------------------------------------------------------------------
\subsection{Radial behaviour of the excitation orbits}
Setting $E=E^+(p_r,p_\theta,p_z,r)$ and inverting, the radial momemtum branches
are found to be
\begin{equation}
	p_r^\pm(r)
	=
	\sqrt{
		p_F^2 -p_z^2 -\frac{\hbar^2 (\mu^2 +1/4)}{r^2}
		\pm 2m \sqrt{
				\left( E +\frac{\hbar^2 \mu}{2m r^2} \right)^2
				-
				|\Delta(r)|^2
			    }
	     } \ ,
\label{eqn:MomentumSegmentsDeltaOfr}
\end{equation}
the $\pm$ corresponding to particle-like or hole-like excitations respectively.
Notice the effect upon $p^\pm_r(r)$ of having $\hbar {\pmb \nabla}\phi/2$ in 
the Hamiltonian is to introduced the two new terms
$\hbar^2/4r^2$ and $\hbar^2 \mu/r^2$.  

\begin{figure}[htbp]
	\begin{center}
		\epsfig{figure=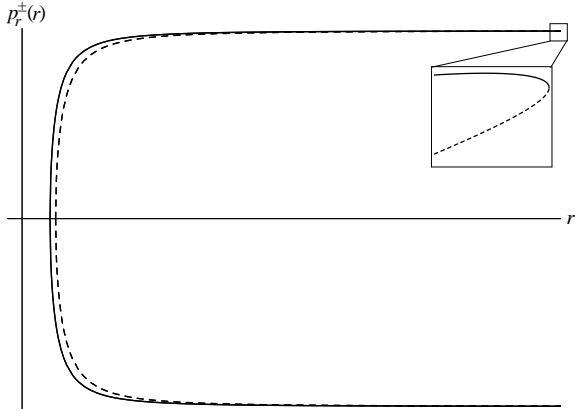,clip=,width=86mm}
		\caption{$p^\pm_r(r)$ as a function of $r$. Solid line
			denotes a particle-like branch, a broken line a
			hole-like branch. The inset shows an enlarged view of
			the turning point where Andreev reflection occurs.}
		\label{fig:radialorbit}
	\end{center}
\end{figure}

FIG.~\ref{fig:radialorbit} shows the radial form of the orbit.
%-------------------------------------------------------------------------
%--------------------------------------------------------------------------
One should notice the following features:
(i) The complete orbit is comprised of both particle-like and hole-like
segments. 
(ii) As $r \rightarrow 0$ both $p_r^\pm(r) \rightarrow 0$ i.e. we have
classical turning points. Since, by Hamiltons equations,
\begin{equation}
	\dot{r}^\pm=\pm\frac{\sqrt{
				\left(
					E+\hbar^2\mu/2mr^2
				\right)^2 
				-
				|\Delta(r)|^2
			       }
			}
			{E}
		\frac{p_r^\pm(r)}{m},
\nonumber
\end{equation}
for $p_r^\pm(r) > 0$ we have $\dot{r}^+>0$ and $\dot{r}^-<0$ i.e. upon the 
positive quasiparticle branch the excitation is moving away from the vortex 
core whilst along the positive quasihole branch it is moving 
towards the vortex core.
(iii) As the excitation propagates into the superconductor the velocity 
becomes zero at a point, $r_d$, defined by
\begin{equation}
	\sqrt{
				\left(
					E+\hbar^2\mu/2mr_d^2
				\right)^2 
				-
				|\Delta(r_d)|^2
			       }
	=0,
\label{eqn:MASTurningPoint}
\end{equation}
and beyond this point $p_r^\pm$ become {\it complex}. Thus $r_d$
is also a turning point. However since 
$p_r^+(r_d)=p_r^-(r_d)$, $r_d$ is the point at which a 
particle-like excitation converts smoothly into a hole-like excitation
(see inset in Fig.~\ref{fig:radialorbit}).
%--------------------------------------------------------------------------- 
\subsection{Calculating the semiclassical spectrum}
Our general quantisation rule for the radial momentum reads:
\begin{equation}
		\oint p^\beta_r(r)dr=2\pi\hbar
		\left(
			n+\frac{m}{4}
		\right),
\label{eqn:VortexQuantisationRule}
\end{equation}
where $\beta=\pm$. $p^\beta_r(r)$ stands for the appropriate branches of the 
momentum along the orbit in Fig.~\ref{fig:radialorbit}. (The Maslov index is 
$m=2$ since the orbit is topologically a circle.)
In general we cannot obtain an analytic solution to the integral in equation
(\ref{eqn:VortexQuantisationRule}).
None the less we can make some headway by using a simple model for the profile 
of $\Delta$ which retains the essential physics. 
%-----------------------------------------------------------------------------
\subsubsection{Model step profile for $|\Delta(r)|$}
The model we adopt replaces
the profile shown in Fig.~\ref{fig:deltaprofile} with a step profile. In
particular we take:
\begin{equation}
	|\Delta(r)|
	=
	\left\{
		\begin{array}{rr}
			0,&\quad r\leq \xi, \\
			\Delta_\infty&\quad r>\xi,
		\end{array}
	\right.
\label{eqn:modeldelta}
\end{equation}
where $\xi$ is a length scale characterising the distance over which 
$|\Delta(r)|$ rises, i.e. represents the vortex core. 
For instance we might take it to be the BCS coherence
length $\xi_0=\hbar v_F/\pi \Delta_\infty$, but we need not do this. (We
have in mind here the work of Gygi and Schl\"uter ~\cite{gygi:91:0} 
who have shown that $\xi\ll \xi_0$
as the temperature approaches zero, and indeed can end up comparable with
the atomic spacing.)

Our model has the following features for the bound states $E<\Delta_\infty$:
(i) when $|\Delta(r)|=\Delta_\infty$ the solution of 
equation (\ref{eqn:MASTurningPoint}) for the turning point is
\begin{equation}
	r_d
	=
	\frac{
		\hbar \mu^{1/2}
	     }
	     {
		\sqrt{	
			2m(\Delta_\infty-E)
		     }
	     }.
\nonumber
\end{equation}
We note that $r_d \rightarrow \infty$ for $\mu\rightarrow \infty$ or 
$E\rightarrow \Delta_\infty$, i.e a bound excitation with in a 
high angular momentum state can in principle 
propagate far inside the superconductor.
(ii) when $|\Delta(r)|=0$ the only natural turning point encountered away
from the core is set by
$\xi$.
Thus all excitations for which $r_d <\xi$ are confined to the vortex core
turning around at $\xi$, whilst those which have $r_d \ge \xi$ 
penetrate into the bulk of the superconductor, turning around at $r_d(\mu)$. 
%---------------------------------------------------------------------------
\subsubsection{Calculation of EBK integrals and Spectrum}
(i) For excitations confined to the core the $p^\pm(r)$ simplify to 
\begin{equation}
	p_r^\pm(r)
	=
	\sqrt{
		p_F^2 -p_z^2 -\frac{\hbar^2 (\mu \mp 1/2)^2}{r^2}
		\pm 2mE
	     } \ .
\label{eqn:MomentumSegmentsDeltaZero}
\end{equation}
Integrals of these functions can be done exactly in terms of elementary
functions. Thus the integral in (\ref{eqn:VortexQuantisationRule})
becomes
\begin{align}
	\oint p^\beta_r(r)dr
	=
	&\sum_{\beta=\pm}
	\beta 2 \hbar
	\left(
		\mu -\beta \frac{1}{2}
	\right)
	\times
\nonumber \\
	&\times
	\left[
		\sqrt{
			\left(
				\xi/L_\beta
			\right)^2
			-1
		     }
		-
		\arctan
		\sqrt{
			\left(
				\xi/L_\beta
			\right)^2
			-1
		     }
	\right],
\nonumber
\end{align}
where the length, $L_\beta$, is defined to be
\begin{equation}
	L_\beta
	=
	\frac{
		\hbar \left(
				\mu -\beta \frac{1}{2}
		      \right)  
	     }
	     {
		\sqrt{
			p_F^2 \sin^2\alpha +\beta 2mE
		     }
	     }.
\nonumber
\end{equation}
Here we introduced the angle $\alpha$ through  $p_z=p_F\cos \alpha$.
($\alpha=0$ or $\pi$ corresponds 
to the excitation travelling along the direction of the vortex axis.)
Excluding excitations with high angular momentum or large  
$p_z$ ($\sin \alpha \rightarrow 0$), we see this length scale is small 
compared to $\xi$.
Thus, for $(\xi/L_\beta)^2 \gg 1$, we expand our result to find
\begin{align}
	\oint p^\beta_r(r)dr
	=
	\sum_{\beta=\pm}
	&\beta 2 \hbar
	\left(
		\mu -\beta \frac{1}{2}
	\right)
\times
\nonumber \\
\times
	&\left[
		\frac{\xi}{L_\beta}
		-\frac{\pi}{2}
		+\frac{L_\beta}{2\xi}
		+{\cal{O}}\left( (L_\beta/\xi)^3\right)
	\right].
\label{eqn:VortexIntegralSmallMuResult}
\end{align}
The energy, entering into this expression through $L_\beta$, is quantised
using equation (\ref{eqn:VortexQuantisationRule}). Taking the first two
terms in (\ref{eqn:VortexIntegralSmallMuResult}), and seeking the energy
in the form $E=E_0 +E_1+\cdots$ we find for $E_0$
\begin{equation}
	E_0
	=
	\frac{
		\pi \hbar v_F |\sin \alpha|
	     }
	     {2\xi}
	\left(
		n+\frac{m}{4}-\frac{1}{2}
	\right).
\label{eqn:ZerothOrderVortexSpectrum}
\end{equation}
Here we have used $E \ll p_F^2\sin^2 \alpha/2m$, so that
our result is valid for states whose kinetic energy along the field 
direction is small. (This is consistent with the above approximations.)

We are interested in bound states for which $E < \Delta_\infty$. To see
how large $E_0$ is we substitute into (\ref{eqn:ZerothOrderVortexSpectrum})
with $\xi_0=\hbar v_F /\pi \Delta_\infty$ and find (remember $m=2$)
\begin{equation}
	E_0
	=
	\frac{\pi^2}{2}\Delta_\infty |\sin \alpha| n.
\label{eqn:E0ofn}
\end{equation}
For states with $\sin \alpha \sim 1$ only the $n=0$ quantum number
corresponds to a bound excitation. (This agrees with the findings of
Bardeen, K\"ummel, Jacobs and Tewordt ~\cite{bardeen:69:0}.) 
Notice our result becomes stronger for 
$\xi <\xi_0$, although for states with a large $p_z$ component
new branches to the spectrum ($n>0$) may appear. 
Thus to describe the low lying excitations in the vortex
we must take $n=0$ and seek the next order correction to the energy, $E_1$.
Including the $L_\beta/2\xi$ terms from 
(\ref{eqn:VortexIntegralSmallMuResult}),
which should be evaluated at $E=E_0=0$, we find
\begin{equation}
	E_1
	=
	\mu
	\frac{\hbar^2}{2m \xi^2}.
\label{eqn:VortexSpectrum}
\end{equation}
In particular, for $\xi=\xi_0$, the excitation spectrum becomes
\begin{equation}
	E
	= \left(
		\frac{\pi}{2}
	  \right)^2
	\mu \frac{\Delta^2_\infty}{\epsilon_F}.
\label{eqn:SmallmuSpectrum}
\end{equation}
We have recovered the famous 
Caroli, de Gennes, Matricon states ~\cite{caroli:64:0}
of low lying excitations confined to a vortex core which however were found
by matching wave functions. 

As well as reproducing the excitation spectrum one can also show
(see appendix \ref{sec:LimitingBehaviour}) that the radial behaviour of the
semiclassical wave functions reproduces the correct $r$-behaviour near the
origin where the exact result was also given by Caroli, de Gennes and
Matricon ~\cite{caroli:64:0}. 

\begin{figure}[htbp]
	\begin{center}
		\epsfig{file=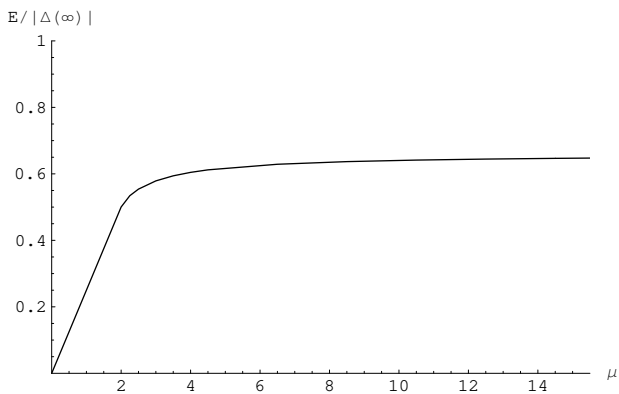,clip=,width=86mm}
		\caption{$E(\mu)$ flattens off as $\mu$ increases.}
		\label{fig:vortex_musmall}
	\end{center}
\end{figure}

\begin{figure}[htbp]
	\begin{center}
		\epsfig{file=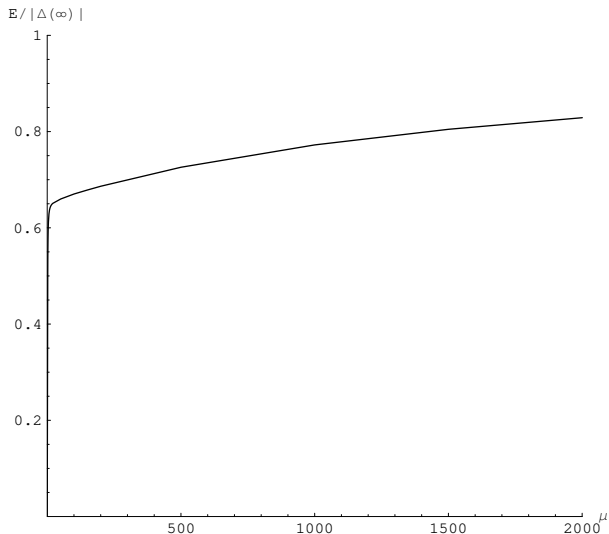,clip=,width=86mm}
		\caption{$E(\mu)$ tends towards $\Delta_\infty$ for very
			large angular momentum.}
		\label{fig:vortex_mularge}
	\end{center}
\end{figure}

(ii) Now we turn our attention to states which penetrate into the bulk of
the superconductor.
For $\xi<r<r_d$ we must add to our integral a further 
contribution with $\Delta(r)=\Delta_\infty$. In this region $p^\pm(r)$
are
\begin{equation}
	p^\pm(r)
	=
	\sqrt{
		p_F^2 -p_z^2 -\frac{\hbar^2 (\mu^2 +1/4)}{r^2}
		\pm 2m \sqrt{
				\left( E +\frac{\hbar^2 \mu}{2m r^2} \right)^2
				-
				\Delta^2_\infty
			    }
	     } \ .
\label{eqn:MomentumSegmentsDeltaInfty}
\end{equation}
The analytic calculation of integrals of nested square root functions is 
very complicated. For example
the Riemann surface upon which $p^+(r)$ 
(as defined by (\ref{eqn:MomentumSegmentsDeltaInfty})) 
is single valued has the topology of
a sphere with 5 handles! 
The problem of evaluating a quantisation rule written as a line integral
can be transformed
into evaluating contour integrals on these Riemann surfaces. In general, and in
the example shown, the answer is not expressible in terms of either elementary
or Elliptic functions. The integrals are however rather easy to 
evaluate numerically. This we have done and quantised the areas
to obtain the spectrum, $E(\mu)$. The result is
shown in figures \ref{fig:vortex_musmall} and \ref{fig:vortex_mularge}. 
%-------------------------------------------------------------------------
%-------------------------------------------------------------------------

Observe that whilst for small $\mu$ the spectrum increases steeply that once
the excitation starts to penetrate the superconducting bulk it `feels' less
confined and hence $E$ flattens off. Only very high angular momentum states
have their energy approaching $\Delta_\infty$.

%-------------------------------------------------------------------------
%-------------------------------------------------------------------------
As stated above, the small $\mu$ behaviour of $E(\mu)$
agrees with the work of Caroli, de Gennes and Matricon ~\cite{caroli:64:0}, 
and now we also see that the global behaviour of $E(\mu)$ is in agreement with
the general findings of Bardeen, K\"ummel, Jacobs and 
Tewordt ~\cite{bardeen:69:0}.
However our method of solution differs substantially from the work of these
authors. 
%-------------------------------------------------------------------
\section{Summary and Discussion}
In this paper we have developed a semiclassical theory for quasiparticles
in superconductors. In doing so we have pushed semiclassical methods to the
limit by constructing a (fictitous) classical mechanics describing the orbits
of quasiparticles propagating in the presence of lines of phase singularities.
Adopting this approach enabled us to bring the full machinery of modern 
semiclassics to the problem. In particular we used torus quantisation to 
construct a generalised EBK quantisation rule for determining the 
semiclassical spectrum. This rule included both the Maslov index, familiar
from modern semiclassics, and a topological integer arising due to the 
global curvature of the space in which the quasiparticles propagate. The
later is a direct consequence of phase singularities of the pairing
potential, i.e., vortices.
The power of this approach, first considered by Azbel' in the superconducting
context, and extended by us here, lies in the general nature of the 
Hamiltonian system we have constructed. Unlike the approach of other 
authors ~\cite{degennes:89:0,andreev:64:0,bardeen:69:0} where one must return
to the BdG equations for each new problem and consider solving the 
differential equation across various length scales, here our starting point
is with the Hamiltonians, (\ref{eqn:hdependentClassicalHamiltonian}), and 
Hamilton's equations of motion. Once the pairing potential, vector potential
and crystal lattice potential have been chosen for the problem at hand one
can immediately investigate the quasiparticle orbits. In the case where the
classical dynamics is integrable one can then proceed, via the generalised
EBK rule, to quantise the quasiparticle motion. To make this possible a
number of technical challenges needed to be overcome. Most important amoungst
these was the observation that `standard' semiclassics (i.e., asymptotics in
the small parameter $\hbar$) does not, in the multicomponent case, lead to the
construction of a generalised EBK rule. We resolved this problem by 
constructing a new semiclassical theory whose corresponding Hamiltonian system
contains $\hbar$-dependent terms. For such a system, as we have shown, a
generalised EBK quantisation rule does exist. We demonstrated the power of 
our approach by solving two well known problems chosen to elucidate those 
aspects of the theory which are new ($\hbar$-dependent Hamiltonians and 
topological phases due to vortices). 
Of course the problems we have considered have integrable classical dynamics.
However there are many situations one may wish to solve where the classical
dynamics is non-integrable. Thus we should examine which aspects of our theory
remain valid in this case.

When our Hamiltonian system exhibits chaos our corresponding (stationary)
Hamilton-Jacobi equation has no solution, i.e., the eigenstates of the 
BdG equations cannot have the `simple' multicomponent WKB form. Since
our `classical' mechanics was derived in the first place using this ansatz
for the wave function one must question whether the Hamiltonian system has
any meaning for non-integrable systems. The answer is yes and the reason is
as follows. If we start with the time-dependent BdG equations, (replace
$E_\lambda$ by $i\hbar \partial/\partial t$ in (\ref{eqn:BdGequations})), and
seek a solution in terms of a time-dependent multicomponent WKB ansatz, then
we obtain, in place of the stationary Hamilton-Jacobi equation, the time
dependent Hamilton-Jacobi equation:
\begin{equation}
	E^\alpha({\pmb \nabla}S({\bf r},t),{\bf r})
	+
	\frac{\partial S}{\partial t}
	=0.
\label{eqn:TimeHamiltonJacobi}
\end{equation}
A solution to this equation always exists unlike for the stationary version.
The Hamiltonian system corresponding to (\ref{eqn:TimeHamiltonJacobi}) is
identical to the one for the stationary equation and thus we have discovered
our (fictitous) classical mechanics is the correct one to describe 
quasiparticle dynamics in superconductors regardless of the type of the
dynamics exhibited. (The same cannot be said of our eigenfunction 
(\ref{eqn:WaveFunctionEffective}), and the generalised EBK rule neither 
of which have any meaning in chaotic systems for which trace formulae must
be derived ~\cite{gutzwiller:90:0}.)

One of our principle motivations for developing the semiclassical theory 
was to construct a microscopic theory of quasiparticles in superconductors
in large magnetic fields. Under such conditions the groundstate of a Type
II superconductor adopts a so called 
Abrikosov flux lattice state ~\cite{abrikosov:88:0}. This is
a regular lattice of vortices, each carrying one superconducting flux 
quantum, $\Phi_0$, and having supercurrents associated with it. 
We have developed a simplified model of this 
state ~\cite{duncan:98:0,duncan:99:0,duncan:2002:0}, 
whose dynamics is integrable, and have used it to give an explaination of the
origin of de Haas-van Alphen oscillations in Type II superconductors.

Another problem for which the above strategy may be usefully 
deployed is that of the mesoscopic metals proximity coupled a 
superconductor.The refined semiclassics we have presented here may
serve as a bases for including `quantum diffraction' corrections 
missing from the naive analysis criticised by Altland, Simons and 
Taras-Semchuk ~\cite{altland:00:0}.

Finally we point out that the theory we have presented is restricted to 
$s$-wave pairing. An ongoing `hot-topic' in superconductivity research is
exotic superconductivity, in particular $d$- and $p$- wave pairing. We are 
in the process of extending our present theory to encompuss exotic pairing
so that we may study quasiparticle orbits in the presence of $d$- and $p$-wave
pairing potentials in magnetic fields.
\appendix
%%  All sections after the \appendix command will be appendixes
%%  Subsections function normally in appendixes.
%% section* will give an untitled appendix
%\section*{}
%This is how an unnamed appendix looks.  You must use the command  \cmd{section*}\texttt{\{\}} for a appendix without a title.
%---------------------------------------------------------------------------
\section[Labeling amplitudes]{Appropriate labels for the zeroth order amplitudes}
\label{app:BetajDependence}
Our zeroth order spinor amplitudes contain $\beta=\pm$ which are determined
uniquely by $j$ as follows:
For a given branch ${\bf p}_0^j({\bf r})$ of the multivalued momentum 
fuction we calculate the left hand side of
\begin{equation}
	\frac{{\bf p}_0^2}{2m} + \frac{1}{2}m{\bf v}_0^2 
	+V({\bf r}) - \epsilon_F
	=
	\beta
	\sqrt{ 
		\left( 
			E_{\bf I}^\alpha
			-
			{\bf p}_0\cdot{\bf v}_0
		\right)^2
		-
		\left| \Delta({\bf r}) \right|^2
	     }.
\label{eqn:DistinguishingBeta}
\end{equation}
Denote this quantity by $T({\bf p}_0^j)$. Then if $T>0$, $\beta=+$, whilst
for $T<0$, $\beta=-$. In this way each ${\bf p}_0^j$ determines a unique
choice of $\beta$. 
To give an interpretation to the sign of $\beta$ we will need the following
definition: If
\begin{equation}
	\left(  u_{0,{\bf I}}^{\alpha,j} ({\bf r})  \right)^2
	-
	\left( v_{0,{\bf I}}^{\alpha,j} ({\bf r})  \right)^2>0 
\nonumber
\end{equation}
we call the excitation quasiparticle-like or simply `particle-like', whilst for
\begin{equation}
	\left(  u_{0,{\bf I}}^{\alpha,j} ({\bf r})  \right)^2
	-
	\left( v_{0,{\bf I}}^{\alpha,j} ({\bf r})  \right)^2<0
\nonumber 
\end{equation}
we call the excitation quasihole-like or simply `hole-like'.
From equation (\ref{eqn:NormalisedAmplitudes}) we have
\begin{equation}
	\left(  
		u_{0,{\bf I}}^{\alpha,j} ({\bf r})  
	\right)^2
	-
	\left( 
		v_{0,{\bf I}}^{\alpha,j} ({\bf r})  
	\right)^2
	=
	\beta \textstyle
	\frac{
		\sqrt{ 
			\left( 
				E_{\bf I}^\alpha
				-
				{\bf p}_0^j\cdot{\bf v}_0
			\right)^2
			-
			\left| \Delta({\bf r}) \right|^2
		     }
	      }
	      {
		E_{\bf I}^\alpha-{\bf p}_0^j\cdot{\bf v}_0
	      }.
\nonumber 
\end{equation}
Noting $\text{sgn} (E_{\bf I}^\alpha-{\bf p}_0^j\cdot{\bf v}_0) = \alpha$ 
we construct Table \ref{tab:interpretingBeta} to interpret the sign of 
$\beta$. 
%----table----
\begin{table}
   \begin{center}
   \begin{math}
   \begin{array}{|c|c|c|} \hline % 3 columns!
% column headings
	\text{Hamiltonian System}
	 & \text{Amplitude index} & \text{Excitation type} \\ \hline
% first row
	\alpha = +
	 & \beta = + & \text{particle-like} \\ %\hline
% second row
	& \beta =- & \text{hole-like}  \\ \hline
% third row
	\alpha = -
	 & \beta = + & \text{hole-like} \\ %\hline
% fourth row
	& \beta =- & \text{particle-like}  \\ \hline
   \end{array}
   \end{math}
   \caption{Identifying the excitation type}
   \label{tab:interpretingBeta}
   \end{center}
\end{table}
%----endtable----
Since, as we have shown, $\beta$ is determined by
${\bf p}_0^j$ we can similarly interpret a given ${\bf p}_0^j$ as corresponding
to `particle-like' or `hole-like' excitation. 
%---------------------------------------------------------------------------
\section[First order quantities]{Derivation of the transport equation and other first order quantities}
\label{sec:appendixA}
Instead of expanding equation (\ref{eqn:UnitaryTransformedBdGequations}) 
directly we begin by taking its expectation value:
{ \setlength{\multlinegap}{0pt}
\begin{multline}
	\left(
		\begin{array}{cc}
			\tilde{u}^*_\lambda ({\bf r}) &
			\tilde{v}^*_\lambda ({\bf r})
		\end{array}
	\right)
	\times
\\
	\left(
		\begin{array}{cc}
			E_\lambda-H
			(
				\hat{\bf p} +
				\frac{\partial S_0}{\partial {\bf r}},
				{\bf r}
			) &
			-|\Delta({\bf r})| e^{i\phi({\bf r})} \\
			-|\Delta({\bf r})| e^{-i\phi({\bf r})} &
			E_\lambda+H^*
			(
				\hat{\bf p} +
				\frac{\partial S_0}{\partial {\bf r}},
				{\bf r}
			)
		\end{array}
	\right)
	\left(
		\begin{array}{c}
			\tilde{u}_\lambda ({\bf r}) \\
			\tilde{v}_\lambda ({\bf r}) 
		\end{array}
	\right)
\\
	=F^\dagger \hat{D} F =0,
\label{eqn:MatrixElementBdGequations}
\end{multline} }
where we have introduced $\hat{D}$ for the matrix differential operator and $F$
to represent the spinor. Then expanding the result, equation
(\ref{eqn:MatrixElementBdGequations}),
upto and including first order in $\hbar$ we find
\begin{equation}
	0=F_0^\dagger D_0 F_0 +F_1^\dagger D_0 F_0 
		+ F_0^\dagger D_0 F_1 +F_0^\dagger \hat{D}_1 F_0.
\label{eqn:FirstOrderMatrices}
\end{equation}
Here the subscripts denote quantities of zeroth or first order in $\hbar$.
$D_0$ is the zeroth order Hamiltonian matrix
\begin{equation}
	D_0 =
	\left(
		\begin{array}{cc}
			E_{\bf I} -H^e_0({\bf p}_0,{\bf r}) &
			-|\Delta ({\bf r})|e^{i\phi({\bf r})} \\
			-|\Delta ({\bf r})|e^{-i\phi({\bf r})}  &
			E_{\bf I}+H^h_0({\bf p}_0,{\bf r})
		\end{array}
	\right),
\label{eqn:DzeroMatrix}	
\end{equation}
$F_0$ is the by now familiar
spinor (\ref{eqn:correctedguess}), $\hat{D}_1$ is the first order matrix
differential operator, 
which we have not yet written down explicitly, but what is $F_1$?
$F_1$ is the spinor obtained by going beyond the first two terms in the 
expansion of $S({\bf r})$ and $\Sigma({\bf r})$, equations
(\ref{eqn:expandedS}) and (\ref{eqn:expandedSigma}). 
Thus the wave function written to include 
the next order terms is
\begin{align}
	\left(
		\begin{array}{c}
			u_{\bf I}({\bf r}) \\
			v_{\bf I}({\bf r})
		\end{array}
	\right)
	& = 
	\left(
		\begin{array}{c}
			\tilde{u}_{\bf I} ({\bf r})
			e^{i \hbar \Sigma_2({\bf r})} \\
			\tilde{v}_{\bf I} ({\bf r})
			e^{-i \hbar \Sigma_2({\bf r})}
		\end{array}
	\right)
	e^{i\hbar S_2({\bf r})}
	e^{iS_0({\bf r})/\hbar} \nonumber \\
	& =
	\left(
		\begin{array}{c}
			\tilde{u}_{\bf I} ({\bf r}) \\
			\tilde{v}_{\bf I} ({\bf r})
		\end{array}
	\right)
	e^{iS_0({\bf r})/\hbar} 
	\! +
	\left( \!\!\!
		\begin{array}{c}
			\tilde{u}_{\bf I} ({\bf r})
			\left(
				+i \hbar \Sigma_2({\bf r}) +i\hbar S_2({\bf r})
			\right) \\
			\tilde{v}_{\bf I} ({\bf r})
			\left(
				-i \hbar \Sigma_2({\bf r}) +i\hbar S_2({\bf r})
			\right)
		\end{array}\!\!\!
	\right)\!
	e^{iS_0({\bf r})/\hbar},
 \nonumber \\
	& =
	F_0 e^{iS_0({\bf r})/\hbar} + F_1 e^{iS_0({\bf r})/\hbar}.
\label{eqn:SecondOrderWaveFunction}
\end{align}
In (\ref{eqn:SecondOrderWaveFunction}) we used 
$\exp(i\hbar\Sigma_2) \approx 1+i\hbar\Sigma_2$, for
$i\hbar\Sigma_2 \ll 1$ and so on.
Turning our attention back to (\ref{eqn:FirstOrderMatrices}) we recognise
the first term as the zeroth order equation
\begin{align}
	{\cal{}O}\left(\hbar^0\right) \qquad 
	0 & =F_0^\dagger D_0 F_0,
\label{eqn:ZerothOrderEquation} \\
\intertext{and because $D_0F_0=0$ the second term is also zero. This leaves}
	0 & =F_0^\dagger D_0 F_1 +F_0^\dagger \hat{D}_1 F_0. 
\nonumber \\
\intertext{Since $D_0^\dagger=D_0$, the first term can, however, be written as}
	F_0^\dagger D_0 F_1
	& =
	\left(
		D^\dagger_0 F_0
	\right)^\dagger
	F_1
	=
	\left(
		D_0 F_0
	\right)^\dagger F_1
	= 0. 
\nonumber \\
\intertext{Consequently the expansion of the matrix elements
(\ref{eqn:MatrixElementBdGequations}) gives the first order equation}
	{\cal O}\left(\hbar\right) \qquad
	0 & =F_0^\dagger \hat{D}_1 F_0.
\label{eqn:FirstOrderEquation}   %note not to use \\ and \label !!
\end{align}
In particular notice that $F_1$, the ${\cal O}\left(\hbar\right)$ 
correction to 
the wave function, does not feature in the expansion of the BdG equations
up to first order in $\hbar$ thus justifying the omission of such terms
from the spinor elements $\tilde{u}_{\bf I} ({\bf r})$ and
$\tilde{v}_{\bf I} ({\bf r})$ in (\ref{eqn:correctedguess}).

To find the explicit form of $\hat{D}_1$  we require the order $\hbar$ terms of
$H\left(
	\hat{\bf p}+\frac{\partial S_0}{\partial {\bf r}},{\bf r}
\right)$. Writing
$H$ out we have
\begin{align}
	H
	\left(
		\hat{\bf p} +
		\frac{\partial S_0}{\partial {\bf r}},{\bf r}
	\right)
	& =
	\frac{1}{2m}
	\left(
		\frac{\hbar}{i}{\pmb \nabla}
		+
		{\bf P}^+
	\right)^2 + V({\bf r})-\epsilon_F, 
\nonumber \\
	& =
	H_0^e({\bf p}_0,{\bf r}) +
	\frac{1}{2m}
	 \left(
			\frac{\hbar}{i}{\pmb \nabla}
			\cdot {\bf P}^+
			+
			{\bf P}^+
			\cdot \frac{\hbar}{i}{\pmb \nabla}
	\right)
	+
	{\cal O}\left(\hbar^2\right),
\nonumber
\end{align}
where ${\bf P}^+=\frac{\partial S_0}{\partial {\bf r}} +e{\bf A}({\bf r})$.
$H^*\left(
	\hat{\bf p}+\frac{\partial S_0}{\partial {\bf r}},{\bf r}
\right)$ is 
written in similar fashion by replacing ${\bf P}^+$ with
${\bf P}^-=\frac{\partial S_0}{\partial {\bf r}} -e{\bf A}({\bf r})$.
Then the matrix $\hat{D}_1$ takes the form
\begin{equation}
	\hat{D}_1=
	\left(
		\begin{array}{c}
			-\frac{1}{2m}
	 		\left(
				\frac{\hbar}{i}{\pmb \nabla}
				\cdot {\bf P}^+
				+
				{\bf P}^+
				\cdot \frac{\hbar}{i}{\pmb \nabla}
			\right) \qquad \qquad 0 \qquad \quad  \\
			\qquad \quad 0 \qquad \qquad
			+\frac{1}{2m}
	 		\left(
				\frac{\hbar}{i}{\pmb \nabla}
				\cdot {\bf P}^-
				+
				{\bf P}^-
				\cdot \frac{\hbar}{i}{\pmb \nabla}
			\right)
		\end{array}
	\right)
\nonumber
\end{equation}
so that equation (\ref{eqn:FirstOrderEquation}) becomes
\begin{multline}
	F_0^\dagger
	\left(
		\begin{array}{cc}
			-\frac{1}{2m}
			\frac{\hbar}{i}{\pmb \nabla}
			\cdot
			{\bf P}^+\!
			& 0 \\
			0 &
			+\frac{1}{2m}
			\frac{\hbar}{i}{\pmb \nabla}
			\cdot
			{\bf P}^-\!
		\end{array}
	\right)
	F_0 +
\nonumber \\
	+F_0^\dagger
	\left(
		\begin{array}{cc}
			-\frac{1}{2m} 
			{\bf P}^+\!
			\cdot
			\frac{\hbar}{i}{\pmb \nabla}
			& 0 \\
			0 &
			+\frac{1}{2m}
			{\bf P}^- \!
			\cdot
			\frac{\hbar}{i}{\pmb \nabla}
		\end{array}
	\right)
	F_0
	= 0,
\nonumber
\end{multline}
and if we then pull out the $\frac{\hbar}{i}{\pmb \nabla}$ from the first
term we obtain
\begin{equation}
% first term
	\frac{\hbar}{i}{\pmb \nabla}
	\cdot
	\left\{
		F_0^\dagger
		{\bf M}
		F_0
	\right\}
% second term
	-
	\frac{\hbar}{i}
	\left\{
		{\pmb \nabla}
		F_0^\dagger
	\right\}
	\cdot
	{\bf M}
	F_0 
% third term
	+
	\frac{\hbar}{i}
	F_0^\dagger
	{\bf M}
	\cdot
	{\pmb \nabla}
	F_0
	= 0,
\label{eqn:RealAndImagFirstOrderEquation}
\end{equation}
where the vector matrix ${\bf M}$ is given by 
\begin{equation}
	{\bf M}
	=
	\left(
		\begin{array}{cc}
			-\frac{{\bf P}^+}{2m} & 0 \\
			0& +\frac{{\bf P}^-}{2m}
		\end{array}
	\right).
\label{eqn:VectorMatrix}
\end{equation}
The first term is purely imaginary since the differential operator acts
on $F_0^\dagger F_0=|F_0|^2$ (${\bf M}$ being diagonal). 
The remaining two terms taken
together are purely real. To see this we rewrite the middle term in 
equation (\ref{eqn:RealAndImagFirstOrderEquation}):
\begin{align}
	\left\{
		{\pmb \nabla}
		F_0^\dagger
	\right\}
	\cdot
	{\bf M}
	F_0 
	&=
	\left[
		({\bf M} F_0)^\dagger
		\cdot
		({\pmb \nabla}F_0)
	\right]^\dagger,
\nonumber \\
	&=
	\left[
		F_0^\dagger{\bf M}^\dagger
		\cdot
		{\pmb \nabla}F_0
	\right]^\dagger,
\nonumber
\end{align}
but since ${\bf M}={\bf M^\dagger}$ the last two terms in 
equation (\ref{eqn:RealAndImagFirstOrderEquation}) take the form 
\begin{equation}
	-\frac{\hbar}{i}
	\left[
		F_0^\dagger{\bf M}
		\cdot
		{\pmb \nabla}F_0
	\right]^\dagger
	+\frac{\hbar}{i}
	\left[
		F_0^\dagger{\bf M}
		\cdot
		{\pmb \nabla}F_0
	\right]
	=
	\frac{\hbar}{i} 
	\;2i \;\text{Im}
		\left[
		F_0^\dagger{\bf M}
		\cdot
		{\pmb \nabla}F_0
	\right],
\nonumber
\end{equation}
which is certainly real. 
Equation (\ref{eqn:RealAndImagFirstOrderEquation}) can 
therefore be separated into two
equations which given explicitly are
\begin{align}
	{\cal O}\left(\hbar\right)\qquad
	{\pmb \nabla}
	\cdot
	\left\{
	\left(
		\begin{array}{cc}
			\tilde{u}^*_{\bf I}  &
			\tilde{v}^*_{\bf I} 
		\end{array}
	\right)
	\left(
		\begin{array}{c}
			-\frac{{\bf P}^+}{2m}
			\tilde{u}_{\bf I}  \\
			+\frac{{\bf P}^-}{2m}
			\tilde{v}_{\bf I} 
		\end{array}
	\right)
	\right\} 
	& = 0,
\label{eqn:app:TransportEquation} \\
	{\cal O}\left(\hbar\right)\qquad
	2\; \mathrm{Im}
	\left(
		\begin{array}{cc}
			\tilde{u}^*_{\bf I}  &
			\tilde{v}^*_{\bf I} 
		\end{array}
	\right)
	\left(
		\begin{array}{c}
			-\frac{{\bf P}^+}{2m}
			\cdot
			{\pmb \nabla}
			\tilde{u}_{\bf I}  \\
			+\frac{{\bf P}^-}{2m}
			\cdot
			{\pmb \nabla}
			\tilde{v}_{\bf I} 
		\end{array}
	\right)
	& = 0,
\label{eqn:app:FirstOrderPhasesEquation}
\end{align}
with ${\bf P}^\pm={\bf p}_0({\bf r})\pm e{\bf A}({\bf r})$.
The first, (\ref{eqn:app:TransportEquation}), contains no phases and is the 
amplitude transport equation as we will show shortly. The other, equation
(\ref{eqn:app:FirstOrderPhasesEquation}), 
determines the phase $S_1^r({\bf r})$.

In order to rewrite equations (\ref{eqn:app:TransportEquation}) and
(\ref{eqn:app:FirstOrderPhasesEquation}) in a more 
tractable form we will need Hamilton's 
equations (\ref{eqn:HamiltonsEquations})
\begin{equation}
	\dot{\bf r}^\alpha
	=
	\left(
		\frac{\partial E_0^\alpha ({\bf p},{\bf r}) }
			{ \partial {\bf p}}
	\right),
	\quad
	\dot{\bf p}^\alpha
	=
	-\left(
		\frac{\partial E_0^\alpha ({\bf p},{\bf r}) }
			{ \partial {\bf r}}
	\right),
\label{eqn:HamiltonsEquationsAppendix}
\end{equation}
and also the follwing relation
\begin{equation}
	{\bf j}^\alpha({\bf r})
	=
	-\left(
		\frac{\partial E_0^\alpha ({\bf p},{\bf r}) }
			{\partial {\bf A}({\bf r}) }
	\right)_{
		 {\bf p}={\bf p}_0^\alpha ({\bf r}),
		 {\bf r}={\bf r}^\alpha (t)
		},
\label{eqn:CurrentFromdEdA}
\end{equation}
where ${\bf j}^\alpha({\bf r})$ is the current density at the point ${\bf r}$.
We require an expression for Hamilton's equations and the current
in terms of the zeroth order quantities already found.
For this we use the zeroth order equations 
(\ref{eqn:RealZerothOrderBdGequations})
\begin{multline}
	\left(
		\begin{array}{cc}
			\tilde{u}^\alpha_{\bf I}  &
			\tilde{v}^\alpha_{\bf I} 
		\end{array}
	\right)^*
	E_0^\alpha
	\left(
		\begin{array}{c}
			\tilde{u}^\alpha_{\bf I}  \\
			\tilde{v}^\alpha_{\bf I} 
		\end{array}
	\right)
	=
\\
	\left(
		\begin{array}{cc}
			\tilde{u}^\alpha_{\bf I}  &
			\tilde{v}^\alpha_{\bf I} 
		\end{array}
	\right)^*
	\left(
		\begin{array}{cc}
			H^e_0({\bf p}_0,{\bf r}) &
			|\Delta ({\bf r})| e^{i\phi({\bf r})} \\
			|\Delta ({\bf r})| e^{-i\phi({\bf r})} &
			-H^h_0({\bf p}_0,{\bf r})
		\end{array}
	\right)
	\left(
		\begin{array}{c}
			\tilde{u}^\alpha_{\bf I}  \\
			\tilde{v}^\alpha_{\bf I}  
		\end{array}
	\right),
\nonumber
\end{multline}
which we differentiate to give
\begin{multline}
	e^{-2S^i_1({\bf r})}
	\left(
		\frac{\partial E_0^\alpha ({\bf p},{\bf r}) }
			{\partial{\bf p}}
	\right)
	= 
\\
	\left(
		\begin{array}{cc}
			\tilde{u}^\alpha_{\bf I} ({\bf r}) &
			\tilde{v}^\alpha_{\bf I} ({\bf r})
		\end{array}
	\right)^*
	\left(
		\begin{array}{cc}
			\frac{\partial H^e_0}{\partial {\bf p} } &
			0 \\
			0 &
			-\frac{\partial H^h_0}{\partial {\bf p} }
		\end{array}
	\right)
	\left(
		\begin{array}{c}
			\tilde{u}^\alpha_{\bf I} ({\bf r}) \\
			\tilde{v}^\alpha_{\bf I} ({\bf r}) 
		\end{array}
	\right),
\label{eqn:AmplitudeTimesdEdp}
\end{multline}
where
\begin{align}
	\left(
		\begin{array}{cc}
			\tilde{u}^\alpha_{\bf I}  &
			\tilde{v}^\alpha_{\bf I} 
		\end{array}
	\right)^*
	\left(
		\begin{array}{c}
			\tilde{u}^\alpha_{\bf I}  \\
			\tilde{v}^\alpha_{\bf I} 
		\end{array}
	\right)
	&=
	e^{-2S^i_1({\bf r})}
	\left(
		(u^\alpha_{0,{\bf I}})^2
		+
		(v^\alpha_{0,{\bf I}})^2
	\right)
	=
	e^{-2S^i_1({\bf r})},
\nonumber 
\end{align}
has been used. We can do the same for
$\frac{\partial E_0^\alpha ({\bf p},{\bf r}) }{\partial {\bf A}({\bf r}) }$:
\begin{equation}
	\frac{\partial E_0^\alpha ({\bf p},{\bf r}) }
		{\partial {\bf A}({\bf r}) }
	=
	\left(
		\begin{array}{cc}
			u^\alpha_{0,{\bf I}}  &
			v^\alpha_{0,{\bf I}}
		\end{array}
	\right)
	\left(
		\begin{array}{cc}
			\frac{\partial H^e_0}{\partial {\bf A} } &
			0 \\
			0 &
			-\frac{\partial H^h_0}{\partial {\bf A} }
		\end{array}
	\right)
	\left(
		\begin{array}{c}
			u^\alpha_{0,{\bf I}} \\
			v^\alpha_{0,{\bf I}}
		\end{array}
	\right),
\label{eqn:dEdA}
\end{equation}
where the phases and amplitude, $e^{-2S^i_1}$, have been cancelled because
they shall not be needed. The explicit derivatives of $H_0^{e/h}$ are
\begin{align}
	\frac{\partial H^e_0}{\partial {\bf p} }
	= &
	\frac{1}{m}
	\left(
		{\bf p}+e{\bf A}
	\right)
	=
	\frac{ {\bf P}^+ }{m}, 
\label{eqn:DerivativeOfHelectron} \\	
	-\frac{\partial H^h_0}{\partial {\bf p} }
	= &
	-\frac{1}{m}
	\left(
		{\bf p}-e{\bf A}
	\right)
	=
	-\frac{ {\bf P}^- }{m},
\label{eqn:DerivativeOfHhole} \\
\intertext{and}
	\frac{\partial H^e_0}{\partial {\bf A} }
	= &
	\frac{1}{m}
	\left(
		{\bf p}+e{\bf A}
	\right)e
	=
	\frac{ {\bf P}^+ }{m} e,
\label{eqn:DerivativeOfHelectronByA} \\
	-\frac{\partial H^h_0}{\partial {\bf A} }
	= &
	-\frac{1}{m}
	\left(
		{\bf p}-e{\bf A}
	\right)(-e)
	=
	-\frac{ {\bf P}^- }{m}(-e),
\label{eqn:DerivativeOfHholeByA}
\end{align}
all evaluated for ${\bf p}={\bf p}_0^\alpha ({\bf r})$.

Now let us use Hamilton's equations and the current relation to 
rewrite the transport equation
(\ref{eqn:app:TransportEquation}) and the equation for $S_1^r$
(\ref{eqn:app:FirstOrderPhasesEquation}). 
Using (\ref{eqn:DerivativeOfHelectron}) and (\ref{eqn:DerivativeOfHhole})
we see that the right hand side of (\ref{eqn:AmplitudeTimesdEdp}) is 
precisely the inner product between the vectors in the order $\hbar$
equation (\ref{eqn:app:TransportEquation}). Thus the transport equation
can be rewritten as 
\begin{equation}
	{\pmb \nabla}
	\cdot
	\left(
		e^{-2S^i_1({\bf r})}
		\left.
			\frac{\partial E_0^\alpha ({\bf p},{\bf r}) }
			     {\partial{\bf p}}
		\right|_{{\bf p}={\bf p}_0^\alpha({\bf r})}
	\right)
	= 0.
\label{eqn:app:AmplitudeTransportEquation}
\end{equation}
Manipulating (\ref{eqn:app:AmplitudeTransportEquation}) into the form
\begin{equation}
	\sum_k
	\left(
		\frac{\partial E}{\partial p_k}
		\frac{
			\partial (e^{-S_1^i})
		     }
		     {\partial x_k}
		+
		\frac{1}{2}
		e^{-S_1^i}
		\frac{\partial^2 E}{\partial p_k \partial x_k}
	\right)
	=0,
\nonumber
\end{equation}
we recognise the time independent transport 
equation of van Vleck ~\cite{vanvleck:28:0}. The solution is given by
the determinant
\begin{equation}
	e^{-S_1^i({\bf r},{\bf I})}
	=c
	\left|
		\det
		\frac{ 
			\partial^2 S_0^{\alpha,j}({\bf r},{\bf I}) 
		     }
		     {
			\partial {\bf r} \partial {\bf I}
		     }
	\right|^{1/2},
\label{eqn:app:VanVleckDet}
\end{equation}
where $c$ is a constant.

All that is left to determine now is $S_1^r({\bf r})$. Returning to 
(\ref{eqn:app:FirstOrderPhasesEquation}) we have
\begin{align}
	0 &=
	2\; \mathrm{Im}
	\left(
		\begin{array}{cc}
			\tilde{u}^\alpha_{\bf I}  &
			\tilde{v}^\alpha_{\bf I} 
		\end{array}
	\right)^*
	\left(
		\begin{array}{c}
			-\frac{{\bf P}^+}{2m}
			\cdot
			{\pmb \nabla}
			\tilde{u}^\alpha_{\bf I}  \\
			+\frac{{\bf P}^-}{2m}
			\cdot
			{\pmb \nabla}
			\tilde{v}^\alpha_{\bf I} 
		\end{array}
	\right)
\nonumber \\
	&=\left(
		\begin{array}{cc}
			\tilde{u}^\alpha_{\bf I}  &
			\tilde{v}^\alpha_{\bf I} 
		\end{array}
	\right)^*
	\left(
		\begin{array}{c}
			-\frac{{\bf P}^+}{m}
			\cdot
			\left(
				{\pmb \nabla} S^r_1 +
				{\pmb \nabla} \Sigma^r_1
			\right)	
			\tilde{u}^\alpha_{\bf I}  \\
			+\frac{{\bf P}^-}{m}
			\cdot
			\left(
				{\pmb \nabla} S^r_1 -
				{\pmb \nabla} \Sigma^r_1
			\right)
			\tilde{v}^\alpha_{\bf I} 
		\end{array}
	\right),
\nonumber
\end{align}
or
\begin{multline}
	0 =
	\left\{
		\frac{e{\bf A} }{m}
		+
		\left(
			(u^\alpha_{0,{\bf I}})^2
			-
			(v^\alpha_{0,{\bf I}})^2
		\right)
		\frac{{\bf p}}{m}
	\right\}
	\cdot
	{\pmb \nabla} S_1^r
\\
	+
	\left\{
		\frac{{\bf p}}{m}
		+
		\left(
			(u^\alpha_{0,{\bf I}})^2
			-
			(v^\alpha_{0,{\bf I}})^2
		\right)
		\frac{e{\bf A} }{m}
	\right\}
	\cdot
	{\pmb \nabla} \Sigma_1^r.
\label{eqn:MessyEqnForGradS}
\end{multline}
Now use equations (\ref{eqn:AmplitudeTimesdEdp}) and
(\ref{eqn:dEdA}), which written out in full become
\begin{align}
	\frac{\partial E_0^\alpha  }
		{\partial{\bf p}}
	&=	
	\frac{e{\bf A} }{m}
	+
	\left(
		(u^\alpha_{0,{\bf I}})^2
		-
		(v^\alpha_{0,{\bf I}})^2
	\right)
	\frac{{\bf p}}{m}, 
\nonumber \\
	\frac{1}{e}
	\frac{\partial E_0^\alpha  }
	{\partial {\bf A} }
	&=
	\frac{{\bf p}}{m}
	+
	\left(
		(u^\alpha_{0,{\bf I}})^2
		-
		(v^\alpha_{0,{\bf I}})^2
	\right)
	\frac{e{\bf A} }{m},
\label{eqn:dEdAInMessyForm}
\end{align}
so that (\ref{eqn:MessyEqnForGradS}) becomes
\begin{align}
	\frac{\partial E_0^\alpha  }
		{\partial{\bf p}}
	\cdot
	{\pmb \nabla}S_1^r
	&=
	-\frac{1}{e}
	\frac{\partial E_0^\alpha  }
	{\partial {\bf A} }
	\cdot
	{\pmb \nabla}\Sigma_1^r,
	\nonumber \\
\intertext{or}
	{\pmb \nabla}S_1^r
	\cdot
	\dot{\bf r}^\alpha
	&=
	-\frac{1}{e}
	{\bf j}^\alpha
	\cdot
	{\pmb \nabla}\Sigma_1^r,
\nonumber
\end{align}
which we integrate along a trajectory of the Hamiltonian system 
from some initial time $t_0$ to $t$
\begin{equation}
	\int^t_{t_0}
	{\pmb \nabla}S_1^r
	\cdot
	\frac{d{\bf r}^\alpha}{dt'}
	dt'
	=
	-\frac{1}{e}
	\int^t_{t_0}
	{\bf j}^\alpha({\bf r})
	\cdot
	{\pmb \nabla}\Sigma_1^r
	dt'.
\nonumber
\end{equation}
Finally we have found an expression for $S_1^r$:
\begin{equation}
	S_1^r({\bf r})
	=
	-\frac{1}{e}
	\int^t_{t_0}
	{\bf j}^\alpha({\bf r})
	\cdot
	{\pmb \nabla}\Sigma_1^r
	dt',
\nonumber
\end{equation}
or
\begin{equation}
	S_1^r({\bf r})
	=
	-\frac{1}{e}
	\int^t_{t_0}
	{\bf j}^\alpha({\bf r})
	\cdot
	\frac{ {\pmb \nabla}\phi({\bf r}) }{2}
	dt'.
\label{eqn:app:FirstOrderPhase}
\end{equation}
We interpret this shift to first order in $\hbar$  of the phase in the
following way. From (\ref{eqn:dEdAInMessyForm}) we have
\begin{equation}
	-\frac{1}{e}
	{\bf j}^\alpha({\bf r})
	=
	\frac{{\bf p}^\alpha}{m}
	+
	\left(
		(u^\alpha_{0,{\bf I}})^2
		-
		(v^\alpha_{0,{\bf I}})^2
	\right)
	\frac{e{\bf A} }{m},
\label{eqn:CurrentInTermsOfe*}
\end{equation}
which is essentially the velocity of an excitation 
with
an effective charge
$e^*({\bf r})=\left(
		(u^\alpha_{0,{\bf I}})^2
		-
		(v^\alpha_{0,{\bf I}})^2
	\right)e$.
This effective charge is determined by the relative amounts of particle
and hole that the excitation is composed of. As the spinor is transported
along the phase space trajectory, figure 
\ref{fig:multivaluedP},
the particle-hole composition slowly changes giving rise to a changing
charge $e^*({\bf r})$ (charge is not a good quantum number in superconductors)
and hence a changing current ${\bf j}^\alpha({\bf r})$. This varying
current interacts with the phase gradient of the order parameter and 
modifies the
phase of the wave function to first order in $\hbar$. Notice in particular
that if no phase gradient exists, i.e. $\phi=constant$, then
$S_1^r=0$.
%----------------------------------------------------------------------------
\section[Analytic continuation]{Technical details of the analytic continuation of the SNS junction eigenfunctions to determine a quantisation condition}
The basic idea of the analytic continuation of solutions is simple: 
our asymptotic solutions
\label{sec:ComplexMethodSNS}
are valid not only along the real axis but for complex coordinates as well,
so long as we stay sufficiently far away from $y_\pm$. Thus we obtain an
approximate solution to the BdG equations throughout the complex plane
(excluding discs, radius $\rho$ say, centred upon $y_\pm$) by continuing 
our asymptotic solutions to complex coordinates.
In particular we can continue a solution valid for $y>y_+ +\rho$ along a
semicircular path, radius $>\rho$, in the plane and arrive at $y<y_+ -\rho$,
thus determining the appropriate amplitudes and phases of the solution for 
$y_- +\rho < y< y_+ -\rho$. Similarly we can circumvent $y_-$ to obtain the
correctly phased solution for $y<y_- -\rho$. The situation is however
complicated by Stokes phenomenon.

Stokes phenomenon is named after its discoverer Sir George Gabriel Stokes
(1819-1903). A general discussion of historical issues can be found in 
Heading ~\cite{heading:62:0}.
Stokes, studying Airy's equation, noticed
that if a general solution, given in terms of the two power series with
arbitrary coefficients, was represented by a certain linear combination of
the asymptotic solutions in one region of the complex plane then in an
adjacent region the same general solution was represented by a completely 
different form
of linear combination of these asymptotic solutions. He discovered
the constants of the linear combination changed discontinuously when
crossing certain lines in the plane - now known as Stokes lines.
(Note it is a change in form not the numerical value which takes place.)

For us, this means that if we choose an asymptotic solution, i.e. we set
the coefficients of other solutions to zero, and then analytically 
continue, we will cross Stokes lines where the coefficients can jump,
in particular from zero to non-zero, so that solutions which were absent
suddenly appear changing the form of our solution. 
What we need to know is how to locate the Stokes
lines and what rule determines the change in the coefficients as the lines
are crossed.
%-------------------------------------------------------------------
\subsection{Stokes phenomenon in the presence of two solutions}
Stokes phenomenon occurs when the exponentially subdominant solution out
of a pair is at its smallest.
Consider the asymptotic (WKB) form for the solutions to Airys equation:
\begin{equation}
	\Psi_\pm = (z)^{-1/4} e^{\pm i\frac{2}{3}z^{3/2}}, 
	\quad z \rightarrow \infty.
\nonumber
\end{equation}
For real $z$ these solutions are oscillatory but for complex $z$ the
solutions aquire the exponential factors $e^{+|w(z)|}$ or $e^{-|w(z)|}$, 
with $w(z)= \frac{2}{3}\text{Re}\{ iz^{3/2} \}$. Thus one solution is 
exponentially dominant, the other exponentially subdominant. Now $e^{+|w(z)|}$
is maximally dominant when $\text{Im} \{iz^{3/2} \} =0$, and it is then, when
the second solution is least visable, that its coefficient can jump.
Thus Stokes lines are defined by
\begin{equation}
	\text{Im}\{ iz^{3/2}\} =0,
\label{eqn:LocationOfStokesLines}
\end{equation}
in the present case.

Anti-Stokes lines are defined by
\begin{equation}
	\text{Re} \{iz^{3/2} \}=0.
\label{eqn:LocationOfAntiStokes}
\end{equation}
These are the lines along which the effect of the Stokes jump becomes
important because both solutions are purely oscillatory. The 
solutions are said to be neutral, neither being subdominant wrt. the other.
Notice that both the Stokes lines and anti-Stokes lines eminate from a
branch point at $z=0$.

The general prescription when crossing a
Stokes-line is given by
\begin{align}
	[{\text{New subdominant coeff.}}]
	=&[{\text{Old subdominant coeff.}}]
\nonumber \\
	&+
	({\text{Stokes constant}}) \times
\nonumber \\
	& [{\text{coeff. of dominant term}}].
\nonumber
\end{align}
By following a given solution along a circuit surrounding
the branch point a number of Stokes constants will be introduced. Requiring
our intial and final solutions to be the same determines the value of these
constants.

The situation we must solve involves four solutions. Before we can proceed
we must therefore generalise the above discussion.
%-------------------------------------------------------------------------
\subsection{Stokes phenomenon in the presence of more than two solutions} 
%---------------footnote----------------------------------------------
\footnote{I am indebted to Chris Howls 
for explaining how Stokes phenomenon is to be understood and 
treated in the case of more than two solutions.}
%-------------endfootnote-------------------------------------------------
Suppose now we have $N$ solutions. Let us change our notation
and write each solution as
\begin{equation}
	\Psi_i(z) = e^{\phi_i(z)},
\label{eqn:GeneralStokeSolution}
\end{equation}
$i=1,\dotsc, N$. Then one might expect that as we move around in the complex
plane a given solution, say $i$, could pick up contributions from some or
all of the other solutions, $j\neq i$. The question we must ask is when 
can Stokes phenomenon occur? The answer is as follows. Introduce the
{\em singulant} for a pair of solutions as the quantity $\phi_i-\phi_j$. Then a 
necessary condition for $\Psi_i$ to experience Stokes phenomenon is
\begin{equation}
	\text{Im} \{ \phi_i-\phi_j \} = 0, \qquad i\neq j,
\label{eqn:ImagPartOfSingulant}
\end{equation}
and sufficiency
%-----------footnote------------------------------------------------------
\footnote{Assuming the Riemann sheets on which the individual solutions live
are not locally disjoint.}
%------------endfootnote--------------------------------------------------
is ensured by
\begin{equation}
	\text{Re} \{ \phi_i \} > \text{Re} \{ \phi_j \} .
\label{eqn:SufficiencyForStokes}
\end{equation}
We can understand the first of these conditions,
 (\ref{eqn:ImagPartOfSingulant}),
as being a generalisation of the Stokes lines for two solutions. Indeed when
we have two solutions so that $\phi_j =-\phi_i$ (\ref{eqn:ImagPartOfSingulant})
reduces to 
\begin{equation}
	\text{Im} \{ 2\phi_i \} = 0,
\nonumber
\end{equation}
which is just the usual equation (\ref{eqn:LocationOfStokesLines}) for 
Stokes lines. When there are more than two solutions, so that in general
$\phi_j \neq \phi_i$, (\ref{eqn:ImagPartOfSingulant}) says that Stokes lines
exist where the imaginary part of the exponents are the same. This ensures
that when comparing two solutions the effect of the phase can be discounted.
Then we can concentrate on the real parts which must satisfy 
(\ref{eqn:SufficiencyForStokes}). Clearly (\ref{eqn:SufficiencyForStokes})
implies
\begin{equation}
	e^{\phi_i} > e^{\phi_j},
\nonumber
\end{equation}
so that $\Psi_i$ is maximal over $\Psi_j$ whose coefficient can then change.

Thus Stokes phenomenon occurs where solutions are pair-wise maximally 
dominant and subdominant. 

Anti-Stokes lines are defined by 
\begin{equation}
	\text{Re} \{ \phi_i - \phi_j \} = 0.
\label{eqn:RealPartOfSingulant}
\end{equation}
This situation will be clarified in the specific treatment of the 
semiclassical SNS junction
eigenfunctions which follows.
%-----------------------------------------------------------------------------
\subsection{Conventions and notation}
It will be useful to introduce the following compact notation for the 
semiclassical solutions with phase reference $y_+$:
\begin{align}
	\Psi_1(z)
	&=
	\frac{ 1 }
	     {
		\sqrt{
			\left(
				\frac{\partial E }
		    		{\partial p_y}
			\right)
		     }
	     }_{ \!\!p_y^+(z)} \!\!\!\!\!\!\!\!\!\!\!
	\left(
		\begin{array}{c}
			u^+_{0,I}(z)e^{+i\phi/2} \\
			v^+_{0,I}(z)e^{-i\phi/2}	
		\end{array}
	\right)
	e^{\textstyle
		\frac{i}{\hbar}\int_z^{y_+} dz' \ p_y^+(z')
	  },
\label{eqn:WaveFunctionOne} \\
	\Psi_2(z)
	&=
	\frac{ 1 }
	     {
		\sqrt{
			\left(
				\frac{\partial E }
		    		{\partial p_y}
			\right)
		     }
	     }_{ \!\!p_y^+(z)} \!\!\!\!\!\!\!\!\!\!\!
	\left(
		\begin{array}{c}
			u^+_{0,I}(z)e^{+i\phi/2} \\
			v^+_{0,I}(z)e^{-i\phi/2}		
		\end{array}
	\right)
	e^{\textstyle
		\frac{i}{\hbar}\int_{y_+}^{z\ } dz' \ p_y^+(z')
	  },
\label{eqn:WaveFunctionTwo} \\
	\Psi_3(z)
	&=
	\frac{ 1 }
	     {
		\sqrt{
			\left(
				\frac{\partial E }
		    		{\partial p_y}
			\right)
		     }
	     }_{ \!\!p_y^-(z)} \!\!\!\!\!\!\!\!\!\!\!
	\left(
		\begin{array}{c}
			u^-_{0,I}(z)e^{+i\phi/2} \\
			v^-_{0,I}(z)e^{-i\phi/2}		
		\end{array}
	\right)
	e^{\textstyle
		\frac{i}{\hbar}\int_z^{y_+} dz' \ p_y^-(z')
	  },
\label{eqn:WaveFunctionThree} \\
	\Psi_4(z)
	&=
	\frac{ 1 }
	     {
		\sqrt{
			\left(
				\frac{\partial E }
		    		{\partial p_y}
			\right)
		     }
	     }_{ \!\!p_y^-(z)} \!\!\!\!\!\!\!\!\!\!\!
	\left(
		\begin{array}{c}
			u^-_{0,I}(z)e^{+i\phi/2} \\
			v^-_{0,I}(z)e^{-i\phi/2}		
		\end{array}
	\right)
	e^{\textstyle
		\frac{i}{\hbar}\int_{y_+}^{z\ } dz' \ p_y^-(z')
	  },
\label{eqn:WaveFunctionFour}
\end{align}
where the $\Psi_i(z)$ are analytic continuations of the $\Psi_i(y)$ to
complex coordinates in a domain suitably cut to ensure single-valued
definitions (see FIG.~\ref{fig:BranchCutPlaneSNS} and the next section).
The same four solutions written with phase reference $y_-$ are 
obtained by replacing $y_+$ by $y_-$ in equations
(\ref{eqn:WaveFunctionOne})-(\ref{eqn:WaveFunctionFour})
and shall by distinguished from $\Psi_i(z)$ by appending a minus sign as a
superscript thus
\begin{equation}
	\Psi_1^-(z)
	=
	\frac{ 1 }
	     {
		\sqrt{
			\left(
				\frac{\partial E }
		    		{\partial p_y}
			\right)
		     }
	     }_{ \!\!p_y^+(z)} \!\!\!\!\!\!\!\!\!\!\!
	\left(
		\begin{array}{c}
			u^+_{0,I}(z)e^{+i\phi/2} \\
			v^+_{0,I}(z)e^{-i\phi/2}	
		\end{array}
	\right)
	e^{\textstyle
		\frac{i}{\hbar}\int_z^{y_-} dz' \ p_y^+(z')
	  },
\label{eqn:WaveFunctionOneMinus}
\end{equation}
etc. Note that
\begin{equation}
	\Psi_1(z) = [1]\Psi_1^-(z),
\nonumber
\end{equation}
where we have introduced $[1]$ as the pure phase factor
\begin{equation}
	[1]
	=
	\exp
	\left(
		\frac{i}{\hbar}\int_{y_-}^{y_+} dz' \ p_y^+(z')
	\right).
\nonumber
\end{equation}
We also have
\begin{align}
	[2] &= [1]^*, 
\nonumber \\
	[3]
	&=
	\exp
	\left(
		\frac{i}{\hbar}\int_{y_-}^{y_+} dz' \ p_y^-(z')
	\right),
\nonumber \\
	[4] &= [3]^*,
\nonumber
\end{align}
where $*$ indicates complex conjugation. Furthermore we use $\phi_i$ and
$\phi_i^-$ ($i=1,\dotsc,4$) to indicate the exponents including the 
asymptotic parameter ($1/\hbar$) thus
\begin{align}
	\phi_1
	&=
	\frac{i}{\hbar}\int_z^{y_+} dz' \ p_y^+(z'),
\label{eqn:PhaseOne} \\
	\phi_1^-
	&=
	\frac{i}{\hbar}\int_z^{y_-} dz' \ p_y^+(z'),
\label{eqn:PhaseOneMinus}
\end{align}
and so on.
%--------------------------------------------------------------------------
\subsection{Ensuring $p_y^\pm(z)$ and the wave functions are single-valued}
\begin{figure}
	\begin{center}
		\epsfig{file=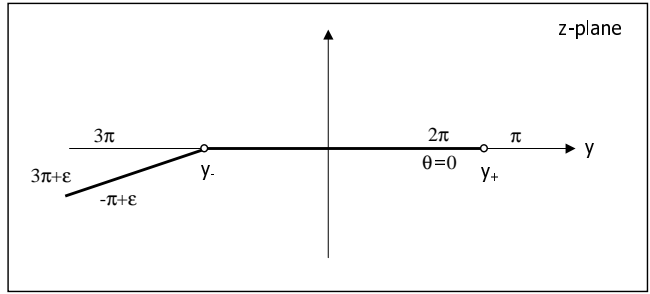,width=86mm,clip=}
		\caption{Branch cuts which ensure the WKB solutions are
			single-valued.}
		\label{fig:BranchCutPlaneSNS}
	\end{center}
\end{figure}

In order to construct a solution valid in the complex plane
we must have single-valued WKB solutions. 
The momentum branches for the superconducting state are 
multivalued functions. Recall $p_y^\pm(y)$ (\ref{eqn:MomentumBranchesForSNS1}):
\begin{equation}
	p_y^\beta(y)
	=
	\sqrt{ 
		p_F^2 -p_x^2 -p_z^2
		+\beta 2m
		\sqrt{
			E^2 - |\Delta(y)|^2
	     	     }
	}.
\nonumber
\end{equation}
In fact $p_y^\pm(y)$ are two branches of one function $p_y(z)$ given by
\begin{equation}
	p_y(z)
	=
	\sqrt{ 
		p_F^2 -p_x^2 -p_z^2
		+ 2m
		\sqrt{
			E^2 - \Delta^2(z)
	     	     }
	},
\label{eqn:PyOfZ}
\end{equation}
which is single-valued on its Riemann surface comprising more than one
Riemann sheet. (Here we use $\Delta(z)$ to represent the analytic continuation
of $|\Delta(y)|$ to complex values.)
Suppose we have the branch of (\ref{eqn:PyOfZ}) which is
equal to $p_y^+(y)$ on the real axis, then we obtain $p_y^-(y)$ by 
analytically continuing the function around one of the zeros of
$\sqrt{ E^2 -|\Delta(y)|^2 }$. We can see this explicitly if we consider
a smooth slowly varying $|\Delta(y)|$ and expand in the vacinity of 
the turning point
\begin{equation}
	|\Delta(y)|
	=
	\Delta_0
	+
	\left.
	\frac{
		\partial |\Delta|
	     }
	     {
		\partial y
	     }
	\right|_{y_\pm}
	\left( y-y_\pm
	\right)
	+ \dotsm
\label{eqn:SmoothDeltaExpansion}
\end{equation}
The function $\sqrt{ E^2 -\Delta^2(z) }$ 
then becomes ($\Delta_0=E$)
\begin{equation}
	\sqrt{ 
		E^2 -\Delta^2(z) 
	     }
	=
	\sqrt{
		2E
		\begin{smallmatrix}
			\left| \frac{
					\partial |\Delta|
	    	     	     	    }
	    	             	    {
					\partial y
			      	     }
			\right|_{y_\pm} \\
		\end{smallmatrix}
		\varrho_\pm
	     }
	\ e^{i\theta_\pm/2},
\label{eqn:2ndSquareRootZCoords}
\end{equation}
where $z_\pm = \varrho_\pm e^{i\theta_\pm/2}$ are the coordinates centred
on $y_\pm$. Clearly $\theta_\pm \rightarrow \theta_\pm + 2\pi$ causes
the function in (\ref{eqn:2ndSquareRootZCoords}) to change sign so that
$p_y(z_\pm)=p_y^+(y)$ becomes $ p_y(z_\pm e^{i2\pi})=p_y^-(y)$.
To make $p_y(z)$ single-valued in the complex plane we insert a branch cut
between $y_+$ and $y_-$. 
Notice that, so long as $p_F^2-p_x^2-p_z^2 > 2m\text{Re} 
\{ \sqrt{E^2-\Delta^2(z)} \}$, the Riemann surface of $p_y(z)$ consists
of just two sheets. However our WKB solutions are constructed from $p^\pm_y$
and $-p_y^\pm$ so that we must consider the four Riemann sheets of $p_y(z)$
and $-p_y(z)$ together. FIG.~\ref{fig:BranchCutPlaneSNS}
%-------------------figure------------------------------------------------
%------------endfigure----------------------------------------------------
shows the plane in which the WKB solutions are single-valued. The second
branch cut emanating from $y_-$ is necessary because from 
(\ref{eqn:OverallAmplitudeSNS}) 
\begin{equation}
	\left(
		 \frac{\partial E}
		      { \partial p_y} 
	\right)^{-1/2}
	\propto
	\frac{
		1
	     }
	     {
		\sqrt[\leftroot{8} 4]{
				E^2 - \Delta^2(z)
			}
	     },
\label{eqn:OverallAmplitudePropTo}
\end{equation}
which in the vicinity of $y_-$ is
\begin{equation}
	\left(
		 \frac{\partial E}
		      { \partial p_y} 
	\right)^{-1/2}
	\propto
	\frac{
		e^{-i\theta_-/4}
	     }
	     {
		\sqrt[\leftroot{8} 4]{
				2E
				\begin{smallmatrix}
					\arrowvert \frac{
						\partial \Delta
	    	     	  			        }
	    	          			        {
							\partial y
		      	   			        } \arrowvert_{y_-} \\
				\end{smallmatrix}
				\varrho_-
	    			 }
	     }.
\label{eqn:OverallAmplitudeMultivalued}
\end{equation}
Now (\ref{eqn:OverallAmplitudeMultivalued}) multiplies each of the WKB
solutions and changes sign when $\theta \rightarrow \theta + 4\pi$ even 
though the momentum is single-valued under this change. The WKB solutions
are then only single-valued if a second branch cut is inserted (as shown).
%---------------------------------------------------------------------------
\subsection{Rules for continuing across branch cuts}
\begin{figure}
	\begin{center}
		\epsfig{file=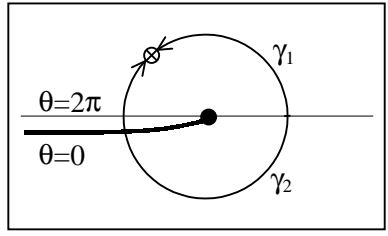,clip=}
		\caption{Alternative contours in complex plane. $\gamma_2$
			 crosses the branch cut.}
		\label{fig:TwoContoursInAppendixA}
	\end{center}
\end{figure}

Let us start with the turning point $y_+$ 
(FIG.~\ref{fig:BranchCutPlaneSNS}), and consider $p_y(z)=p_y^+(z)$.
By inserting the branch cut between $y_+$ and $y_-$ 
we have prevented $p_y(z)$ taking
the value $p_y^-(z)$ at each $z$.
To display all of the values of $p_y(z)$ we take two copies of the complex
plane, label them sheet 1 and sheet 2 respectively, and assign sheet 1 as the
domain of $p_y^+$ and sheet 2 the domain for $p_y^-$.
Look at Fig.~\ref{fig:TwoContoursInAppendixA}. If a contour respects the 
branch cut, $\gamma_1$ for instance, it stays on the first sheet. However
if we follow a contour like $\gamma_2$ when we cross the branch cut of sheet
1 $p_y(z)$ is changing smoothly and so we appear on sheet 2 and $p_y(z)$ has
become $p_y^-(z)$. What happens to the eigenfunctions?

Consider
\begin{equation}
	\Psi_1(z)
	=
	\frac{ 1 }
	     {
		\sqrt{
			\left(
				\frac{\partial E }
		    		{\partial p_y}
			\right)
		     }
	     }_{ \!\!p_y^+(z)} \!\!\!\!\!\!\!\!\!\!\!
	\left(
		\begin{array}{c}
			u^+_{0,I}(z)e^{+i\phi/2} \\
			v^+_{0,I}(z)e^{-i\phi/2}	
		\end{array}
	\right)
	e^{\textstyle
		\frac{i}{\hbar}\int_z^{y_+} dz' \ p_y^+(z')
	  },
\label{eqn:WaveFunctionOneAppendix}
\end{equation}
where
\begin{equation}
	\left(
		 \frac{\partial E}
		      { \partial p_y} 
	\right)^{-1/2}
	\propto
	\frac{
		E^{1/2}
	     }
	     {
		\sqrt[\leftroot{8} 4]{
				E^2 - \Delta^2(z)
			}
	     }
	\frac{
		1
	     }
	     {
		\sqrt{
			p_y^+(z)
		     }
	     },
\label{eqn:OverallAmplitudePropTo.AppendixA}
\end{equation}
and the spinor is given by (see equations
(\ref{eqn:NormaliseUinE})-(\ref{eqn:NormaliseVinE}))
\begin{equation}
	\left(
		\begin{array}{cc}
			u^\beta_{0,I}(z)e^{+i\phi/2} \\
			v^\beta_{0,I}(z)e^{-i\phi/2}
		\end{array}
	\right)
	=
	\left(
		\begin{array}{cc}
			\sqrt{
				\frac{1}{2}
				\left(
					1+\frac{
					\beta \sqrt{ E^2-\Delta^2(z) } 
			     		       }
			       		       { E }
				\right)
	     		     } e^{+i\phi/2}\\
			\sqrt{
				\frac{1}{2}
				\left(
					1-\frac{
					\beta \sqrt{ E^2-\Delta^2(z) } 
			  		        }
			 	                { E }
				\right)
	    		     } e^{-i\phi/2}
		\end{array}
	\right),
\label{eqn:SpinorInAppendixA}
\end{equation}
$\beta=+$ corresponds to $p_y^+$. Now suppose we follow the two
contours shown in FIG.~\ref{fig:TwoContoursInAppendixA}. The first,
$\gamma_1$, does not cross the branch cut.
%-------------------figure------------------------------------------------
%------------endfigure----------------------------------------------------
Let us use $z_{\text{above}}=re^{i\theta_{\text{above}}}$ 
to label the $z$-coordinate in the plane from
following this contour. If instead we follow $\gamma_2$ we do cross the 
branch cut. Let the coordinate from following this path be $z_{\text{below}}$.
We then have $z_{\text{above}}=z_{\text{below}}e^{+i2\pi}$. What we require
is $\Psi_1(z_{\text{below}})$ written in terms of $z_{\text{above}}$. Let us
consider in turn each of the terms entering into $\Psi_1(z)$.
%----------------------subsubsection--------------------------------------
\subsubsection{The change in $\left( \partial E/\partial p_y \right)^{-1/2}$}
Firstly
consider (\ref{eqn:OverallAmplitudePropTo.AppendixA}) and note that
\begin{equation}
	\sqrt{
		E^2 -\Delta^2(z)
	     }
	\approx
	\sqrt{
		2E|\begin{smallmatrix}
			\frac{\partial \Delta}
			     {\partial y}
		  \end{smallmatrix}|_{y_+}
		r
	     } \
	e^{i\theta/2},
\label{eqn:ApproximateSecondSqRoot}
\end{equation}
in the vacinity of $y_+$ so that (\ref{eqn:OverallAmplitudePropTo.AppendixA})
becomes (using $z_{\text{below}}$)
\begin{equation}
	\left(
		 \frac{\partial E}
		      { \partial p_y} 
	\right)^{-1/2}_{p_y^+}
	\propto
	\frac{
		E^{1/2}e^{-i\theta_{\text{below}}/4 }
	     }
	     {
		\sqrt[\leftroot{8} 4]{
					2E|\begin{smallmatrix}
						\frac{\partial \Delta}
			    			 {\partial y}
		  			   \end{smallmatrix}|_{y_+}
				  	   r
				     }
	     }
	\frac{
		1
	     }
	     {
		\sqrt{
			p_y^+\left(
					z_{\text{below}}
			     \right)
		     }
	     }.
\label{eqn:OverallAmplitudeAppendixA}
\end{equation}
We then have
\begin{align}
	\left(
		 \frac{\partial E}
		      { \partial p_y} 
	\right)^{-1/2}_{p_y^+(z_{\text{below}} )}
	&=
	\left(
		 \frac{\partial E}
		      { \partial p_y} 
	\right)^{-1/2}_{p_y^+(z_{\text{above}}e^{-i2\pi})}
\nonumber \\
	&=
	\frac{
		E^{1/2}e^{-i(\theta_{\text{above}} -2\pi)/4 }
	     }
	     {
		\sqrt[\leftroot{8} 4]{
					2E|\begin{smallmatrix}
						\frac{\partial \Delta}
			    			 {\partial y}
		  			   \end{smallmatrix}|_{y_+}
				  	   r
				     }
	     }
	\frac{
		1
	     }
	     {
		\sqrt{
			p_y^+\left(
					z_{\text{above}}e^{-i2\pi}
			     \right)
		     }
	     },
\nonumber \\
	&=
	e^{i\pi/2}
	\left(
		 \frac{\partial E}
		      { \partial p_y} 
	\right)^{-1/2}_{p_y^-(z_{\text{above}})},
\label{eqn:OverallAmplitudeInTermsOfzabove}
\end{align}
where $p_y^+( z_{\text{above}} e^{-i2\pi} )=p_y^-(z_{\text{above}})$ has 
been used. We find this amplitude
aquires a factor $+i$. 
%----------------------subsubsection--------------------------------------
\subsubsection{Change in the spinor}
Here we also have the spinor 
(\ref{eqn:SpinorInAppendixA}) which using (\ref{eqn:ApproximateSecondSqRoot})
becomes
\begin{equation}
	\left(
		\begin{array}{cc}
			u^+_{0,I}(z_{\text{below}})e^{+i\phi/2} \\
			v^+_{0,I}(z_{\text{below}})e^{-i\phi/2}
		\end{array}
	\right)
	=
	\left(
		\begin{array}{cc}
			u^-_{0,I}(z_{\text{above}}) e^{+i\phi/2}\\
			v^-_{0,I}(z_{\text{above}})e^{-i\phi/2}
		\end{array}
	\right).
\label{eqn:SpinorBelowInTermsOfSpinorAbove}
\end{equation}
%----------------------subsubsection--------------------------------------
\subsubsection{Change in the phase integral}
Finally consider the phase
\begin{align}
	\phi_1(z_{\text{below}})
	=&
	\frac{i}{\hbar}
	\int^{y_+}_{z_{\text{below}}}
	p_y^+(z')dz',
\nonumber \\
	=&
	\frac{i}{\hbar}
	\int^{y_+}_{z_{\text{above}}e^{-i2\pi}}
	p_y^+(z')dz'.
\label{eqn:PhaseOneInTermsOfzabove}
\end{align}
Changing the limit $z_{\text{above}}e^{-i2\pi}$ to $z_{\text{above}}$ will
mean the phase of $z'$ along a contour is decreased by $2\pi$, we
must therefore explicitly include $e^{-i2\pi}$ to maintain
the value of (\ref{eqn:PhaseOneInTermsOfzabove}) thus
\begin{align}
	\phi_1(z_{\text{below}})
	&=
	\frac{i}{\hbar}
	\int^{y_+}_{z_{\text{above}}}
	p_y^+(z''e^{-i2\pi})dz'',
\nonumber \\
	&=
	\frac{i}{\hbar}
	\int^{y_+}_{z_{\text{above}}}
	p_y^-(z'')dz'',
\nonumber \\
	&=
	\phi_3(z_{\text{above}}).
\label{eqn:PhiOneInTermsOfPhiThree}
\end{align}
%----------------------subsubsection--------------------------------------
\subsubsection{How the eigenfunctions change} 
Putting (\ref{eqn:OverallAmplitudeInTermsOfzabove}), 
(\ref{eqn:SpinorBelowInTermsOfSpinorAbove}) and 
(\ref{eqn:PhiOneInTermsOfPhiThree}) together we find the rule for crossing
the branch cut to be
\begin{align}
	\Psi_1(z_{\text{below}})
	&=
	+i
	\frac{1}
	     {
		\sqrt{
			\left(
				\frac{\partial E }
		    		{\partial p_y}
			\right)
		     }
	     }_{\!\!p_y^-(z_{\text{above}})}
		\!\!\!\!\!\!\!\!\!\!\!\!\!\!\!\!\!\!\!\!\!\!
	\left(
		\begin{array}{cc}
			u^-_{0,I}(z_{\text{above}}) \\
			v^-_{0,I}(z_{\text{above}})
		\end{array}
	\right)
	e^{\phi_3(z_{\text{above}})},
\nonumber \\
	&=
	+i \Psi_3(z_{\text{above}}).
\label{eqn:Psi1InTermsOfPsi3}
\end{align}
Clearly we also have
\begin{align}
	\Psi_2(z_{\text{below}})
	&
	=+i
	\Psi_4(z_{\text{above}}),
\label{eqn:Psi2InTermsOfPsi4} \\
	\Psi_3(z_{\text{below}})
	&=
	+i
	\Psi_1(z_{\text{above}}),
\label{eqn:Psi3InTermsOfPsi1} \\
	\Psi_4(z_{\text{below}})
	&=
	+i
	\Psi_2(z_{\text{above}}).
\label{eqn:Psi4InTermsOfPsi2}
\end{align}
These rules apply when crossing the branch cut in the clockwise sense around
$y_+$. When moving in the positive sense $+i$ is replaced by $-i$. It is also
easy to see that $\Psi_1^-=+i\Psi_3^-$ etc. for solutions with phase reference
$y_-$.
%----------------------subsubsection--------------------------------------
\subsubsection{A rule for the pure phase factors}
We have previously introduced the definition
\begin{equation}
	[1]
	=
	\exp
	\left(
		\frac{i}{\hbar}\int_{y_-}^{y_+} dz' \ p_y^+(z')
	\right).
\nonumber
\end{equation}
The integral may be taken above or below the cut connecting $y_+$ to $y_-$.
Since $p_y^+(z_{\text{above}})=p_y^-(z_{\text{below}})$ we have
\begin{equation}
	[1]_{\text{above}}
	=
	[3]_{\text{below}},
\nonumber
\end{equation}
or
\begin{equation}
	[1] \rightarrow [3].
\label{eqn:PurePhaseFactor1To3}
\end{equation}
Similarly:
\begin{align}
	[2]& \rightarrow [4], 
\label{eqn:PurePhaseFactor2To4} \\
	[3]& \rightarrow [1],
\label{eqn:PurePhaseFactor3To1} \\
	[4]& \rightarrow [2].
\label{eqn:PurePhaseFactor4To2}
\end{align}
We now have all the rules needed for crossing the branch cut.

Our task is now to find the location of the Stokes and anti-Stokes lines as
defined by (\ref{eqn:ImagPartOfSingulant}) and 
(\ref{eqn:RealPartOfSingulant}).
%---------------------------------------------------------------------------
\subsection{Locating the Stokes and anti-Stokes lines}
\label{sec:LocationOfStokesLines}
\begin{figure}
	\begin{center}
		\epsfig{file=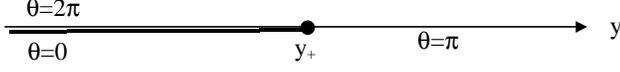,width=86mm,clip=}
		\caption{Branch cut in complex plane.}
		\label{fig:BranchCutFromyMinusInAppendix}
	\end{center}
\end{figure}

In order to locate the Stokes and anti-Stokes lines in the vicinity of $y_+$
and $y_-$ we require the form of the momentum branches there. These are
\begin{equation}
	p_y^\pm(\tilde{z})
	=
	\sqrt{
		p_F^2 -p_x^2 -p_z^2
		\pm 2m\sqrt{
				2E\left|
					\frac{\partial \Delta}
					     {\partial y}
				  \right|_{y_\pm}
				\tilde{z}
			   }
	     },
\label{eqn:MomentumBranchesSNSInAppendix}
\end{equation}
where we have used the expansion for $\Delta(\tilde{z})$ (equation
(\ref{eqn:SmoothDeltaExpansion})), and $\tilde{z}=re^{i\theta}$
is the coordinate in the complex plane centred on either $y_+$ or $y_-$
(which of these it is should be clear from the context). Firstly consider
the vicinity of $y_+$, FIG.~\ref{fig:BranchCutFromyMinusInAppendix}.
%-------------------figure------------------------------------------------
%------------endfigure----------------------------------------------------
We require the exponents of the WKB solutions. Let us start by considering
$\phi_1$
\begin{align}
	\phi_1(z)
	&=
	\frac{i}{\hbar}
	\int^{y_+}_z
	p_y^+(z')dz',
\nonumber \\
	&=
	\frac{i}{\hbar}
	\int^{y_+-z}_0
	p_y^+(\tilde{z})d\tilde{z}.
\label{eqn:PhaseOneExplicit}
\end{align}
To calculate the form of this integral we note that 
$p_F^2 -p_x^2 -p_z^2
\gg 2m\sqrt{2E
		|\begin{smallmatrix}
			\frac{\partial \Delta}{\partial y} 
		\end{smallmatrix}| 
		\tilde{z}
	 }
$ so that 
(\ref{eqn:MomentumBranchesSNSInAppendix}) can be further approximated by
expanding in powers of $\tilde{z}^{1/2}$. 
(\ref{eqn:MomentumBranchesSNSInAppendix}) then becomes
\begin{equation}
	p_y^\pm(z)
	=
	p^{(0)}
	\pm
	p^{(1)} \tilde{z}^{1/2}
	+\cdots,
\label{eqn:MomentumBranchesFurtherExpanded}
\end{equation}
where 
$p^{(0)}=
	\sqrt{ 
		p_F^2 -p_x^2 -p_z^2
	     }
$ 
and $p^{(1)}=\left|
			\frac{\partial p_y^\pm }{\partial z^{1/2}}
	     \right|
	    =
	     \frac{ m\sqrt{2E
				|\begin{smallmatrix} 
					\frac{\partial \Delta}{\partial y}
				\end{smallmatrix}| 
			  }
		  }
		  { p^{(0)} }
$.
Only the first two terms of (\ref{eqn:MomentumBranchesFurtherExpanded})
shall be needed. Inserting the $p_y^+$ expansion into 
(\ref{eqn:PhaseOneExplicit}) we have
\begin{align}
	\phi_1(z)
	&=
	\frac{i}{\hbar}
	\int^{y_+-z}_0
	d\tilde{z}
	\left(	
		p^{(0)} + p^{(1)}\tilde{z}^{1/2}
	\right),
\nonumber \\
	&=
	\frac{i}{\hbar}
	\left\{
		p^{(0)}\left(y_+-z\right) 
		+ \frac{2}{3}p^{(1)}\left(y_+-z\right)^{3/2}
	\right\},
\nonumber
\end{align}
or using $\tilde{z}=re^{i\theta}$ ($\theta=0$ along bottom of the branch
cut in FIG.~\ref{fig:BranchCutFromyMinusInAppendix})
\begin{equation}
	\phi_1(z)
	=
	\frac{i}{\hbar}
	\left\{
		p^{(0)}re^{i\theta} 
		+\frac{2}{3} p^{(1)} r^{3/2}e^{i3\theta/2}
	\right\}.
\label{eqn:PhaseOneUsingExpandedpy}
\end{equation}
If we write 
\begin{equation}
	\{ \}_\pm
	=
	\left\{
		p^{(0)} re^{i\theta} 
		\pm \frac{2}{3} p^{(1)}
		r^{3/2}
		e^{i3\theta/2}
	\right\},
\label{eqn:CompactBracket}
\end{equation}
then the four
exponents of the WKB solutions with phase reference $y_+$ can be written
compactly as
\begin{align}
	\phi_1(z)
	&=
	+\frac{i}{\hbar}
	\{ \}_+,
\label{eqn:CompactPhaseOne} \\
	\phi_2(z)
	&=
	-\frac{i}{\hbar}
	\{ \}_+,
\label{eqn:CompactPhaseTwo} \\
	\phi_3(z)
	&=
	+\frac{i}{\hbar}
	\{ \}_-,
\label{eqn:CompactPhaseThree} \\
	\phi_4(z)
	&=
	-\frac{i}{\hbar}
	\{ \}_-.
\label{eqn:CompactPhaseFour}
\end{align}
%---------------------------------------------------------------------------
\subsubsection{Stokes and anti-Stokes lines around $y_+$}
For the Stokes lines we require $\text{Im} \{ \phi_i - \phi_j \} = 0$
for each distinct pairing of 
(\ref{eqn:CompactPhaseOne})-(\ref{eqn:CompactPhaseFour}). We consider
one example and then the remaining pairs have their results summarised
in table \ref{tab:SummaryOfStokesAndAntiStokes}.

Consider $\phi_1 - \phi_2$:
\begin{align}
	\phi_1 - \phi_2
	&=
	\frac{i}{\hbar}
	\{ \}_+ 
	-
	\frac{-i}{\hbar}
	\{ \}_+,
\nonumber \\
	&=
	\frac{2i}{\hbar}
	\{ \}_+.
\nonumber
\end{align}
The imaginary part is then 
\begin{equation}
	\text{Im}
	\{ \phi_1 -\phi_2 \}
	=
	\frac{2}{\hbar}
	\left(
		p^{(0)} r \cos \theta
		+\frac{2}{3}
		p^{(1)} r^{3/2} \cos \left(
						\frac{3\theta}{2}
				     \right)
	\right),
\label{eqn:ImCompactPhaseDifference}
\end{equation}
whilst the real part is
\begin{equation}
	\text{Re}
	\{ \phi_1 -\phi_2 \}
	=
	-\frac{2}{\hbar}
	\left(
		p^{(0)} r \sin \theta
		+\frac{2}{3}
		p^{(1)} r^{3/2} \sin \left(
						\frac{3\theta}{2}
				     \right)
	\right).
\label{eqn:ReCompactPhaseDifference}
\end{equation}
The Stokes lines are determined by the angles $\theta$ for which $\text{Im}
\{\phi_1 - \phi_2\}$ vanishes. It will be sufficient to locate these lines
approximately. Close to $y_+$ (i.e. $r \rightarrow 0$) we can in the first
approximation neglect the second term in (\ref{eqn:ImCompactPhaseDifference}).
Then
\begin{equation}
	p^{(0)} r\cos \theta^0 =0,
\nonumber
\end{equation}
gives $\theta^0=\pi/2, 3\pi/2$. 
Returning to (\ref{eqn:ImCompactPhaseDifference}) and substituting 
$\theta = \theta^0 +\delta \theta$ we can determine how the second term
shifts $\theta$. Since $\delta \theta$ is small all we require is its sign.
For $\theta^0=\pi/2$
we find $\delta \theta = -|\delta \theta|$ and the Stokes line is then
located at $\theta = \pi/2 -|\delta \theta|$. For $\theta^0=3\pi/2$ we 
also have $\delta \theta =-|\delta \theta|$ as can be checked.
Thus $\theta = 3\pi/2 -|\delta \theta|$. 

To decide which function $\Psi_1$
or $\Psi_2$ is dominant and which is subdominant on the Stokes line we 
consider the sign of $\text{Re} \{ \phi_1 - \phi_2 \}$. For this purpose
it suffices to ignore the second term in (\ref{eqn:ReCompactPhaseDifference}).
For 
$\theta=\pi/2-|\delta\theta|$, $\sin (\pi/2 - |\delta \theta|) >0$, 
$\text{Re} \{ \phi_1 -\phi_2 \}<0$ 
so that 
$\Psi_2 > \Psi_1$. At $\theta=3\pi/2 -|\delta \theta |$ 
the opposite is true ($\Psi_1>\Psi_2$).

We have now located the Stokes lines for the pair of functions $\Psi_1$ and
$\Psi_2$, and we know the dominancy on each line. Calculating the anti-Stokes
lines proceeds similarly.
The results are summerised in table \ref{tab:SummaryOfStokesAndAntiStokes}.
\begin{sidewaystable}
   \renewcommand{\arraystretch}{2.5} % For Single line spacing Interrow spacing times by 2.5
   \begin{center}
   \begin{math}
   \begin{array}{|r r||c|c|c|c|c|c|} \hline % 8 columns!
% column headings
	\multicolumn{2}{|r||}{\phi_i-\phi_j}
	 & \phi_1-\phi_2 & \phi_1 -\phi_3 & \phi_4-\phi_2 &
	\phi_1-\phi_4 & \phi_3-\phi_2 & \phi_3 -\phi_4 \\ \hline
% first row
	\multicolumn{2}{|r||}{\phi_i-\phi_j=}
	 & \frac{i}{\hbar}\{ \}_+ -\frac{-i}{\hbar}\{ \}_+ &
	\multicolumn{2}{c|}{\frac{i}{\hbar}\{ \}_+ -\frac{i}{\hbar}\{ \}_-} &
	\multicolumn{2}{c|}{\frac{i}{\hbar}\{ \}_+ -\frac{-i}{\hbar}\{ \}_-} &
	\frac{i}{\hbar}\{ \}_- -\frac{-i}{\hbar}\{ \}_- \\ \hline
% second row
	\multicolumn{2}{|r||}{$Im$\{ \phi_i-\phi_j \}\propto} &
	\cos \theta+\text{fn}(r)\cos \left( \frac{3\theta}{2} \right) &
	\multicolumn{2}{c|}{ r^{3/2} \cos \left( \frac{3\theta}{2} \right)} &
	\multicolumn{2}{c|}{ r \cos \theta} &
	\cos \theta-\text{fn}(r)\cos \left( \frac{3\theta}{2} \right) \\ \hline
% third row
	&\theta_1 &
	\frac{\pi}{2}-|\delta \theta|, \ \Psi_2>\Psi_1 &
	\multicolumn{2}{c|}{
		\frac{\pi}{3}, \ \Psi_3>\Psi_1, \ \Psi_2>\Psi_4 } &
	\multicolumn{2}{c|}{
		\frac{\pi}{2}, \ \Psi_4>\Psi_1, \ \Psi_2>\Psi_3 } &
	\frac{\pi}{2}+|\delta \theta|, \ \Psi_4>\Psi_3 \\
% fourth row
	&\theta_2 &
	\frac{3\pi}{2}-|\delta \theta|, \ \Psi_1>\Psi_2 &
	\multicolumn{2}{c|}{
		\pi, \ \Psi_1>\Psi_3, \ \Psi_4>\Psi_2 } &
	\multicolumn{2}{c|}{
		\frac{3\pi}{2}, \ \Psi_1>\Psi_4, \ \Psi_3>\Psi_2 } &
	\frac{3\pi}{2}+|\delta \theta|, \ \Psi_3>\Psi_4 \\
% fifth row
	&\theta_3 & -
	& \multicolumn{2}{c|}{
		\frac{5\pi}{3}, \ \Psi_3>\Psi_1, \ \Psi_2>\Psi_4 } &
	\multicolumn{2}{c|}{-} & - \\ \hline
% sixth row
	\multicolumn{2}{|r||}{
	$Re$\{\phi_i-\phi_j\} \propto} &
	\sin \theta+\text{fn}(r)\sin \left( \frac{3\theta}{2} \right) &
	\multicolumn{2}{c|}{ r^{3/2} \sin \left( \frac{3\theta}{2} \right)} &
	\multicolumn{2}{c|}{ r \sin \theta} &
	\sin \theta-\text{fn}(r)\sin \left( \frac{3\theta}{2} \right) \\ \hline
% seventh row
	&\theta_1 & 0 &
	\multicolumn{2}{c|}{0} &
	\multicolumn{2}{c|}{0} &
	0 \\
% eighth row
	&\theta_2 &  \pi-|\delta \theta|,  &
	\multicolumn{2}{c|}{\frac{2\pi}{3} } &
	\multicolumn{2}{c|}{\pi } &
	\pi+|\delta \theta| \\
% nineth row
	&\theta_3 &2\pi &
	\multicolumn{2}{c|}{\frac{4\pi}{3} } &
	\multicolumn{2}{c|}{2\pi } &
	2\pi \\
% tenth row
	\begin{rotate}{+90}\hspace*{50.0mm}
		\parbox[t]{29mm}{\par\vspace*{+4mm} 
		$\theta_{\text{Im}(\phi_i-\phi_j)=0}$, and 
		Re$\{\phi_{\ell}\} >$Re$\{\phi_m\}$ 
		$\Rightarrow \Psi_\ell>\Psi_m$}
	\end{rotate}
	\begin{rotate}{+90}\hspace*{0.0mm} 
		\parbox[t]{42mm}{\par\vspace*{+4mm}
		$\theta_{\text{Re}(\phi_i-\phi_j)=0}$,
		$\!\!\! \Rightarrow \!\!\! (\Psi_i,\Psi_j)$ \\
		(see column heading
		\\ for neutral pair)}
	\end{rotate}
% the above rotated text is thus in the first column of row 10, the rest of
% this row is below.
	&\theta_4 & - &
	\multicolumn{2}{c|}{2\pi} &
	\multicolumn{2}{c|}{-} &
	- \\ \hline
   \end{array}
   \end{math}
   \caption{Summary of Stokes lines and anti-Stokes lines located 
		around the turning point $y_+$.}
   \label{tab:SummaryOfStokesAndAntiStokes}
   \end{center}
\end{sidewaystable}

\begin{figure}
	\begin{center}
		\epsfig{file=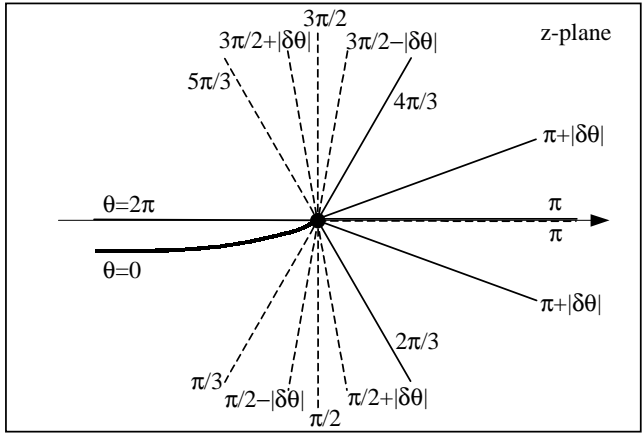,width=86mm,clip=}
		\caption{Stokes and anti-Stokes lines around $y_+$.}
		\label{fig:ManyStokesEtcAroundy+}
	\end{center}
\end{figure}

FIG.~\ref{fig:ManyStokesEtcAroundy+} shows the Stokes and anti-Stokes lines
around $y_+$.
%-------------------figure------------------------------------------------
%------------endfigure----------------------------------------------------
Notice that the anti-Stokes lines at $+\pi-|\delta\theta|$ and 
$+\pi+|\delta\theta|$ are shifted off the real axis ($|\delta\theta|\neq 0$)
by the second term in (\ref{eqn:ReCompactPhaseDifference}). These
lines are where $\phi_1-\phi_2$ and $\phi_3-\phi_4$ are neutral but since
$\phi_2 = -\phi_1$, and $\phi_4=-\phi_3$ this neutrality condition reduces
to $\Psi_1,\Psi_2,\Psi_3$ and $\Psi_4$ being individually neutral (i.e.,
oscillatory) on the appropriate Stokes line. It is these Stokes lines no
longer coinciding with the real $y$-axis that ensures the existence of
evanescent solutions for $y \rightarrow +\infty$, i.e on the real axis.
%-------------------------------------------------------------------------
\subsubsection{Stokes and anti-Stokes lines around $y_-$}
\begin{figure}
	\begin{center}
		\epsfig{file=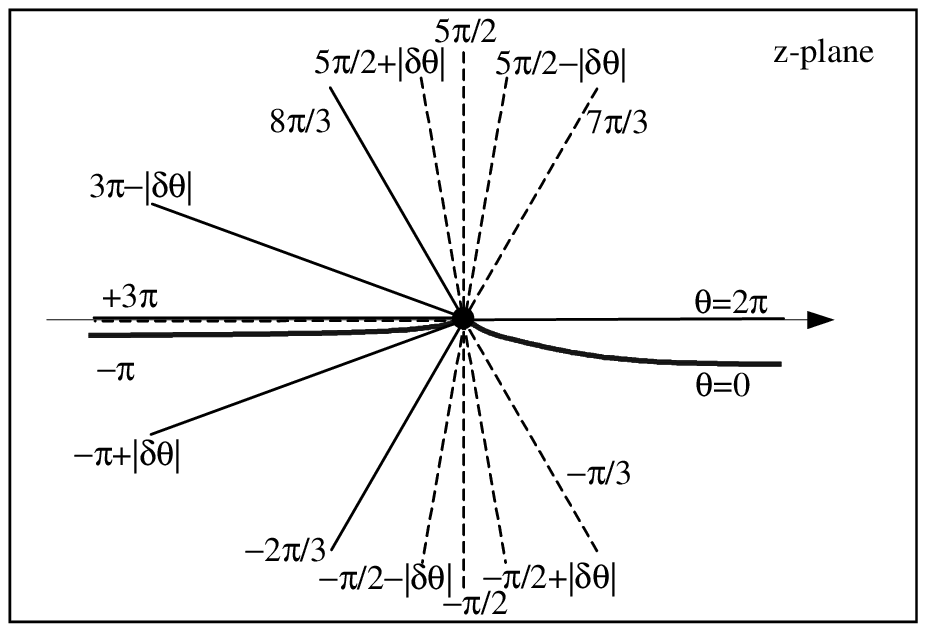,width=86mm,clip=}
		\caption{Stokes and anti-Stokes lines around $y_-$}
		\label{fig:ManyStokesEtcAroundy-}
	\end{center}
\end{figure}

\begin{figure}[htbp]
	\begin{center}
		\epsfig{file=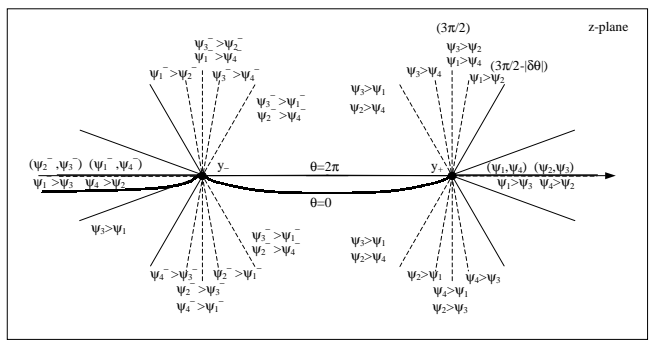,width=86mm,clip=}
		\caption[Location of the Stokes and anti-Stokes lines
			for the SNS junction]
			{Location of the Stokes and anti-Stokes lines
			for the SNS junction problem when there are four
			WKB solutions to consider.}
		\label{fig:StokesLinesForTwoTurningPointsSNS}
	\end{center}
\end{figure}

So far we have only considered the turning point at $y_+$. The situation is
very similar at $y_-$. Note however that the phases are given by
\begin{align}
	\phi_1^-(z)
	&=
	\frac{i}{\hbar}
	\int^{y_-}_z
	p_y^+(z')
	dz',
\nonumber \\
	&=
	-\frac{i}{\hbar}
	\int^{z-y_-}_0
	p_y^+(\tilde{z})
	d\tilde{z},
\nonumber \\
	&=
	-\frac{i}{\hbar}
	\left\{
		p^{(0)}\left(z-y_-\right) 
		+ \frac{2}{3}p^{(1)}\left(z-y_-\right)^{3/2}
	\right\},
\label{eqn:PhaseOneMinusExpanded}
\end{align}
where now $z-y_- = re^{i\theta}$ ($\theta=0$ below the cut again), i.e.,
\begin{align}
	\phi_1^-(z)
	&=
	-\frac{i}{\hbar}
	\{ \}_+,
\label{eqn:CompactPhaseOneMinus}
\intertext{and}
	\phi_2^-(z)
	&=
	+\frac{i}{\hbar}
	\{ \}_+,
\label{eqn:CompactPhaseTwoMinus} \\
	\phi_3^-(z)
	&=
	-\frac{i}{\hbar}
	\{ \}_-,
\label{eqn:CompactPhaseThreeMinus} \\
	\phi_4^-(z)
	&=
	+\frac{i}{\hbar}
	\{ \}_-,
\label{eqn:CompactPhaseFourMinus}
\end{align}
(compare (\ref{eqn:CompactPhaseOne})-(\ref{eqn:CompactPhaseFour})). Using
these the Stokes and anti-Stokes lines around $y_-$ can be calculated. The
results are summarised in table \ref{tab:SummaryOfStokesAndAntiStokesy-},
and drawn in FIG.~\ref{fig:ManyStokesEtcAroundy-}.
%-------------------figure------------------------------------------------
%------------endfigure----------------------------------------------------
\begin{sidewaystable}
   \renewcommand{\arraystretch}{2.5} % For Single line spacing Interrow spacing times by 2.5
   \begin{center}
   \begin{math}
   \begin{array}{|r r||c|c|c|c|c|c|} \hline % 8 columns!
% column headings
	\multicolumn{2}{|r||}{\phi_i^--\phi_j^-}
	 & \phi_2^--\phi_1^- & \phi_3^- -\phi_1^- & \phi_2^--\phi_4^- &
	\phi_4^--\phi_1^- & \phi_2^--\phi_3^- & \phi_4^- -\phi_3^- \\ \hline
% first row
	\multicolumn{2}{|r||}{\phi_i^--\phi_j^-=}
	 & \frac{i}{\hbar}\{ \}_+ -\frac{-i}{\hbar}\{ \}_+ &
	\multicolumn{2}{c|}{-\frac{i}{\hbar}\{ \}_- -\frac{-i}{\hbar}\{ \}_+} &
	\multicolumn{2}{c|}{\frac{i}{\hbar}\{ \}_+ -\frac{-i}{\hbar}\{ \}_-} &
	\frac{i}{\hbar}\{ \}_- -\frac{-i}{\hbar}\{ \}_- \\ \hline
% second row
	\multicolumn{2}{|r||}{$Im$\{ \phi_i-\phi_j \}\propto} &
	\cos \theta+\text{fn}(r)\cos \left( \frac{3\theta}{2} \right) &
	\multicolumn{2}{c|}{ r^{3/2} \cos \left( \frac{3\theta}{2} \right)} &
	\multicolumn{2}{c|}{ r \cos \theta} &
	\cos \theta-\text{fn}(r)\cos \left( \frac{3\theta}{2} \right) \\ \hline
% third row
	&\theta_1 &
	-\frac{\pi}{2}+|\delta \theta|, \ \Psi_2^->\Psi_1^- &
	\multicolumn{2}{c|}{
		-\frac{\pi}{3}, \ \Psi_3^->\Psi_1^-, \ \Psi_2^->\Psi_4^- } &
	\multicolumn{2}{c|}{
		\frac{\pi}{2}, \ \Psi_4^->\Psi_1^-, \ \Psi_2^->\Psi_3^- } &
	-\frac{\pi}{2}-|\delta \theta|, \ \Psi_4^->\Psi_3^- \\
% fourth row
	&\theta_2 &
	\frac{5\pi}{2}+|\delta \theta|, \ \Psi_1^->\Psi_2^- &
	\multicolumn{2}{c|}{
		-\pi, \ \Psi_1^->\Psi_3^-, \ \Psi_4^->\Psi_2^- } &
	\multicolumn{2}{c|}{
		+\frac{5\pi}{2}, \ \Psi_1^->\Psi_4^-, \ \Psi_3^->\Psi_2^- } &
	\frac{5\pi}{2}-|\delta \theta|, \ \Psi_3^->\Psi_4^- \\
% fifth row
	&\theta_3 & -
	& \multicolumn{2}{c|}{
		\frac{5\pi}{3}, \ \Psi_3^->\Psi_1^-, \ \Psi_2^->\Psi_4^- } &
	\multicolumn{2}{c|}{-} & - \\ \hline
% sixth row
	\multicolumn{2}{|r||}{
	$Re$\{\phi_i^--\phi_j^-\} \propto} &
	\sin \theta+\text{fn}(r)\sin \left( \frac{3\theta}{2} \right) &
	\multicolumn{2}{c|}{ -r^{3/2} \sin \left( \frac{3\theta}{2} \right)} &
	\multicolumn{2}{c|}{ -r \sin \theta} &
	\sin \theta-\text{fn}(r)\sin \left( \frac{3\theta}{2} \right) \\ \hline
% seventh row
	&\theta_1 & 0 &
	\multicolumn{2}{c|}{0} &
	\multicolumn{2}{c|}{0} &
	0 \\
% eighth row
	&\theta_2 &  -\pi+|\delta \theta|,  &
	\multicolumn{2}{c|}{-\frac{2\pi}{3} } &
	\multicolumn{2}{c|}{-\pi } &
	3\pi-|\delta \theta| \\
% nineth row
	&\theta_3 &2\pi &
	\multicolumn{2}{c|}{\frac{8\pi}{3} } &
	\multicolumn{2}{c|}{2\pi } &
	2\pi \\
% tenth row
	\begin{rotate}{+90}\hspace*{50.0mm}
		\parbox[t]{29mm}{\par\vspace*{+4mm} 
		$\theta_{\text{Im}(\phi_i^--\phi_j^-)=0}$ and 
		Re$\{\phi_{\ell}^-\} >$Re$\{\phi_m^-\}$ 
		$\Rightarrow \Psi_\ell>\Psi_m$}
	\end{rotate}
	\begin{rotate}{+90}\hspace*{-2.0mm}
		\parbox[t]{42mm}{\par\vspace*{+4mm} 
		$\theta_{\text{Re}(\phi_i^- \! -\phi_j^-)=0},\!
		\Rightarrow$$(\Psi_i^-\! ,\Psi_j^-)$\\ (see column heading \\
		for neutral pair)\\ \ }
	\end{rotate}
% the above rotated text is thus in the first column of row 10, the rest of
% this row is below.
	&\theta_4 & - &
	\multicolumn{2}{c|}{-} &
	\multicolumn{2}{c|}{3\pi} &
	- \\ \hline
   \end{array}
   \end{math}
   \caption{Summary of Stokes lines and anti-Stokes lines located 
		around the turning point $y_-$.}
   \label{tab:SummaryOfStokesAndAntiStokesy-}
   \end{center}
\end{sidewaystable}
The $z$-plane containing both $y_+$ and $y_-$, and all the Stokes-lines and
dominancy changes is shown in Fig.~\ref{fig:StokesLinesForTwoTurningPointsSNS}.
%-------------------figure------------------------------------------------
%------------endfigure----------------------------------------------------
This figure contains no fewer than 18 Stokes lines and 11 anti-Stokes lines.
When there are more than two solutions it becomes difficult to label a 
solution with dominance or subdominance. Dominant with respect to which of
the other solutions? The approach we have taken here is to write on the 
figure either $\Psi_i>\Psi_j$ (where we use the $>$ 
symbol not just to represent
greater than but rather to indicate maximal dominance of $\Psi_i$ over
$\Psi_j$) or ($\Psi_i,\Psi_j$) which indicates that the pair of solutions
are neutral with respect to each other. Clearly $\Psi_i>\Psi_j$ is needed
at each Stokes line and ($\Psi_i,\Psi_j$) at each anti-Stokes line. Thus if
a $\Psi_1>\Psi_2$ Stokes line is crossed the coefficient of $\Psi_2$
undergoes Stokes phenomenon unless $\Psi_1$ is not present. Notice that 
along the real axis for $y>y_+$ and $y<y_-$ we have coincident Stokes and
anti-Stokes lines. At first this seems like a contradiction but it is not
for the following reason. The anti-Stokes line means that the pair
($\Psi_1,\Psi_4$) is neutral, as is ($\Psi_2,\Psi_3$). But according to the
Stokes line $\Psi_1>\Psi_3$ and $\Psi_4>\Psi_2$. Taken together these 
statements imply ($\Psi_1,\Psi_4$) $>$ ($\Psi_2,\Psi_3$), or in words, the 
neutral pair
($\Psi_1,\Psi_4$) is maximally dominant and ($\Psi_2,\Psi_3$) maximally
subdominant. 
Explicitly the solution along $y>y^+$ which decays is
\begin{align}
	\Psi_{y>y_+}(y)
	&=
	A \Psi_2 + B \Psi_3,
\label{eqn:DecayingSolution} \\
	&=
		\frac{ A }
		     {
			\sqrt{
				\left(
					\frac{\partial E }
			    		{\partial p_y}
				\right)
			     }
		     }_{\!\!p_y^+(y)} \!\!\!\!\!\!\!\!\!\!\!
		\left(\!\!\!
			\begin{array}{c}
				u^+_{0,I}(y)e^{+i\phi/2} \\
				v^+_{0,I}(y)e^{-i\phi/2}		
			\end{array}\!\!\!
		\right)\!
		e^{\! \textstyle
			+\frac{i}{\hbar}
			\int_{y_+}^y p_y^{r}(y')dy'
		  }\!
	e^{\textstyle
	 	-\frac{1}{\hbar} \left|
					\int_y^{y_+} \! p_y^{i}(y')dy'
				 \right|
	  }
\nonumber \\
	&+
		\frac{ B }
		     {
			\sqrt{
				\left(
					\frac{\partial E }
		    			{\partial p_y}
				\right)
			     }
		     }_{\!\!p_y^-(y)} \!\!\!\!\!\!\!\!\!\!\!
		\left(\!\!\!
			\begin{array}{c}
				u^-_{0,I}(y) e^{+i\phi/2}\\
				v^-_{0,I}(y) e^{-i\phi/2}	
			\end{array}\!\!\!
		\right)\!
		e^{\! \textstyle
			-\frac{i}{\hbar}
			\int_{y_+}^y p_y^{r}(y')dy'
		  }\!
	e^{\textstyle
	 	-\frac{1}{\hbar} \left|
					\int_y^{y_+}\! p_y^{i}(y')dy'
				 \right|
	  },
\label{eqn:ExplicitDecayingSolution}
\end{align}
where the modulus in the exponent ensures an evanescent solution. In 
(\ref{eqn:ExplicitDecayingSolution}) we have used $p_y^r$ and $p_y^i$
to represent the real and imaginary parts respectively of the momentum.
Note that $p_y^+=(p_y^-)^*$ so that the real parts are the same. Now we
can understand fully the more general definitions of the Stokes lines and
anti-Stokes lines. The pair ($\Psi_2,\Psi_3$) are neutral because they
have a common damping factor, neither is more damped than the other. It
is only the real part of the exponents, $\phi_i,\phi_j$,
which control this hence
$\text{Re}\{ \phi_i - \phi_j \}=0$ is the general anti-Stokes condition.
The exponentially growing solution is
\begin{align}
	\Psi_{y>y_+}(y)
	&=
	C \Psi_1 + D \Psi_4,
\label{eqn:GrowingSolution} \\
	&=
		\frac{ C }
		     {
			\sqrt{
				\left(
					\frac{\partial E }
			    		{\partial p_y}
				\right)
			     }
		     }_{\!\!p_y^+(y)} \!\!\!\!\!\!\!\!\!\!\!
		\left(\!\!\!
			\begin{array}{c}
				u^+_{0,I}(y)e^{+i\phi/2} \\
				v^+_{0,I}(y)e^{-i\phi/2}		
			\end{array}\!\!\!
		\right)\!
		e^{\! \textstyle
			-\frac{i}{\hbar}
			\int_{y_+}^y p_y^{r}(y')dy'
		  }\!
	e^{\textstyle
	 	+\frac{1}{\hbar} \left|
					\int_y^{y_+} \! p_y^{i}(y')dy'
				 \right|
	  }
\nonumber \\
	&+
		\frac{ D }
		     {
			\sqrt{
				\left(
					\frac{\partial E }
		    			{\partial p_y}
				\right)
			     }
		     }_{\!\!p_y^-(y)} \!\!\!\!\!\!\!\!\!\!\!
		\left(\!\!\!
			\begin{array}{c}
				u^-_{0,I}(y) e^{+i\phi/2}\\
				v^-_{0,I}(y) e^{-i\phi/2}	
			\end{array}\!\!\!
		\right)\!
		e^{\! \textstyle
			+\frac{i}{\hbar}
			\int_{y_+}^y p_y^{r}(y')dy'
		  }\!
	e^{\textstyle
	 	+\frac{1}{\hbar} \left|
					\int_y^{y_+} \! p_y^{i}(y')dy'
				 \right|
	  }.
\label{eqn:ExplicitGrowingSolution}
\end{align}
Comparing this with (\ref{eqn:ExplicitDecayingSolution}) 
we see that $\Psi_1$ and
$\Psi_3$ have the same oscillatory part, likewise $\Psi_4$ and $\Psi_2$.
Again attention is focused on the real part, 
$\text{Re}\{ \phi_{\text{1,4}} \} > \text{Re}\{ \phi_{\text{2,3} } \}$,
because we can discount the common phase. The condition for discounting the
phase is the general Stokes-line statement $\text{Im}\{\phi_i - \phi_j \}=0$.

Suppose we start along $y>y_+$ with the decaying solution 
(\ref{eqn:DecayingSolution}). If we follow the solution around $y_+$ in the
upper half plane (see FIG.~\ref{fig:ManyStokesEtcAroundy+} 
page \pageref{fig:ManyStokesEtcAroundy+})
we expect the first Stokes jump to occur at $\theta=+3\pi/2$. 
Continuing around to $\theta=+2\pi$
we expect to have had four Stokes jumps and hence four Stokes constants are
introduced. All four solutions are then present on the real $y$-axis between
the turning points. The solution along $y<y_-$ would in principle contain
10 Stokes constants. Fortunately there
\label{page:ManyStokesConstantsAreZero}
is a quite remarkable simplification as will be seen in the next section
when we calculate the Stokes constants.

This concludes our discussion about Stokes and anti-Stokes lines.
%---------------------------------------------------------------------------
\subsection{Calculation of Stokes constants}
\label{sec:CalculationOfStokesConstants}
Stokes constants can be calculated by following the changes of a solution
when passing all the way around a turning point and back to the starting
point. 

We will now calculate the Stokes constants corresponding to the Stokes lines
in figures \ref{fig:ManyStokesEtcAroundy+} and \ref{fig:ManyStokesEtcAroundy-}.
We shall follow the changes in a 
solution, evanescent along $y>y_+$, as we move around $y_+$ in the complex
plane. We will need to know what to do when crossing the branch cut along
$y<y_+$.

\begin{figure}
	\begin{center}
		\epsfig{file=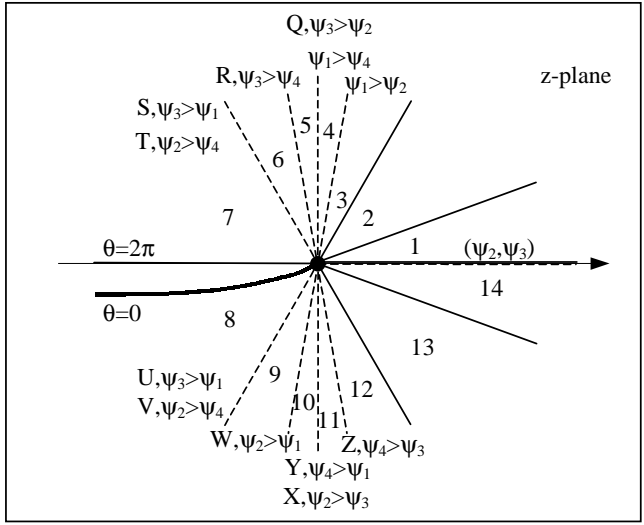,width=86mm,clip=}
		\caption{Stokes - - - and anti-Stokes {\bf -----} 
			lines around $y_+$.}
		\label{fig:StokesLinesAndConstantsAroundy+}
	\end{center}
\end{figure}

FIG.~\ref{fig:StokesLinesAndConstantsAroundy+} shows the various
sectors in the complex plane, and we have also included a Stokes constant
%-------------------figure------------------------------------------------
%------------endfigure----------------------------------------------------
corresponding to each condition $\Psi_i>\Psi_j$ which is relevant when
starting from the evanescent pair ($\Psi_2,\Psi_3$). Notice that while we
require the solution for $y>y_+$ to decay in order to satisfy the physical
boundary conditions any arbitrary combination of $\Psi_2$ and $\Psi_3$ 
have this property. Thus we consider $\Psi=A\Psi_2+B\Psi_3$ ($A,B$ arbitrary).
In what follows the numbers indicate the sector under consideration and next
to them we have written the form of the solution in that sector. We have:
\begin{equation}
	\begin{array}{rl}
	1-4:  & A\Psi_2+B\Psi_3 \\
	5:    & (A+BQ)\Psi_2+B\Psi_3 \\
	6:    & (A+BQ)\Psi_2+B\Psi_3 +BR\Psi_4 \\
	7:    & SB\Psi_1 +(A+BQ)\Psi_2+B\Psi_3 \\
	      &	+(BR+(A+BQ)T)\Psi_4 \\
	8:   &-iSB\Psi_3 -i(A+BQ)\Psi_4-iB\Psi_1 \\
	      &	-i(BR+(A+BQ)T)\Psi_2 \\
	9:   &-iB(1+SU)\Psi_1 +\text{same}\ \Psi_2,\Psi_3,\\
	     &-i( A+BQ+(BR+(A+BQ)T)V)\Psi_4 \\
	10:  &-i(B(1+SU)+W(BR+(A+BQ)T))\Psi_1 \\
	     &+\text{same}\ \Psi_2,\Psi_3,\Psi_4 \\
	11:  &-i(B(1+SU)+W(BR+(A+BQ)T) \\
		&+Y( A+BQ+(BR+(A+BQ)T)V))\Psi_1 \\
             & +\text{same}\ \Psi_2,\Psi_4,\\
	     &-i(SB+X(BR+(A+BQ)T))\Psi_3 \\
	12-14:&-i (B(1+SU)+W(BR+(A+BQ)T) \\
		&\quad  +Y( A+BQ+(BR+(A+BQ)T)V))\Psi_1 \\
		&-i(BR+(A+BQ)T)\Psi_2 \\
		&-i(SB+X(BR+(A+BQ)T) \\
		&\quad +Z( A+BQ+(BR+(A+BQ)T)V))\Psi_3 \\
		&-i( A+BQ+(BR+(A+BQ)T)V)\Psi_4
	\end{array}
\label{eqn:PsiInNeutralRegionFory+}
\end{equation}
Comparing the solution in sector 14 with that in 1 fixes the 
Stokes constants. To help us here we are allowed to treat each equation
obtained by equating coefficients as two equations because the parts
depending upon $A$ and $B$ can be matched separately (remember $A$ and $B$
were arbitrary so we can vary either independently). Thus for instance
matching the coefficient of $\Psi_2$ gives
\begin{equation}
	A=-i(BR+(A+BQ)T)
\nonumber
\end{equation}
or
\begin{align}
	1 &=-iT \qquad (\text{from varying }A),
\label{eqn:EquationGivingT} \\
	0 &=R+QT \qquad (\text{from varying }B).
\label{eqn:EquationRelatingRandQ}
\end{align}
(\ref{eqn:EquationGivingT}) gives $T=+i$. 
Proceeding similarly we find
\begin{align}
	S=T=U=V=+i,
\nonumber \\
	Q=R=W=X=0.	
\nonumber
\end{align}
$Y$ and $Z$ are undetermined but this does not matter because we can
avoid crossing the corresponding Stokes lines when solving the two turning
point problem.

\begin{figure}
	\begin{center}
		\epsfig{file=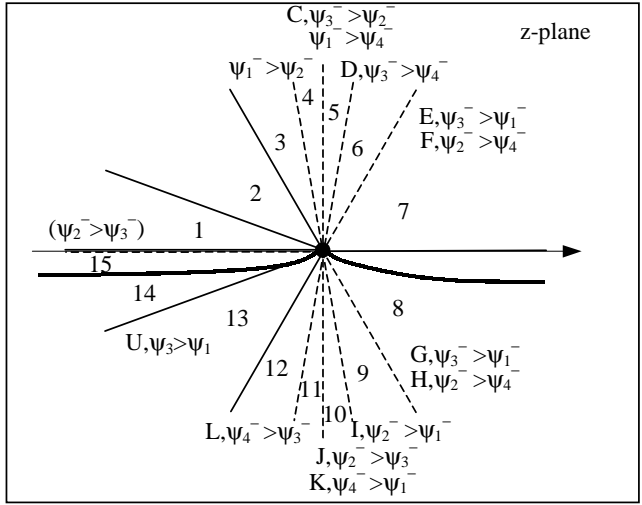,width=86mm,clip=}
		\caption{Stokes - - - and anti-Stokes {\bf -----}
			lines around $y_-$.}
		\label{fig:StokesLinesAndConstantsAroundy-}
	\end{center}
\end{figure}

Calculating Stokes constants around $y_-$ proceeds in exactly the same way.
FIG.~\ref{fig:StokesLinesAndConstantsAroundy-}
%-------------------figure------------------------------------------------
%------------endfigure----------------------------------------------------
shows the various constants. (The extra branch cut between sectors 14 and 15
means $\Psi_i^-(z_{14})=-\Psi_i^-(z_{15})$.)
Note however that moving
through the sectors $1 \rightarrow 14$ we are moving around
 $y_-$ in the negative (clockwise) sense and we require Stokes constants 
in the positive sense. Taking this into account we find
\begin{align}
	E=F=G=H=+i,
\nonumber \\
	C=D=I=J=0.	
\nonumber
\end{align}
$K$ and $L$ are undetermined, but again they will not be needed.
%-------------------------------------------------------------------------
\subsection{Discussion of Stokes constants}
We have found that all the constants are zero apart from those corresponding
to $\theta=+\pi/3,+5\pi/3$ around $y_+$ and $\theta=-\pi/3,+7\pi/3$
around $y_-$. This is surprising since it appeared that 
the sufficient condition for Stokes phenomenon to occur was given 
by (\ref{eqn:SufficiencyForStokes}). If we consider again going around
$y_+$ we find the following is true
\begin{center}
	$\Psi_3$ only `sees' $\Psi_1$, \\
	$\Psi_2$ only `sees' $\Psi_4$.
\end{center}
Return to the definitions of $\Psi_i$, equations
(\ref{eqn:WaveFunctionOne})-(\ref{eqn:WaveFunctionFour}), then we observe that
the exponent of $\Psi_1$ depends upon $+p_y^+$, and $\Psi_3$ upon $+p_y^-$,
whilst $\Psi_2$ and $\Psi_4$ depend upon $-p_y^+$,and $-p_y^-$ respectively.
This is the resolution of the problem. When analytically continuing around
the complex plane $p_y^+$ can be continued to $p_y^-$ and likewise $-p_y^+$
to $-p_y^-$, but there is no way (at least locally in the plane) for
$p_y^+$ to be continued to $-p_y^+$. This would require the real part of
$p_y^+$ to pass through zero, but we have assumed
$p_F^2-p_x^2-p_z^2$ is large so that 
$\pm2m\sqrt{E^2 - \Delta^2(z)}$ cannot reduce the momentum to zero.
However one might imagine that $|\Delta(y)|$ analytically continued into
the complex plane may have singularities and might therefore reduce $p_y^+$
to zero somewhere. For the moment at least we had better say that 
%%%%%----------must find a way of centering this and aligning it-----------
\begin{quote}if {\it locally} in the complex plane solutions cannot be 
analytically continued one into another then there will be 
no Stokes phenomenon between them.
\end{quote}
To put it another way, solutions on disjoint pieces of a Riemann surface
do not experience Stokes phenomenon. Note that $p_y(z)$ has two sheets
(corresponding to $p_y(z=y)=p^+_y(y)$ and $p_y(z=y)=p^-_y(y)$) and $-p_y(z)$
also has two sheets $-p^+_y$ and $-p^-_y$, but although all four sheets make up
the Riemann surface for the solution the sheets exist in mutually disjoint
pairs.

\subsection{Derivation of a generalised Bohr-Sommerfeld quantisation rule}
\begin{figure}
	\begin{center}
		\epsfig{file=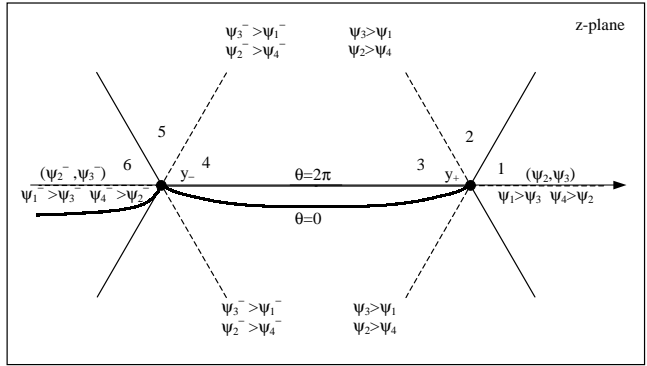,width=86mm,clip=}
		\caption[Location of Stokes lines after simplification]
			{Location of Stokes lines and anti-Stokes lines
			once the simplification due to disjoint Riemann
			sheets has been taken into account.}
		\label{fig:SimplifiedProblemStokesLines}
	\end{center}
\end{figure}

Returning to the problem at hand it is clear that a significant simplification
of FIG.~\ref{fig:StokesLinesForTwoTurningPointsSNS} is possible. We
need only consider those Stokes lines (shown in figure 
\ref{fig:SimplifiedProblemStokesLines}) whose Stokes constants are non-zero.
%-------------------figure------------------------------------------------
%------------endfigure----------------------------------------------------
These Stokes constants are all $+i$ when the lines are crossed in the positive
sense. Finally then, let us follow the evolution of an 
evanescent solution constructed
out of $\Psi_2$ and $\Psi_3$ through sectors $1\rightarrow 6$. We have:
\begin{align}
	1-2:& \ A\Psi_2 + B\Psi_3 
\nonumber \\
	3:& \ iB\Psi_1 + A\Psi_2 + B\Psi_3 +iA\Psi_4
\nonumber \\
	4:& \ iB[1]\Psi_1^- + A[2]\Psi_2^- + B[3]\Psi_3^- +iA[4]\Psi_4^-
\nonumber \\
	5-6:& \ iB\left( [1]+[3] \right)\Psi_1^-
		 + A[2]\Psi_2^- + B[3]\Psi_3^- 
\nonumber \\
	    &  +iA\left( [4]+[2] \right)\Psi_4^-.
\label{eqn:EvanescentSolutionToLeft}
\end{align}
Now for $y<y_-$ the evanescent solution must consist of $\Psi_2^-$ and
$\Psi_3^-$ only so the coefficients of $\Psi_1^-$ and $\Psi_4^-$ must both
be zero. For $\Psi_1^-$ this yields
\begin{align}
	&[1]+[3] =0,
\nonumber \\
	&\frac{[1]}{[3]}=-1,
\nonumber
\end{align}
or
\begin{equation}
	e^{
			\frac{i}{\hbar}
			\int^{y_+}_{y-} p^+_y(y') \ dy'
			-\frac{i}{\hbar}
			\int^{y_+}_{y-} p^-_y(y') \ dy'
	  }
	=
	e^{i\pi(2n+1)}.
\nonumber
\end{equation}
We have derived the quantisation condition
\begin{equation}
	\frac{1}{\hbar}	
	\int^{y_+}_{y-} p^+_y(y') - p^-_y(y')\ dy'
	=
	2\pi \left(n + \frac{1}{2} \right),
\label{eqn:DerivedBohrSommerfeldRuleSNS}
\end{equation}
used earlier
(equation (\ref{eqn:BohrSommerfeldRuleSNS})) to derive the Andreev
spectrum but we have also found the turning point correction $\gamma = 1/2$,
i.e. the Maslov index $m=2$. This is what is expected for the Maslov index
of a Lagrangian manifold which is topologically a circle. In the present
case, from our approach, we see each turning point
introduces a phase change of $\pi/2$, i.e. contributes 1 to the Maslov index
(see (\ref{eqn:DerivedBohrSommerfeldRuleSNS})), whenever the local topology
of the Riemann sheets in the vicinity of the turning point consists of
two sheets. This concludes our derivation of the quantisation condition.

Before leaving this section we wish to comment that in the
main body of this paper the explicit solutions are written down 
for each region. Since
we require these for $z=y$, between the turning points, and our solutions
in sector 3 and 4 are for $z=ye^{i2\pi}$, we must take the functional forms
$\Psi_i(ye^{i2\pi})$ and rewrite them in terms of $\Psi_i(y)$ i.e.,
we must follow the solutions in sector 3 across the branch cut connecting
$y_+$ and $y_-$. Again we use the 
rules (\ref{eqn:Psi1InTermsOfPsi3})-(\ref{eqn:Psi4InTermsOfPsi2}), and
(\ref{eqn:PurePhaseFactor1To3})-(\ref{eqn:PurePhaseFactor4To2}). 
Thus for example the solution $\Psi_B(y)$, equation 
(\ref{eqn:TheExplicitOscillatorySolution}), was obtained by following 
the solution in sector 3:
\begin{equation}
	\Psi_B(z_{\text{above}})
	=
	iB\Psi_1 + A\Psi_2 + B\Psi_3 +iA\Psi_4,
\nonumber
\end{equation}
across the cut to obtain
\begin{equation}
	\Psi_B(z_{\text{below}})
	=
	B\Psi_3 -iA\Psi_4 -iB\Psi_1 +A\Psi_2,
\nonumber
\end{equation}
and then explicitly evaluating this at $z_{\text{below}}=y$.

The prefactors
\begin{equation}
	\left(
				\frac{\partial E }
		    		     {\partial p_y}
	\right)^{-1/2}_{p_y^\beta}
	\left(
		\begin{array}{c}
			u^\beta_{0,I}(y)e^{+i\phi/2} \\
			v^\beta_{0,I}(y)e^{-i\phi/2}		
		\end{array}
	\right),
\nonumber
\end{equation}
in the regions $y<y_-$ and $y_+<y$ are complex and can be shown to be
\begin{equation}
	\left(
				\frac{\partial E }
		    		     {\partial p_y}
	\right)^{-1/2}_{p_y^\beta}
	\left(
		\begin{array}{c}
			u^\beta_{0,I}(y)e^{+i\phi/2} \\
			v^\beta_{0,I}(y)e^{-i\phi/2}		
		\end{array}
	\right)
	=
	a(y)
	\left(
		\begin{array}{c}
			e^{+i\beta \delta(y)+i\phi/2} \\
			e^{-i\beta \delta(y)-i\phi/2}
		\end{array}
	\right)
	e^{+i\beta \theta(y)-i\phi/4},
\label{eqn:ComplexPrefactors}
\end{equation}
where
\begin{align}
	a(y)
	=\ &
	\frac{1}{\sqrt{2}}
	\left(
		\frac{m^{1/2}}
		     {  \left(
				p_F^2 -p_\bot^2
			\right)^{1/4}
		     }
	\right)
	\left(
		1- \frac{E^2}
			{|\Delta(y)|^2}
	\right)^{-1/4}
	\left(
		1+\frac{|\Delta(y)|^2-E^2}
		       {\epsilon_\bot^2}
	\right)^{-1/8}\!\!\!,
\nonumber \\
	\theta(y)
	=\ &
	-\frac{1}{4}
	\arctan \frac{
			\sqrt{|\Delta(y)|^2-E^2}
		     }
		     { \epsilon_\bot },
\nonumber \\
	\delta(y)
	=\ &
	\frac{1}{2}
	\arctan \frac{
			\sqrt{|\Delta(y)|^2-E^2}
		     }
		     { E }.
\nonumber
\end{align}
%--------------------------------------------------------------------------
\section[Limiting behaviour of the wave functions]{Limiting behaviour of both semiclassical wave functions and the
exact solution at the origin of a vortex}
\label{sec:LimitingBehaviour}
We now show that the limiting, $r \rightarrow 0$, behaviour of both the
semiclassical wave function presented in section 
\ref{sec:SemiclassicalFirstOrderQuantities} and the wave function of the
effective semiclassical theory derived in section 
\ref{sec:EffectiveSemiclassicalFirstOrderTheory}
agree with the known exact solution of the Bogoliubov-de Gennes equation
at the origin of a s-wave vortex.
\subsection{Asymptotic behaviour of the exact solution}
It is well known ~\cite{degennes:89:0} that in the $r \rightarrow 0$
limit the BdG equations can be solved exactly. 
Writing the wave function in the form
\begin{equation}
	\left(
		\begin{array}{c}
			u_{\bf I} ({\bf r}) \\
			v_{\bf I} ({\bf r}) 
		\end{array}
	\right)
	=
	\hat{f}(r) e^{ik_z z}e^{i\mu \theta} e^{-i\sigma_z \theta/2},
\nonumber 
\end{equation}
the BdG equations are reduced to a radial equation for $\hat{f}(r)$:
\begin{equation}
	\sigma_z
	\frac{-\hbar^2}{2m}
	\left(
		\frac{1}{r}\frac{\partial}{\partial r}
		r \frac{\partial}{\partial r}
		-
		\frac{1}{r^2}
		\left(
			\mu 
			-
			\sigma_z
			\frac{1}{2}
		\right)^2
		+
		k_\rho^2
		+
		\sigma_z \frac{2mE}{\hbar^2}
	\right)
	\hat{f}(r)
	=
	0,
\label{eqn:Caroli_et_alrZeroLimit}
\end{equation}
(since $|\Delta(r)| \rightarrow 0$ as $r \rightarrow 0$.) 
Here $k_\rho^2=k_F^2-k_z^2$, $k_F$ the Fermi wave vector, and $\sigma_z$
is a Pauli spin matrix.
This equation has the $J_{\nu}(z)$ Bessel functions as solutions:
\begin{equation}
	\hat{f}(r)
	=
	\left(
		\begin{array}{c}
			A_+ J_{\mu - 1/2} \left(
							(k_\rho + q)r
					  \right)
			\\
			A_- J_{\mu + 1/2} \left(
							(k_\rho - q)r
					  \right)
		\end{array}
	\right),
\label{eqn:BesselSolutions}
\end{equation}
where, $\mu\pm1/2$ are integers, and following de Gennes et al, 
we use $k^2_\rho \pm 2mE/\hbar^2 \approx
k^2_\rho
\left( 
	1 \pm q/k_\rho
\right)^2 $, with $q=2mE/\hbar^2 k_\rho$.
Using the asymptotic form ~\cite{abramowitz:70:0} for the 
$J_\nu(z)\sim \left(\frac{1}{2}z\right)^{\nu}/\Gamma(\nu+1)$ we see that
the $r \rightarrow 0$ limit for $\hat{f}(r)$ is
\begin{equation}
	\hat{f}(r)
	=
	\left(
	\begin{array}{c}
			A_+^\nu r^\nu
			\\
			A_-^\nu r^{\nu+1}
		\end{array}
	\right),
\label{eqn:BesselAtOrigin}
\end{equation}
where $\nu=\mu-1/2$, i.e. there is integer power law decay of the 
particle and hole wave functions approaching the origin. 
%-------------------------subsection---------------------------
\subsection{Asymptotic behaviour of the semiclassical wave functions}
Our two semiclassical wave functions are:
\begin{multline}
	\left(
		\begin{array}{c}
			u_{\bf I} ({\bf r}) \\
			v_{\bf I} ({\bf r})
		\end{array}
	\right)_{\!\!\alpha}
	=
	\sum_j
	A^\alpha_j
	\left|
		\det
		\frac{ 
			\partial^2 S_0^{\alpha,j}({\bf r},{\bf I}) 
		     }
		     {
			\partial {\bf I} \partial {\bf r}
		     }
	\right|^{1/2}
	\left(
		\begin{array}{c}
			u^{\alpha,j}_{0,{\bf I}}({\bf r})
			e^{+i\phi({\bf r})/2} \\
			v^{\alpha,j}_{0,{\bf I}}({\bf r})
			e^{-i\phi({\bf r})/2}
		\end{array}
	\right)
\\
	\times
	e^{\textstyle
			\frac{i}{\hbar}S^{\alpha,j}_0({\bf r}) 
			+ i S^{\alpha,j}_1({\bf r})
	  }, 
	\quad \hbar \rightarrow 0,
\nonumber 
\end{multline}
and
\begin{multline}
	\left(
		\begin{array}{c}
			u_{\bf I} ({\bf r}) \\
			v_{\bf I} ({\bf r})
		\end{array}
	\right)_{\!\!\alpha}
	=
	\sum_j
	A^\alpha_j
	\left|
		\det
		\frac{ 
			\partial^2 S^{\alpha,j}({\bf r},{\bf I}) 
		     }
		     {
			\partial {\bf I} \partial {\bf r}
		     }
	\right|^{1/2}
	\left(
		\begin{array}{c}
			u^{\alpha,j}_{0,{\bf I}}({\bf r};\hbar)
			e^{+i\phi({\bf r})/2} \\
			v^{\alpha,j}_{0,{\bf I}}({\bf r};\hbar)
			e^{-i\phi({\bf r})/2}
		\end{array}
	\right)
\\
	\times
	e^{\textstyle
			\frac{i}{\hbar}S^{\alpha,j}({\bf r};\hbar)
	  }, 
	\quad \hat{{\bf p}} \rightarrow 0.
\nonumber
\end{multline}
For the single vortex problem these become:
\begin{align}
	\left(
		\begin{array}{c}
			u_{\bf I} ({\bf r}) \\
			v_{\bf I} ({\bf r})
		\end{array}
	\right)
	&=
	\sum_j
	\hat{f}^j(r)
	e^{
		+ ik_z z + i \mu \theta -i\sigma_z \theta
	  }, 
	\quad \hbar \rightarrow 0,
\nonumber \\
	\left(
		\begin{array}{c}
			u_{\bf I} ({\bf r}) \\
			v_{\bf I} ({\bf r})
		\end{array}
	\right)
	&=
	\sum_j
	\hat{f}^j(r;\hbar)
	e^{
		+ ik_z z + i \mu \theta -i\sigma_z \theta
	  }, 
	\quad \hat{{\bf p}} \rightarrow 0.
\nonumber
\end{align}
with
\begin{align}
	\hat{f}^j(r) 
	& =
	\frac{A^j}
	     {
		\sqrt{ 
			r\frac{\partial E_0({\bf p},r)}{\partial r } 
		     }
	     }
	\left(
		\begin{array}{c}
			u^{j}_{0,{\bf I}}(r)
			\\
			v^{j}_{0,{\bf I}}(r)
		\end{array}
	\right)
	e^{
			\frac{i}{\hbar}S_0^{j}(r)
			+iS^j_1(r)
	  }, 
\label{eqn:RtoZeroSemiclassical} \\
	\hat{f}^j(r;\hbar) 
	& =
	\frac{A^j}
	     {
		\sqrt{ 
			r\frac{\partial E({\bf p},r)}{\partial r }
		     }
	     }
	\left(
		\begin{array}{c}
			u^{j}_{0,{\bf I}}(r;\hbar)
			\\
			v^{j}_{0,{\bf I}}(r;\hbar)
		\end{array}
	\right)
	e^{
			\frac{i}{\hbar}S^{j}(r;\hbar)
	  }.
\label{eqn:RtoZeroEffective}
\end{align}
Let us investigate these in turn.
%-------------subsubsection-----------------
\subsubsection{The $r\rightarrow 0$ form for $\hat{f}^j(r)$}
The spinor amplitudes in (\ref{eqn:RtoZeroSemiclassical}) are given by 
equation (\ref{eqn:NormalisedAmplitudes}). Expanding as
$r \rightarrow 0$, and noting that ${\bf v}_0=e{\bf A}/m=0$, we obtain
\begin{align}
	\left(
		\begin{array}{c}
			u^+_{0,{\bf I}}(r)
			\\
			v^+_{0,{\bf I}}(r)
		\end{array}
	\right)
	=
	\left(
		\begin{array}{c}
			1 \\ 0
		\end{array}
	\right)
	+ {\cal{O}}(r)
	\left(
		\begin{array}{c}
			0 \\ 1
		\end{array}
	\right),
\nonumber \\
	\left(
		\begin{array}{c}
			u^-_{0,{\bf I}}(r)
			\\
			v^-_{0,{\bf I}}(r)
		\end{array}
	\right)
	=
	\left(
		\begin{array}{c}
			0 \\ 1
		\end{array}
	\right)
	+ {\cal{O}}(r)
	\left(
		\begin{array}{c}
			1 \\ 0
		\end{array}
	\right).
\nonumber
\end{align}
(Here we have assumed $|\Delta(r)|\propto r$.) 
We also have:
\begin{align}
	p_r^\pm(r)
	&=
	\sqrt{
		p_F^2 -p_z^2 -\frac{\hbar^2 \mu^2}{r^2}
		\pm 2m \sqrt{
				E^2
				-
				|\Delta(r)|^2
			    }
	     } \ ,
\nonumber \\
	&=
	i\hbar\frac{|\mu|}{r}
	+
	{\cal{O}}(r),
\label{eqn:SemiclassicalPrExpanded}
\end{align}
from which it follows that
\begin{equation}
	\frac{1}
	     {
		\sqrt{ 
			r\frac{\partial E_0({\bf p},r)}{\partial r } 
		     }
	     }
	=
	\frac{e^{-i\pi/4}}
	     {
		(\hbar\mu)^{1/2}
	     }
	+
	{\cal{O}}(r^2),
\nonumber 
\end{equation}
and since
\begin{equation}
	\frac{i}{\hbar}\int^{r_{a,b}}_r p_r^\pm(r)dr
	=
	\ln \left(\frac{r}{r_{a,b}}\right)^{|\mu|}
	+
	{\cal{O}}(r^2),
\nonumber 
\end{equation}
we have
\begin{equation}
	e^{
		\frac{i}{\hbar}S^\pm_0(r)
	  }
	=
	\left(
		\frac{r}{r_{a,b}}
	\right)^{|\mu|}
	+
	{\cal{O}}(r^{|\mu|+2}).
\nonumber
\end{equation}
In the last two equations the phase reference point $r_a$ ($r_b$) 
is chosen to be the classical turning point defined by $p^+(r_a)=0$
($p^-(r_b)=0$).
We have not yet calculated $S_1(r)$, but if we assume it can be
neglected the appropriate superposition as
$r \rightarrow 0$, call it $\hat{f}_\hbar(r)$, takes the form
\begin{equation}
	\hat{f}_\hbar(r)
	=
	\frac{e^{-i\pi/4}}
	     {
		(\hbar\mu)^{1/2}
	     }
	\left(
		\begin{array}{c}
			A^+ (r/r_a)^{\nu+1/2}
			\\ 
			A^- (r/r_b)^{\nu+1/2}
		\end{array}
	\right)
	+
	{\cal{O}}(r^{\nu +5/2}).
\nonumber
\end{equation}
In particular we note the {\it half} integer power law decay of the 
particle and hole wave functions in contrast to the exact result 
(\ref{eqn:BesselAtOrigin}). This is completely unsatisfactory. One then
hopes that $S_1^j(r)$ corrects this deficency. That this is indeed the case
can be shown since for this problem $S_1^j(r)$ can be calculated. Using
$-e^{-1}\text{j}_\theta(r)=p_\theta/mr$
We have:
\begin{align}
	iS_1^\pm(r)
	&= 
	\int^{t_{a,b}}_t 
	\frac{ip_\theta}{2mr^2} dt,
\nonumber \\
	&=
	\int^{r(t_{a,b})}_{r(t)} 
	\frac{p_\theta}{2mr^2} \frac{dt}{dr^\pm}dr^\pm,
\nonumber 
\end{align}
and using $\frac{dr^\pm}{dt}=\pm (ip_\theta/mr)(1 +{\cal{O}}(r^2))$
\begin{align}
	iS_1^\pm(r)
	&=
	\ln \left(
		   \frac{r}{r_{a,b}}
	    \right)^{\mp 1/2}
	+{\cal{O}}(r^2),
\nonumber \\
	\Rightarrow
	e^{iS_1^\pm(r)}
	&=
	\left(
		   \frac{r}{r_{a,b}}
	\right)^{\mp 1/2}
	+ {\cal{O}}(r^{2\mp 1/2}).
\label{eqn:SemiclassicalAtOrigin}
\end{align}
Then $\hat{f}_\hbar(r)$ becomes
\begin{equation}
	\hat{f}_\hbar(r)
	=
	\frac{e^{-i\pi/4}}
	     {
		(\hbar\mu)^{1/2}
	     }
	\left(
		\begin{array}{c}
			A^+ (r/r_a)^{\nu}
			\\ 
			A^- (r/r_b)^{\nu+1}
		\end{array}
	\right)
	+
	{\cal{O}}(r^2).
\nonumber
\end{equation}
which has the correct integer power law decay as required. By following 
through this analysis we have discovered the importance of the first order
phase, $S_1^j(r)$. We have also checked that the semiclassical theory 
carried out correctly gives not only a quantisation rule but also the wave
function as well. Does our effective semiclassical theory do as well?
%----------------subsubsection--------------------------------
\subsubsection{The $r\rightarrow 0$ form for $\hat{f}^j(r;\hbar)$}
The spinor amplitudes in (\ref{eqn:RtoZeroEffective}) are given by 
equation (\ref{eqn:NormalisedAmplitudesEffective}). Expanding these as
$r \rightarrow 0$ we obtain
\begin{align}
	\left(
		\begin{array}{c}
			u^+_{0,{\bf I}}(r;\hbar)
			\\
			v^+_{0,{\bf I}}(r;\hbar)
		\end{array}
	\right)
	=
	\left(
		\begin{array}{c}
			1 \\ 0
		\end{array}
	\right)
	+ {\cal{O}}(r^3)
	\left(
		\begin{array}{c}
			0 \\ 1
		\end{array}
	\right),
\nonumber \\
	\left(
		\begin{array}{c}
			u^-_{0,{\bf I}}(r;\hbar)
			\\
			v^-_{0,{\bf I}}(r;\hbar)
		\end{array}
	\right)
	=
	\left(
		\begin{array}{c}
			0 \\ 1
		\end{array}
	\right)
	+ {\cal{O}}(r^3)
	\left(
		\begin{array}{c}
			1 \\ 0
		\end{array}
	\right).
\nonumber
\end{align}
The $r^3$ rather than $r$ correction arises since ${\bf v}_s=-\hbar/2mr$ so
that ${\bf p}\cdot{\bf v}_s=-\hbar p_\theta/2mr^2$. 

Now
\begin{align}
	p_r^\pm(r;\hbar)
	&=
	\sqrt{
		p_F^2 -p_z^2 -\frac{\hbar^2 (\mu^2 +1/4)}{r^2}
		\pm 2m \sqrt{
				\left( E +\frac{\hbar^2 \mu}{2m r^2} \right)^2
				-
				|\Delta(r)|^2
			    }
	     },
\nonumber \\
	&=
	i\hbar\frac{|\mu \mp 1/2|}{r}
	+
	{\cal{O}}(r).
\label{eqn:EffectivePrExpanded}
\end{align}
Notice the appearance of $|\mu \mp 1/2|$ here rather than $|\mu|$ in 
equation (\ref{eqn:SemiclassicalPrExpanded}). It follows 
from this that
\begin{equation}
	\frac{1}
	     {
		\sqrt{ 
			r\frac{\partial E({\bf p},r)}{\partial r } 
		     }
	     }
	=
	\frac{e^{-i\pi/4}}
	     {
		\hbar^{1/2}(\mu \mp 1/2)^{1/2}
	     }
	+
	{\cal{O}}(r^2),
\nonumber 
\end{equation}
and 
\begin{equation}
	e^{
		\frac{i}{\hbar}S^\pm(r)
	  }
	=
	\left(
		\frac{r}{r_{a,b}}
	\right)^{|\mu \mp 1/2|}
	+
	{\cal{O}}(r^{|\mu \mp 1/2|+2}).
\nonumber
\end{equation}
Thus the appropriate combination, 
call it $\hat{f}^j_{\hat{\bf p}}(r;\hbar)$, takes the form
\begin{equation}
	\hat{f}_{\hat{\bf p}}(r;\hbar)
	=
	\frac{e^{-i\pi/4}}
	     {
		(\hbar\mu)^{1/2} 
	     }
	\left(
		\begin{array}{c}
			A^+ \frac{(\nu-1/2)}{\nu}  (r/r_a)^{\nu}
			\\ 
			A^- \frac{(\nu+1/2)}{\nu} (r/r_b)^{\nu+1}
		\end{array}
	\right)
	+
	{\cal{O}}(r^2).
\nonumber
\end{equation}
which has the correct integer power law decay agreeing with the previous two 
solutions. In the present case the inclusion of the $1/4$ in equation
(\ref{eqn:EffectivePrExpanded}) was crucial to obtain the dependence
upon $|\mu \pm 1/2|$. This $1/4$ appears in the $p_r^\pm$ due to the 
inclusion in the Hamiltonian, equation (\ref{eqn:SingleVortexHamiltonian}),
of $\hbar^2$ terms. This confirms that our procedure, differing from that
proposed by Littlejohn and Flynn ~\cite{littlejohn:91:0}, is the correct
one for a semiclassical theory of superconductors.

We have successfully verified both versions of the semiclassical
wave function derived in this paper.

%--------------------------------------------------------------------------

\begin{acknowledge}
We would like to thank Michael Berry and Chris Howls for advice concerning
Stokes phenomenon, and John Hannay for many enlightening discussions about the 
semiclassical theory. This work was supported by EPSRC, grant No. GR/M53844.
\end{acknowledge}
%\begin{noteinproof}
%A note added in proof, if there is one, should be the final text before the references.
%\end{noteinproof}
%\bibliography{SemiclassicsForScBib}

\begin{thebibliography}{10}

\bibitem{degennes:89:0}
P.~G. {de Gennes}.
\newblock {\em Superconductivity of Metals and Alloys}.
\newblock Addison-Wesley, 1989.

\bibitem{andreev:64:0}
A.~F. Andreev.
\newblock {\em Sov. Phys. JETP}, 19(5):1228--1238, 1964.

\bibitem{andreev:66:0}
A.~F. Andreev.
\newblock {\em Sov. Phys. JETP}, 22(2):455--458, 1966.

\bibitem{azbel:71:0}
M.~{Ya. Azbel'}.
\newblock {\em Sov. Phys. JETP}, 32(1):159, 1971.

\bibitem{bardeen:69:0}
J.~Bardeen, R.~{K\"ummel}, A.E. Jacobs, and L.~Tewordt.
\newblock {\em Phy. Rev.}, 187(2), 1969.

\bibitem{gutzwiller:90:0}
M.~C. Gutzwiller.
\newblock {\em Chaos in Classical and Quantum Mechanics}.
\newblock Springer-Verlag, 1990.

\bibitem{ozorio:88:0}
A.~M. {Ozorio de Almeida}.
\newblock {\em Hamiltonian Systems: Chaos and Quantisation}.
\newblock Cambridge University Press, 1988.

\bibitem{brack:97:0}
M.~Brack and R.~K. Bhaduri.
\newblock {\em Semiclassical Physics}.
\newblock Addison-Wesley, 1997.

\bibitem{maslov:81:0}
V.~P. Maslov and M.~V. Fedoriuk.
\newblock {\em Semi-Classical Approximation in Quantum Mechanics}.
\newblock Riedel Publishing Company, 1981.

\bibitem{vafek:00:0}
O.~Vafek, A.~Melikyan, M.~Franz, and Z.~Te$\check{\text{s}}$anovi\'c.
\newblock {\em Phy. Rev. B}, 63, 2000.

\bibitem{pismen:99:0}
L.~M. Pismen.
\newblock {\em Vortices in Nonlinear Fields}.
\newblock Oxford University Press, 1999.

\bibitem{anderson:99:0}
P.~W. Anderson.
\newblock cond-mat/9812063.

\bibitem{arnold:89:0}
V.~I. Arnold.
\newblock {\em Mathematical methods of classical mechanics}.
\newblock Springer, 2nd edition, 1989.

\bibitem{vanvleck:28:0}
J.~H. van Vleck.
\newblock {\em Proc. Nat. Acad. Sci. USA}, 14(2):178--188, 1928.

\bibitem{berry:84:0}
M.~V. Berry.
\newblock {\em Proc. R. Soc. Lond A}, 392, 1984.

\bibitem{robbins:97:0}
J.~M. Robbins.
\newblock {\em Encyclopedia of Applied Physics}, 21:549--584, 1997.

\bibitem{littlejohn:91:0}
R.~G. Littlejohn and W.~G. Flynn.
\newblock {\em Phys. Rev. A}, 44(8):5239--5256, 1991.

\bibitem{yabana:86:0}
K.~Yabana and H.~Horiuchi.
\newblock {\em Prog. Theor. Phys.}, {{\bf 75}}:592--618, 1986.

\bibitem{emmrich:96:0}
C.~Emmrich and A.~Weinstein.
\newblock {\em Comm. Math. Phys.}, 176:701, 1996.

\bibitem{goldstein:80:0}
H.~Goldstein.
\newblock {\em Classical Mechanics}.
\newblock Addison-Wesley, 2 edition, 1980.

\bibitem{bolte:99:0}
J.~Bolte and S.~Keppeler.
\newblock {\em Ann. Phys. (N.Y)}, 274:125, 1999.

\bibitem{dirac:31:0}
P.~A.~M. Dirac.
\newblock {\em Proc. R. Soc. A}, 133:60--72, 1931.

\bibitem{berry:80:0}
M.V.Berry.
\newblock {\em Eur. J. Phys}, 1:240--244, 1980.

\bibitem{kulik:70:0}
I.~O. Kulik.
\newblock {\em Sov. Phys. JETP}, 30(5):944--950, 1970.

\bibitem{bardeen:72:0}
J.~Bardeen and J.~L. Johnson.
\newblock {\em Phys. Rev. B}, 5(1):72--78, 1972.

\bibitem{sipr:97:0}
{O. $\check{\text{S}}$ipr and B. L. Gy\"orffy}.
\newblock {\em J. Low Temp. Phys.}, 106(3/4):315, 1997.

\bibitem{abrikosov:88:0}
A.~A. Abrikosov.
\newblock {\em Fundamentals of the Theory of Metals}.
\newblock North-Holland, 1988.

\bibitem{blonder:82:0}
G.~E. Blonder, M.~Tinkham, and T.~M. Klapwijk.
\newblock {\em Phys. Rev. B}, 25(7), 1982.

\bibitem{caroli:64:0}
C.~Caroli, P.~G. {de Gennes}, and J.~Matricon.
\newblock {\em Phys. Letters}, 9:307, 1964.

\bibitem{gygi:91:0}
F.~Gygi and M.~Schl\"uter.
\newblock {\em Phys. Rev. B}, 43(10), 1991.

\bibitem{duncan:98:0}
K.~P. Duncan and B.~L. Gy\"orffy.
\newblock unpublished, 1998.

\bibitem{duncan:99:0}
K.~P. Duncan.
\newblock {\em The semiclassical theory of the de Haas-van Alphen oscillations
  in type-II superconductors}.
\newblock PhD thesis, University of Bristol, 1999.

\bibitem{duncan:2002:0}
K.~P. Duncan and B.~L. {Gy\"orffy}.
\newblock to be published.

\bibitem{altland:00:0}
A.~Altland, B.~Simons, and D.~{Taras-Semchuk}.
\newblock {\em Adv. Phys.}, 49(3):321--394, 2000.

\bibitem{heading:62:0}
J.~Heading.
\newblock {\em An introduction to phase integral methods}.
\newblock London:Methuen, 1962.

\bibitem{abramowitz:70:0}
M.~Abramowitz and I.~A. Stegun, editors.
\newblock {\em Handbook of mathematical functions}.
\newblock Dover, 1970.

\end{thebibliography}

\end{document}